\documentclass[twocolumn]{aastex631}

\usepackage{natbib}
\usepackage{enumitem}
\usepackage{soul}

\defcitealias{peroux2020}{PH20}
\defcitealias{madau2014}{MD14}
\defcitealias{lehner2022}{L22}
\defcitealias{bellstedt2020}{B20}
\defcitealias{lehner2016}{L16}
\usepackage{amsmath}
\usepackage{appendix}

\newcommand{\HI}{\ensuremath{\mbox{\ion{H}{1}}}}
\newcommand{\HII}{\ensuremath{\mbox{\ion{H}{2}}}}

\newcommand{\OVI}{\ensuremath{\mbox{\ion{O}{6}}}}

\newcommand{\CII}{\ensuremath{\mbox{\ion{C}{2}}}}
\newcommand{\CIV}{\ensuremath{\mbox{\ion{C}{4}}}}
\newcommand{\MgII}{\ensuremath{\mbox{\ion{Mg}{2}}}}

\newcommand{\NHI}{\ensuremath{N(\mbox{\ion{H}{1}})}\relax}
\newcommand{\logNHI}{\ensuremath{\log N_{\rm{HI}}}}

\newcommand{\z}{\ensuremath{z}}

\newcommand{\column}{cm$^{-2}$}

\newcommand{\kms}{\ensuremath{{\rm km\,s}^{-1}}}

\newcommand{\lstar}{\ensuremath{L^*}}

\shorttitle{A Global Census of Metals in the Universe}
\shortauthors{Deepak et al.}

\begin{document}

\title{A Global Census of Metals in the Universe}

\author[0000-0003-4203-6223]{Saloni Deepak}
\affiliation{Department of Physics and Astronomy, University of Notre Dame, Notre Dame, IN 46556}

\author[0000-0002-2591-3792]{J. Christopher Howk}
\affiliation{Department of Physics and Astronomy, University of Notre Dame, Notre Dame, IN 46556}

\author[0000-0001-9158-0829]{Nicolas Lehner}
\affiliation{Department of Physics and Astronomy, University
of Notre Dame, Notre Dame, IN 46556}

\author[0000-0002-4288-599X]{C\'eline P\'eroux}
\affiliation{European Southern Observatory, Karl-Schwarzschild-Str. 2, 85748 Garching-bei-M\"unchen, Germany; email: cperoux@eso.org }
\affiliation{Aix Marseille Universit\'e, CNRS, LAM (Laboratoire d'Astrophysique de Marseille) UMR 7326, 13388, Marseille, France}

\begin{abstract}
We present a census of the mass density of metals and their evolution with cosmic time on a global scale throughout the Universe, synthesizing robust estimates of metals in stars, hot intra-cluster gas, and gaseous absorbers tracing neutral gas as well as ionized gas in the circumgalactic and intergalactic media. We observe an order of magnitude increase in the stellar metal mass density from $z\sim2.5$ to 0.7, over which time stars emerge as the most important metal reservoir at low redshifts, housing $\sim30\%$ of the total expected metal density at $z\sim0.1$. Hot virialized intracluster/intragroup gas accounts for $\sim15\%$ and 10\% of metals at $z\sim0.1$ and 0.7, respectively. Using metallicity measurements from CCC, KODIAQ-Z, and HD-LLS surveys covering redshifts $z<1$ to $z\sim2$--3.5, we investigate the global distribution of metals in extragalactic cool ionized gas as a function of \HI\ column density. During the period from $z\approx3$ to $z<1$, the global metal density of cool ($T\sim10^{4-5}$ K) gas has doubled. However, the fractional contribution of the ionized gas to the total expected metal density decreased from $\sim20\%$ at $z\sim3$ to $\sim4\%$ at $z<1$. The cosmic metal density of all gas phases has increased with cosmic time, reflecting an ``inside-out'' metal dispersion by feedback mechanisms and galactic outflows. 

\end{abstract}

\section{Introduction}
\label{sec:intro}

The Universe is composed of three main constituents---baryonic or visible matter ($\sim4.6\%$), dark matter that gravitationally interacts with visible matter ($\sim25\%$), and the most exotic and least understood dark energy ($\sim70\%$) \citep{planck2020}. Together, these components account for almost all the mass-energy budget of the Universe. Although baryons form the smallest fraction of the total mass, their interactions with electromagnetic radiation allow us to study their physical properties extensively. 

Baryons play a crucial role in the formation and evolution of all large-scale structures in the Universe. As the Universe underwent inflation, primordial density fluctuations led to the formation of dark matter halos, which gravitationally confined baryonic matter in the form of gas \citep{black1981,ostriker1986,bromm2009,bromm2011}. In regions where radiative cooling was efficient, these baryons condensed to form proto-galactic halos, while, in regions of inefficient cooling, they coalesced into hot halos \citep{cen2006}. Through accretion and mergers, these halos evolved into galaxies \citep{mo2010,romeel2010,mcquinn2016}. The halos are connected by cosmic filaments and sheets of collapsed baryons that form the cosmic web -- the intergalactic medium within which halos are embedded \citep{lanzetta1995,bond1996,tejos2012,tejos2016,chen2017,burchett2020}. Within galaxies, the same physical processes (gravitational instabilities and radiative cooling) cause baryons to condense and form stars, where metals are eventually synthesized \citep{glover2012,girichidis2020}. Subsequently, these stars die, and feedback from supernovae and outflows drives the dynamics and chemical evolution of the multi-phase gas in the interstellar, circumgalactic and, intergalactic media. The processes that lead to the formation of structures in the Universe also regulate the cycling of gas between the different phases -- cold molecular, cool atomic (neutral and ionized), hot ionized, etc \citep{draine2011,neeleman2015,klessen2016}. Thus, by tracing the global evolution of the physical properties of baryons, we gain insight into the dominant processes that govern the astrophysics of galaxies and how they change over time. Baryons are therefore the key drivers of galaxy evolution, influencing the structure, composition, and overall dynamics of galaxies throughout cosmic time. 

One important characteristic of baryonic matter is the degree to which they are enriched by metals. The production of metals is intimately connected to the formation and subsequent death of stars. \citet{peeples2014} note that a majority of metals are produced within stellar cores via nucleosynthesis or by supernova events at the end of a star's lifetime (80--85\% of metals by mass). A significant fraction is also produced in degenerate supernovae, neutron star mergers, and asymptotic giant branch (AGB) stars ($\sim10\%$). Some synthesized metals are forever trapped in stellar remnants, including white dwarfs, neutron stars, and black holes \citep{venkatesan1999,fields2000}. Of the metals released by stars, their path to inclusion in subsequent generations of stars can be circuitous. While some are rapidly mixed into the galactic interstellar medium (ISM), a substantial portion is expelled from the star-forming regions of galaxies by feedback from stellar winds, supernovae, and active galactic nuclei (AGN) in the form of galactic outflows \citep{kirkpatrick2011,choi2017,choi2020,sharda2024}. These metals are dispersed into the extended reservoirs of gas in the outer regions of galaxies---the circumgalactic medium (CGM) (e.g., \citealt{meiring2009,tumlinson2011Sci,tumlinson2013,werk2014,crighton2015,quiret2016,tumlinson2017,lehner2018a,qu2022,qu2023,chen2024}), or even into the intergalactic medium (IGM) (e.g., \citealt{simcoe2002,simcoe2004,aguirre2004,cooksey2013,savage2014,shull2014,kim2021,dodorico2022}). Some of these dispersed metals may return to a galaxy to enrich star-forming regions billions of years later \citep{angles-alcazar2017}.

Since metals are mostly produced in stars within galaxies, any other metal-enriched material must be polluted by feedback-driven outflows. In other words, the presence of metals---or lack thereof---offers quantitative and qualitative indications of the degree to which a parcel of gas has previously interacted with a galaxy (modulated by factors such as phase change, metal survival, distance to the galaxy, and overdensity). Ultimately, the extent of metal enrichment in cosmic gas can be used to test theories of star formation and feedback in galaxies. On a global scale, the relative chemical enrichment of diverse metal reservoirs---such as stars, the dense star-forming gas in galaxies, and the diffuse gas surrounding galaxies---can be used to trace the circulation of matter as part of the cosmic baryon cycle that drives galaxy evolution. 

Constraining the global cycling of metals in the Universe requires compiling a comprehensive metal budget. Early studies in this area identified a ``missing metals problem'' \citep{pettini1999,pagel1999}, where the total cataloged metal density of the Universe at $z\sim2.5$ was an order of magnitude smaller than estimates of the total metal density formed by stars \citep{prochaska2003}. Subsequent updates to this budgetary shortfall were presented by \cite{ferrara2005}, \cite{bouche2005, bouche2006, bouche2007}, and \cite{shull2014}. Many of these works suggested that the bulk of metals produced in galaxies at $z\sim2$ are dispersed into the diffuse, ionized gas of the CGM and IGM \citep{bouche2007,peeples2014,lehner2014, shull2014}. Our current understanding of the degree of this disparity is limited by two main factors: the relatively large uncertainties inherent in the current nucleosynthetic yields \citep{peeples2014}, and the uncertainties in the amount of metals in the photoionized and collisionally ionized gas \citep{lehner2014,lehner2022,fumagalli2016}.

A more recent census of global metals was conducted by \cite{peroux2020} (hereafter \citetalias{peroux2020}), emphasizing the redshift evolution of the metal budget. \citetalias{peroux2020} showed a majority of the metals produced by stars have now been cataloged in many epochs of the Universe. This is especially true at high redshifts ($z>3$), where \citetalias{peroux2020} find that virtually all the expected metals are found in cold neutral gas. By contrast, the metals at low redshift ($z<1$) are distributed among a diverse set of repositories, with significant contributions from long-lived stars and the intra-cluster medium of massive galaxy clusters. While a significant fraction of metals remains unaccounted for at these low redshifts, several works have suggested that they may plausibly be contained within the highly-ionized, warm-hot ($T\sim10^5$--$10^7$ K) circumgalactic gas or in metals distributed in the IGM traced by the Lyman $\alpha$-forest in quasar spectra \citep{songaila2001, songaila2005,fox2011,anderson2013,lehner2014,peeples2014}. One of the reasons to reappraise the \citetalias{peroux2020} survey is that they focus on those contributors for which robust estimates of the uncertainties could be made at the time. Thus, they did not wholly include contributions from circumgalactic gas at $z<1$ and $z\sim3$, since a thorough assessment of the uncertainties of their metal content was beyond the scope of \citetalias{peroux2020}.

We expand on the global-scale metal budget of the Universe compiled by \citetalias{peroux2020} by incorporating new, up-to-date data from the literature, providing robust uncertainty estimates for metal densities in major reservoirs, and constraining the metal content in the cool CGM and IGM. We prioritize robustness over completeness, excluding potentially significant metal reservoirs when their metal densities are poorly constrained (absence of a large statistical sample or issues relating to modeling the physics) or are prone to biases or are not amenable to rigorous uncertainty analysis. We also pay particular attention to avoiding ``double counting'' problems that might bias our assessments of the global metal budget. That said, we do note best estimates and related references for these reservoirs (see Section \ref{subsec:others}).

Our work is presented as follows. In Section~\ref{sec:analysis}, we describe the reservoirs of metals included in our census. In Section~\ref{subsec:stars}, we describe our calculations for the global metal content of stars.  Section~\ref{subsec:icmigrm} and~\ref{sec:coldgas} detail our calculations of the global metal density estimates for hot ($T\ge10^6$ K) virialized gas and cool ($T\sim10^4$ K) gas, respectively. We discuss other contributions to the global metal budget and the issues associated with their inclusion or exclusion in detail in section~\ref{subsec:others}. We analyze the evolution of the global metal budget and discuss its implications in section~\ref{sec:budget}. Section~\ref{sec:summary} summarizes our work.

We adopt a concordance $\Lambda\rm{CDM}$ cosmology, notably $H_0 = 70.0$ \kms /Mpc, $h_{70}=H_0/70.0$ \kms /Mpc, $\Omega_m = 0.3$ and $\Omega_{\Lambda} = 0.7$. We use the solar abundances from \cite{asplund2021}, in particular, the bulk (proto-solar) mass fraction of heavy elements, $Z_\odot=0.0154$. Wherever applicable, we have used the \cite{chabrier2003} initial mass function. We use comoving coordinates throughout this work. 

\section{Cosmic metal budget}
\label{sec:analysis}
\subsection{Cosmic Metal Inventory}
\begin{figure*}
\begin{centering}
\includegraphics[scale=0.65]{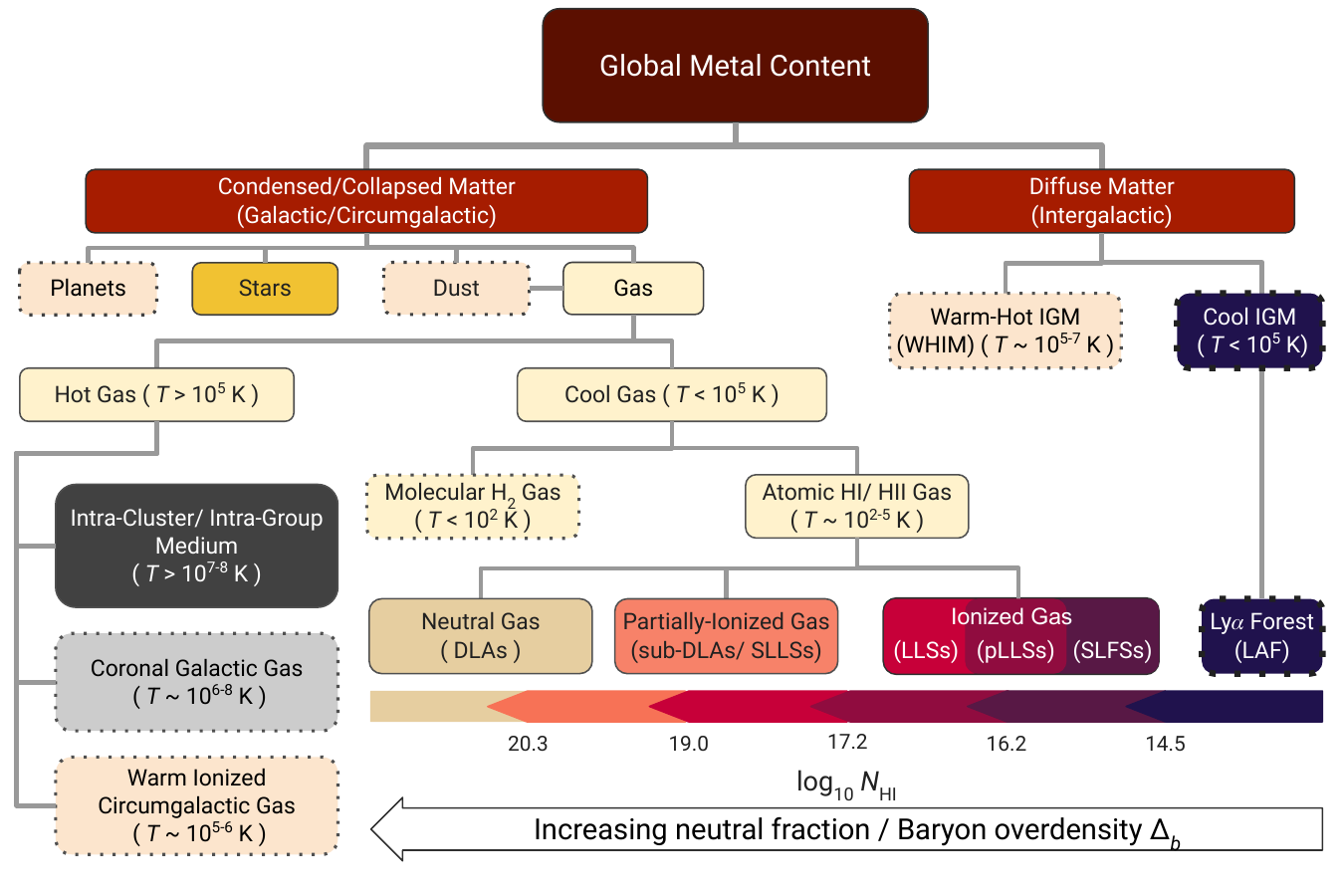}
\par\end{centering}
\begin{centering}
\caption
{A categorization of the various contributions to the global metal budget, based on temperature and density. The contributors that are not included in the census derived in this work are in boxes with dotted borders. Under the ``Cool gas" branch, the gas becomes increasingly neutral with increasing \HI\ column density from right to left as indicated by the color bar. Neutral gas refers to Damped Lyman $\alpha$ (DLA) absorbers, partially-ionized gas refers to sub-DLAs or super Lyman limit systems (SLLSs), and ionized gas includes Lyman limit systems (LLSs), partial LLSs (pLLSs), and strong Lyman forest systems (SLFSs). }
\label{contributors}
\par\end{centering}
\end{figure*}

Our breakdown of the many reservoirs of metals is summarized in Figure~\ref{contributors}. The solid boxes are reservoirs that we include in our census. The dotted boxes indicate contributions that are not included. Broadly, we identify two classes of metal reservoirs: (i) condensed galactic and circumgalactic matter  \citep[e.g.,][]{shull2012},\footnote{This includes galactic and circumgalactic matter---stars, planets, dust, the cool atomic HI gas, and the molecular $\rm{H}_2$ gas. The gas phases probe the interstellar medium and the circumgalactic medium of galaxies.} and (ii) diffuse intergalactic gas. While we have categorized them discretely, the boundaries between these two components are fuzzy and sometimes overlapping. Our calculations focus on the contributions from the galactic and circumgalactic sources. As we aim to comprehend the complex dynamics of the cosmic baryon and metal cycles, condensed matter emerges as the focal point for metal production and redistribution. This is where star formation and, consequently, metal synthesis occurs. We note that at $z\lesssim0.4$, based on previously compiled baryon budgets (\citealt{fukugita2004}; \citealt{shull2012}, \citealt{walter2020}; \citetalias{peroux2020}), $\sim1/5$ of cosmic baryons reside as condensed matter (stars, neutral gas, molecular gas, as well as the CGM of galaxies), while the rest exist as the warm-hot IGM (WHIM) and the cool IGM traced by the Lyman $\alpha$ forest (LAF). Thus, though they are excluded from our census, it is imperative to discuss intergalactic gas in the context of the cosmic baryon and metal cycles, which we do in Section~\ref{subsec:others}. We also discuss the contribution from dust \citep{peroux2023}. The cosmic mass density of planets is negligible compared to cosmic gas densities \citep{fukugita2004}\footnote{The global mass density of planets is $10^{-4}$ times the mass density of the intergalactic medium at $z=0$.}, and we do not discuss it further.

For the hot gas reservoirs in Figure~\ref{contributors}, we identify three contributors based on their temperature and occurrence sites. We discuss these in detail in Sections~\ref{subsec:icmigrm} and~\ref{subsec:others}. Under cool gas, we list molecular hydrogen ($\rm{H}_2$) and atomic hydrogen regimes. The metallicity of the cold molecular phase ($T\lesssim10^2$ K) is difficult to constrain; few metal lines give direct access to the metal content without highly uncertain ionization, fractionation, and dust depletion corrections. Thus, we do not fully include the metals associated with the molecular gas in the Universe in our census, though we do provide an estimate of the metal content for this phase under some limiting assumptions in Section~\ref{subsection:molecular}. The atomic phase spans a wide range in temperature, including the cold neutral medium (CNM) with $T\sim10^{2-3}$ K and the warm neutral medium (WNM) with $T\gtrsim10^{3}$ K \citep{heiles2001,heiles2003,kalberla2018}. Historically, these terms were used for the cool and warm ISM probed using the 21-cm line \citep{davies1975,dickey1990,wolfe2005b,roy2013}.
The predominantly-neutral phases may coexist with cold molecular gas and ionized atomic hydrogen gas in some environments.\footnote{The boundaries between these phases are fuzzy and signify transitions. For example, the diffuse ISM (and DLAs) has both \HI\, and $\rm{H}_2$ with an atomic-to-molecular transition at \HI\/ column density $\sim10^{21}$ $ \rm{cm}^{-2}$ \citep{krumholz2009,draine2011}. Similarly, sub-DLAs or SLLSs are regions of \HI\, to \HII\, transitions.} We categorize the condensed atomic gaseous reservoirs based on their neutral hydrogen (\HI\/) column density (Section~\ref{sec:coldgas}). By relying on \HI-selected absorber surveys (and not metal-selected absorbers or absorbers known to be associated with galaxies), we avoid any bias (concerning luminosity, mass, star formation rate of galaxies, etc.) in deriving the global metal densities. This means that we are agnostic to the environment of the absorbers, such as the ISM (CNM or WNM), CGM, or IGM. It also means that our categorization of the nature of the absorbers is ambiguous: we do not know whether a specific absorber is tracing a galactic disk, an outflowing wind, a diffuse CGM, or a patch of the IGM.

In theory, the various \HI\/ column density regimes can be associated with specific types of regions of the Universe---the CGM, IGM, etc \citep{battisti2012,hafen2017,berg2023,hamanowicz2020}. This classification of gas reservoirs can potentially be related to cosmic environments through cosmological simulations that track the temperature and density evolution of baryons \citep{cen2006,oppenheimer2006,shull2012}. Temperature-overdensity ($T-\delta_b$) phase diagrams show distinct regimes (see Figure 8 in \citealt{oppenheimer2006}): (i) a cool ($T<10^5$ K), diffuse photoionized phase ($\delta_b<10^2$) which may be associated with the LAF and lower \HI\, column density absorbers, (ii) a cool condensed phase ($\delta_b\sim10^{2}$--$10^4$) with overdensities similar to Lyman limit systems and Damped Lyman--$\alpha$ absorbers (DLAs), and (iii) a ``plume" of warm-hot gas ($T>10^5$ K, $\delta_b>10$) that emerges naturally from shock-heating during structure formation. This motivates our categorization of the cool condensed gas, the cool IGM, and the WHIM in Figure~\ref{contributors}.

However, this classification is not consistent globally and the boundaries that demarcate these classes are fuzzy. For example, at high $z$ we often cannot associate gaseous absorbers with galaxies and instead identify them as patchy overdensities in the Universe \citep{nasir2021}. Even at low $z$, similar column density absorbers may have different origins \citep{berg2023,weng2023,weng2023b}. Thus, by grouping absorbers based on the neutral hydrogen column densities instead of their environments, we avoid potential double-counting issues while compiling our metal budget. 

We note in Figure~\ref{contributors} that although strong Lyman forest systems (SLFSs) are classified as condensed, cool circumgalactic gas and LAF is connected to diffuse IGM matter, they are physically very similar, and many SLFSs trace the high-density IGM. Low column density absorbers are also found in the CGM of galaxies \citep{savage2014, Manuwal2021}. DLAs are neutral, but they can also contain a substantial amount of highly ionized gas (traced by \ion{C}{4}, \ion{O}{6}) \citep{fox2007,fox2009,rahmani2010,lehner2014,masribas2017} and occasionally cooler molecular hydrogen gas \citep{balashev2018,balashev2019}. Thus, we outline schematically a general classification scheme based on densities and temperatures, emphasizing the ambiguities in the physical nature of gaseous absorbers.

Our goal is to compile a budget of the metal abundance in and around galaxies as a function of redshift and to show how their distribution changes across a wide range of densities. This includes the metals residing in stars, the metals ejected out into the ISM, as well as those dispersed in the CGM of galaxies and the IGM. The diverse data that we consider for this study have been assimilated from an expansive range of sources, each of which has a different diagnostic or tracer of metallicity. To homogenize these results, we do a case-by-case study of the different components in Sections \ref{subsec:stars}--\ref{subsec:others}.

\begin{figure}
\begin{centering}
\includegraphics[width=\columnwidth]{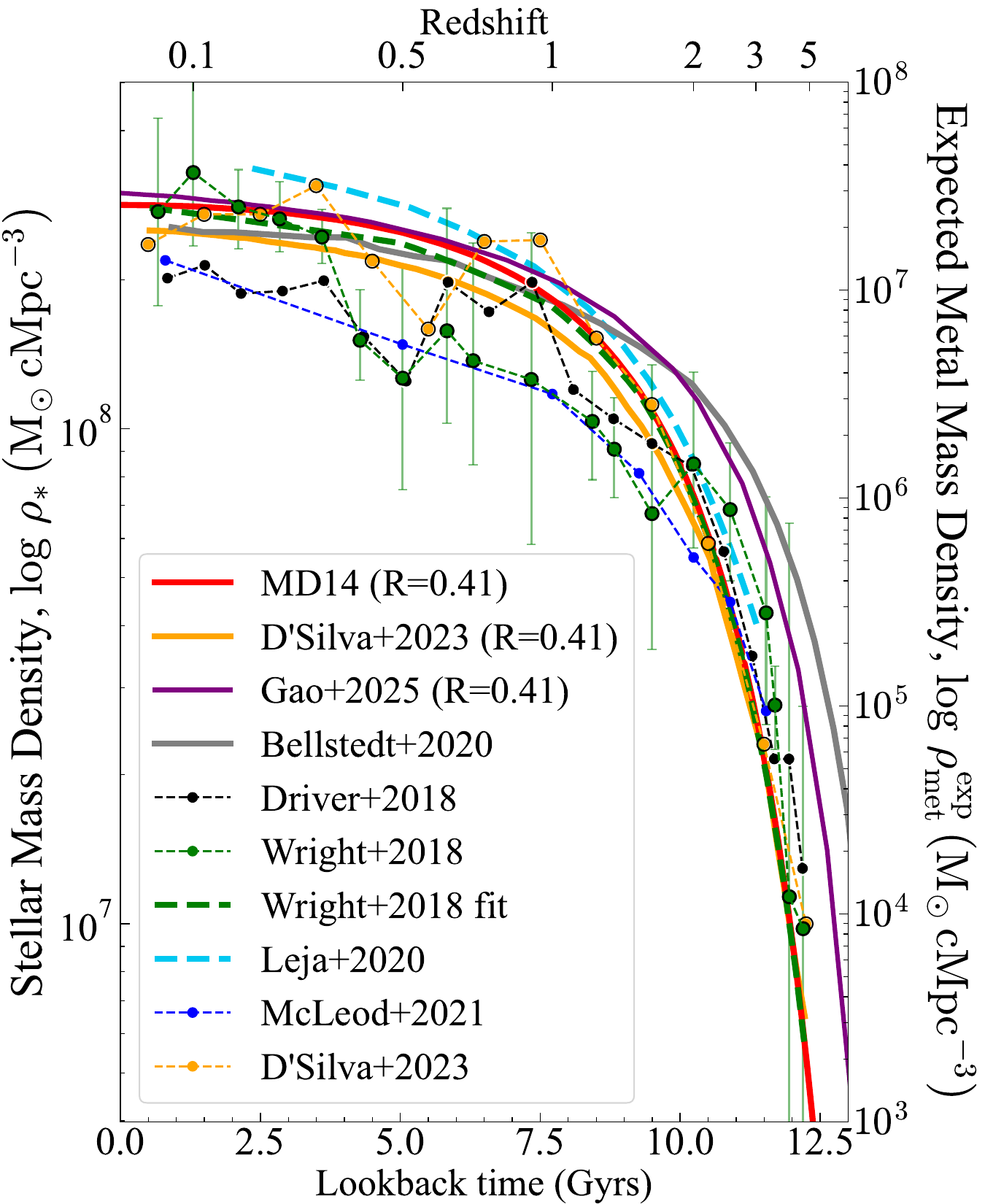}
\par\end{centering}
\begin{centering}
\caption
{Evolution of the stellar mass density of the Universe derived using the star formation rate density (solid lines: \citetalias{madau2014}; \citealt{bellstedt2020,dsilva2023,gao2025}) and the measured galactic stellar mass functions (dashed lines: \citealt{wright2018,mcleoud2021,leja2020,dsilva2023}). The total expected metal mass density $\rho_{\rm{met}}^{\rm{exp}}$ is shown on the right axis and is derived by scaling $\rho_*$ with the integrated metal yield $y=0.033$ \citep{peeples2014}.} 
\label{SMD_comp}
\par\end{centering}
\end{figure}

 \subsection{Expected Global Metal Density of the Universe} 
 \label{subsec:total_metals}
 In parallel to summing the observable metal contributors, one can assess the total amount of metals expected to be produced by stellar populations. The total expected comoving metal mass density, $\rho^{\rm{exp}}_{\rm{met}}$, can be estimated by scaling the total mass density of stars with the integrated yield of metals \citepalias{peroux2020}:
 \begin{equation}
     \rho^{\rm{exp}}_{\rm{met}}(z)=y\rho_{*}(z),
 \end{equation}
 where $\rho_{*}$ is the stellar mass density (SMD)---the total mass remaining in long-lived stars and stellar remnants \citep{madau2014}, and $y$ is the integrated stellar yield, i.e., the ratio of the total mass of metals produced by stars to the total mass of stars formed \citep{peeples2014}. We then define the expected total metal density parameter as:
 \begin{equation}\label{eq:expmet}
    \Omega^{\rm{exp}}_{\rm{met}}\left(z\right) = \frac{\rho^{\rm{exp}}_{\rm{met}}\left(z\right)}{\rho_{c}},
\end{equation}
where $\rho_{c} = 3H_0^2/\left(8\pi G\right) = 1.36\times10^{11}\;\text{M}_{\odot}\;\text{Mpc}^{-3}{h_{70}}^2$ is the critical density at $z=0$.
We can estimate $\rho_{*}$ at a given redshift by integrating the star formation rate density via:
\begin{equation}\label{eq:sfrsmd}
    \rho_*^{\rm{SF}}\left(z\right)=\left(1-R\right)\int^{\infty}_z\psi\left(z\right)\left|\frac{\text{d}t}{\text{d}z}\right|\text{d}z,
\end{equation}
where $R$ is the return fraction, i.e., the fraction of stellar mass that is returned to the gas when massive stars explode, and $\psi(z)$ is the star formation rate density (SFRD) defined as the mass of stars formed per unit comoving volume per unit time \citep{tinsley1980}. For our purposes, we adopt a \cite{chabrier2003} initial mass function (IMF) and the form of the SFRD derived by \cite{madau2014} (henceforth \citetalias{madau2014}) to calculate $\rho_*^{\rm{SF}}\left(z\right)$. For a Chabrier IMF, the $\rho_*^{\rm{SF}}\left(z\right)$ derived by integrating Equation~\ref{eq:sfrsmd} matches the measured stellar mass density when $R=0.41$ \citepalias{madau2014}, under the assumption of instantaneous recycling for stars with mass greater than $1\,\text{M}_\odot$.

Alternatively, $\rho_{*}$ can be derived using measurements of the frequency distribution of stellar mass $m_*$ in galaxies or the galaxy stellar mass function (GSMF) $\Phi_z(m_*)$ at a redshift $z$:
\begin{equation}\label{smd} 
    \rho_*^{\rm{SMF}}\left(z\right)=\int^{m_*^{\rm{max}}}_{m_*^{\rm{min}}}\Phi_z(m_*)m_*dm_*,
\end{equation}
where $m_*^{\rm{min}}$ and $m_*^{\rm{max}}$ are the limits of integration. The stellar mass density derived by \citetalias{madau2014} using Equation~\ref{eq:sfrsmd} is in good agreement with $\rho_*$ determined by \cite{wright2018} and \cite{mcleoud2021} using Equation~\ref{smd} at $z\sim0$ (see Figure~\ref{SMD_comp} and \citetalias{madau2014} for references). At higher redshift, we observe some discrepancy between the two calculations. The $\rho^{\rm{SMF}}_*$ derived at any redshift may be affected by cosmic variance resulting in the inhomogeneous sampling of observed fields \citep{dsilva2023}. This can result in artificially low or high values of $\rho^{\rm{SMF}}_*$ in some surveys \citep{wright2018,mcleoud2021}. On the other hand, the SFRD is empirically determined using much larger datasets. At any redshift, it is better determined as it averages over survey-specific sampling biases, mitigating the effects of cosmic invariance. Thus, we use the integral of the SFRD---Equation~\ref{eq:sfrsmd} to derive $\rho_*(z)$. 

To estimate $\Omega^{\rm{exp}}_{\rm{met}}$, we adopt an integrated yield $y=0.033\pm0.010$ following \cite{peeples2014}. The uncertainty in $y$ plays an important role in our estimate of $\Omega^{\rm{exp}}_{\rm{met}}$ and our accounting of the global metal budget. \citeauthor{peeples2014} made the instantaneous recycling approximation and assumed that the metal yield is independent of the stellar mass and metallicity. Although these approximations hold well for $\alpha$-elements produced via core-collapse supernovae, they fail for elements produced via Type Ia supernovae (iron peak elements). For example, \cite{andrews2017} demonstrated that the yields of $\alpha$-elements (O, Mg, Si, S, Ca, Ti) that undergo hydrostatic burning phases increase almost monotonically with stellar mass and show a weak dependence on metallicity. On the other hand, yields of iron peak elements that undergo delayed enrichment and require more complex modeling show a more complicated relationship with both stellar mass and metallicity. These factors introduce uncertainties that are difficult to quantify.

Another variable is the assumed yield set derived from chemical evolution models \citep{buck2021}. The treatment of stellar rotation and stellar winds in the chemical evolution models greatly influences the derived yields. For instance, \cite{chiappini2008} demonstrated that efficient mixing enhanced by stellar rotation can lead to the diffusion of CNO elements from the inner helium burning zone to the outer stellar zones, even in low metallicity stars. This process ultimately results in highly enriched stellar winds. \cite{maeder1992} showed that stellar winds can expel substantial amounts of helium during initial evolutionary stages, reducing its consumption in the synthesis of heavier elements. This mechanism increases the helium-to-metal yield ratios. A similar effect is observed with carbon and oxygen: carbon yield is enhanced by mass loss via stellar winds, while oxygen yield decreases since the expelled carbon would otherwise have been converted to oxygen \citep{vincenzo2016}. These examples illustrate how assumed stellar physics directly affects derived elemental yields, with net yields derived from different chemical models (with and without stellar rotation) varying by up to a factor of 1.5 \citep{vincenzo2016}.

Stellar yields (and return fractions) derived from models assuming different IMFs may differ by a factor of a few. \cite{vincenzo2016} derive the net yield for all metals for two yield sets \citep{romano2010,nomoto2013} and three IMFs \citep{salpeter1955,kroupa2001,chabrier2003}. They found that the highest yields result from the model that assumes a Chabrier IMF (since it contains the highest fraction of massive stars) and the yield sets from \citeauthor{romano2010} (which include effects of stellar rotation), while the lowest yields correspond to the Salpeter IMF and \citeauthor{nomoto2013} (without stellar rotation). The derived yields also depend on the assumed black hole mass cutoff (or the initial stellar mass above which all stars collapse to black holes) in chemical evolution models. A lower cutoff means more stars end up as black holes, which consume most metals synthesized by the star, resulting in a lower metal yield. For example, \citetalias{madau2014} show that setting the black hole cutoff to 40 M$_{\odot}$ results in a net stellar yield of 0.016 for Salpeter IMF and 0.032 for Chabrier IMF (return fractions of 0.27 and 0.41). If instead the black hole mass limit is set to 60 M$_{\odot}$, the resulting yields are 0.023 (R=0.29) and 0.048 (R=0.44) for Salpeter and Chabrier IMFs, respectively. We note here that, for the estimation of $\Omega_{\rm met}^{\rm exp}$, the quantity of interest is the product $y\times(1-R)$. We find that the product of $y$ and $1-R$ varies by a factor of up to 2.5 between the smallest and largest values of $y\times(1-R)$ resulting from different IMFs and yield sets.  

Here, we adopt the \citeauthor{peeples2014} estimate, as it represents approximately the mean of several other predictions in the literature \citep{woosley1995,portinari1998,chieffi2004,hirchi2005,romano2010,nomoto2013,vincenzo2016}. The uncertainty quoted at the 68\% confidence level is consistent with more recent estimates by \citetalias{madau2014} and \cite{vincenzo2016}, which incorporate different sets of yields and IMFs. 

\section{Metal Mass Density of the Stellar Phase}
\label{subsec:stars}
Though stars constitute only $(5.3\pm0.4)\%$ (see \S  \ref{gsmf}) of the total baryons at $z=0$ \citep{persic1992,madau2014,wright2018,driver2018,walter2020}, they are the only source of metals and can lock away a substantial fraction of the metal mass of the Universe. \citetalias{peroux2020} find that as much as 50\% of all metals reside in stars at $z\sim0.1$. Determining the global metal density of stars requires measuring the mass-weighted metal content of stars in galaxies, notably the stellar mass-metallicity relation, $Z_*\left(m_*\right)$, where $Z$ represents the fractional mass of stars composed of metals. To estimate the global stellar metal density, $Z_*\left(m_*\right)$ must be weighted by the galaxy stellar mass function $\Phi_z\left(m_*\right)$. The global stellar metal mass density can then be estimated via:
\begin{equation} \label{eqn:stellar_met}
    {\rho_{\rm{met,*}}}=\int^{m_*^{\rm{max}}}_{m_*^{\rm{min}}}\Phi_z\left(m_*\right)\, Z_*(m_*)\, m_*\, dm_*,
\end{equation}
\noindent
where $m_*^{\rm{min}}$ and $m_*^{\rm{max}}$ are the limits of integration. The choice of the galaxy stellar mass function, $\Phi_z\left(m_*\right)$, and the stellar mass-metallicity relation, $Z_*\left(m_*\right)$, depends on several factors including the dominant galaxy types at each cosmic epoch, the assumed initial mass functions and population synthesis models. In the following subsections, we discuss these in detail.

\begin{deluxetable*}{lCCCCCC}
\label{table:schec}
\tablecaption{Schechter function parameters from \cite{wright2018}}
\tablehead{\colhead{$z$}
& \colhead{$\phi_1$} & \colhead{$\phi_2$} & \colhead{$\alpha_1$} & \colhead{$\alpha_2$} & \colhead{$M^*$} & \colhead{$C$\tablenotemark{a}}}
\startdata
 $0.1$ & $-2.377^{+0.093}_{-0.103}$ & $-3.037^{+0.205}_{-0.711}$ & $-0.612^{+0.133}_{-0.288}$ & $-1.457^{+0.101}_{-0.110}$ & $10.800^{+0.044}_{-0.043}$ & $0.891$\\
 $0.7$ & $-2.840^{+0.136}_{-0.172}$ & $-3.977^{+0.507}_{-0.356}$ & $-0.858^{+0.062}_{-0.154}$ & $-1.838^{+0.254}_{-0.126}$ & $10.881^{+0.040}_{-0.087}$ & $1.650$\\
 $2.5$ & $-4.039^{+0.277}_{-0.389}$ & $-3.620^{+0.062}_{-0.094}$ & $-0.336^{+0.064}_{-0.074}$ & $-1.580^{+0.033}_{-0.048}$ & $11.075^{+0.098}_{-0.098}$ & $0.664$\\ 
\enddata
\tablenotetext{a}{We scale the GSMF $\Phi_z$ with the re-normalization factor $C$ to match $\rho^{\rm SMF}_*$ with $\rho^{\rm SF}_*$.}
\end{deluxetable*}

\subsection{Galaxy Stellar Mass Function} \label{gsmf}

For the galactic stellar mass function, we adopt the two-component variant of the Schechter mass function \citep{schechter1976} and use the parameter values estimated by \cite{wright2018}. The functional form of $\Phi_z(m_*)$ is:
\begin{equation}
\Phi_z\left(m_*\right)=\frac{1}{M^*}e^{-\frac{m_*}{M^*}}\left[\phi^*_1\left(\frac{m_*}{M^*}\right)^{\alpha_1}+\phi^*_2\left(\frac{m_*}{M^*}\right)^{\alpha_2}\right],
\label{schechter}
\end{equation}
with fit parameters $M^*$, $\alpha_1$, $\alpha_2$, $\phi_1$ and $\phi_2$. In Figure~\ref{SMD_comp} we compare estimates of $\rho_*^{\rm{SF}}$ \citep{madau2014,bellstedt2020,dsilva2023,gao2025} and $\rho_*^{\rm{SMF}}$ \citep{driver2018,wright2018,leja2020,mcleoud2021} reported in the literature. There can be disparities between the $\rho_*$ derived from the integrated SFRD and the integral of the stellar mass functions from specific surveys at some redshifts due to cosmic variance affecting the (smaller area) stellar mass function surveys \citep{dsilva2023} and uncertainties in or differing assumptions about the IMF \citep{wilkins2019,bellstedt2020}. An example of this is seen in Figure~\ref{SMD_comp} when comparing the \citet{wright2018} points with the integral of the \citetalias{madau2014} SFRD at $z\approx0.5$. The mass density derived from the \citeauthor{wright2018} mass functions is lower than the values derived from the SFRD integral. This deficit is due to the presence of a relatively under-dense region of the Universe at these redshifts in the  Galaxy and Mass Assembly (GAMA) survey, which played an important role in the \citeauthor{wright2018} analysis (see Section 3.3 in \citealt{dsilva2023}).

For our purposes, the total stellar mass density is important for understanding the total expected metal content of the Universe, while the shape of the galaxy stellar mass function is important when weighted by the stellar mass-metallicity relationship for assessing the contribution of stars to our metal mass census of the Universe. In what follows, we adopt the integral of the \citetalias{madau2014} SFRD to provide the total stellar mass density with redshift (the red curve in Figure~\ref{SMD_comp}). Integrating over the cosmic-averaged star formation rate density smooths out the local variations resulting from uneven sampling in redshift or mass bins \citep{leja2020,dsilva2023}. For deriving the metal mass in stars we adopt the shape of the galaxy stellar mass functions from \citet{wright2018} at each redshift. However, there needs to be consistency between the two calculations. We follow \cite{dsilva2023} in re-normalizing the integrals of the \citeauthor{wright2018} mass functions at each redshift to match the total mass density provided by the integral of the \citetalias{madau2014} SFRD (see Table \ref{table:schec}). This ensures that the stellar mass densities derived from the galaxy stellar mass functions are consistent with those derived using the SFRD. We show in Figure~\ref{SMD_comp} the re-normalized results of \cite{wright2018} as the thick green curve. Table \ref{table:schec} lists the Schechter function parameters from \cite{wright2018} and our adopted renormalization factor $C$ at each redshift.

We note that a substantial systematic uncertainty arises in both the inferred galactic stellar mass functions \citep{Gottumukkala2024} and the star formation rate density \citep{bouwens2020,matthews2024,sharma2024,liu2025}---and thus the stellar mass density---due to dust-obscured galaxies \citep{Barrufet2024}. Specifically, UV/optical surveys may miss a large fraction of massive dusty star-forming galaxies at all redshifts. For instance, a recent JWST survey by \cite{Gottumukkala2024} suggests that existing surveys of the galactic stellar mass function may be missing as much as $\sim\,20$--30\% of the galaxies with $m_*>10^{10.5}\rm{M}_\odot$ at $z>3$. Their results suggest that the integrated SMD (for $m_*>10^{9.25}$ M$_\odot$) may have been underestimated by up to 15--20\% at $z\sim 3$--6. Similar trends are observed at low redshifts ($0.2<z<1.3$) in radio and combined UV+IR observations \citep{whitaker2017,matthews2024}. For example, \citeauthor{matthews2024} found that UV+IR observations capture only 2/3 of the SFRD inferred from radio observations.
Ultimately, these uncertainties in the inferred stellar mass density of the Universe add to the uncertainty in the stellar and total expected metal mass density of the Universe.

\subsection{Stellar Mass-Metallicity Relation}

To calculate the global stellar metal density, we need to make an appropriate choice of the stellar mass-metallicity relation, $Z_*(m_*)$. This choice is often guided by the dominant galaxy population at any given epoch. Ideally, to calculate the global metal density in stars, we must use a mass-metallicity relation derived using a representative sample of galaxies. However, many studies treat star-forming (SF) and quiescent galaxies separately, in part due to the nature of their spectra. Such studies aim to understand the impact of the two populations on the global star formation history, and the change in relative number densities with redshift \citep{behroozi2019,mcleoud2021,weaver2022}. We use the information imparted from the comparison of the galaxy stellar mass functions of star-forming and quiescent galaxies to make our choice of $Z_*(m_*)$ and the combined $\Phi_z(m_*)$ to determine the limits of integration. 
At high redshifts ($z\gtrsim3$), most of the total stellar mass density is concentrated in star-forming galaxies (see Figure 12 in \citealt{weaver2022}), and a mass-metallicity relation derived using metallicities of star-forming (SF) galaxies provides a good estimate of the global metal density in stars. On the other hand, at lower redshifts ($z<2$), high-mass quiescent galaxies contribute significantly to the stellar mass density \citep{mcleoud2021}. Thus, at $z<2$, we use a sample composed of an integrated population of star-forming and quiescent galaxies to derive $Z_*(m_*)$.

Unlike the galaxy stellar mass function, $Z_*(m_*)$ does not have a standard functional form at all redshifts. At $z\le0.7$, we adopt the mass-metallicity relation derived by \cite{gallazzi2014}:
\begin{eqnarray} \label{eq:gallazzi}
    \log\left(\frac{Z_*}{Z_\odot}\right)=C+\log\left[1+\left(\frac{\overline{M}}{10^{11.5}\,\text{M}_\odot}\right)\right]^\gamma \nonumber \\
    -\log\left[1+\left(\frac{\overline{M}}{m_*}\right)\right]^\gamma,
\end{eqnarray}
where $C$, $\gamma$, $\overline{M}$ are separately derived for $z=0.1$ and $0.7$ in \cite{gallazzi2014}. \citeauthor{gallazzi2014} estimated their function using a sample of star-forming and quiescent galaxies, and the metallicities derived are optical luminosity-weighted. We note that their results are consistent with other local studies \citep{kirby2013,zahid2017,sextl2023,leung2024}. 

At $z\sim2-3$, we perform a linear regression to the mass-metallicity data derived from the Ly$\alpha$ Tomography IMACS Survey (LATIS) by \cite{chartab2023}. This results in a $Z_*(m_*)$ of the form,
\begin{eqnarray}
    \label{eqn:chartab}
    \log\left(\frac{Z_*}{Z_{\odot}}\right)=(0.33\pm0.05)\times\log\left(\frac{m_*}{10^{10}\,\text{M}_{\odot}}\right) \nonumber \\
    -0.82\pm0.02.
\end{eqnarray}
This relationship is consistent with other studies at this redshift, notably \cite{cullen2019} who use the VANDELS survey \citep{mclure2018,pentericci2018} and \cite{kashino2022} who use the zCOSMOS-deep redshift survey \citep{lilly2007,kashino2021}. All three surveys employ the same method to derive $Z_*(m_*)$---fitting the composite spectra of star-forming galaxies binned by $m_*$ with stellar population synthesis models. These stellar metal abundances primarily trace the iron abundance $\left[\rm{Fe/H}\right].$\footnote{ We express the logarithmic abundance relative to solar of a specific element $\rm{X}$ as:
    $[\rm{X}/\rm{H}]=\log \{N({\rm{X}})/ N({\rm{H}})\} - \log \{{\rm X/H}\}_{\odot}$
} The metallicities derived by all three are FUV-luminosity-weighted. The key difference between these studies lies in the choice of stellar synthesis codes. While both \cite{kashino2022} and \cite{chartab2023} use the Binary Population and Spectral Synthesis (BPASS) code \citep{eldridge2017}, \cite{cullen2019} use the Starburst99 (SB99) models \citep{Leitherer1999}. 
In this work, we adopt the \cite{chartab2023} mass-metallicity relation since it utilizes a larger, higher-resolution dataset and employs the BPASS v2.2.1 code, which incorporates binary systems in the stellar synthesis model. This is important because the metallicities are derived by fitting the composite spectra to the SED models. Binary systems, which constitute $\sim70\%$ of the stellar populations, \citep{sana2012,li2013,sana2014,eldridge2017}, cause the resulting spectra to be bluer and, thus, affect the estimated metallicities. For instance, \cite{cullen2019} report metallicities derived using BPASS models with binary systems to be $\sim0.1\,{\rm dex}$ lower than those derived using Starburst99 models.
\begin{figure}
\begin{centering}
\includegraphics[trim= 10 30 0 0,clip,width=\columnwidth]{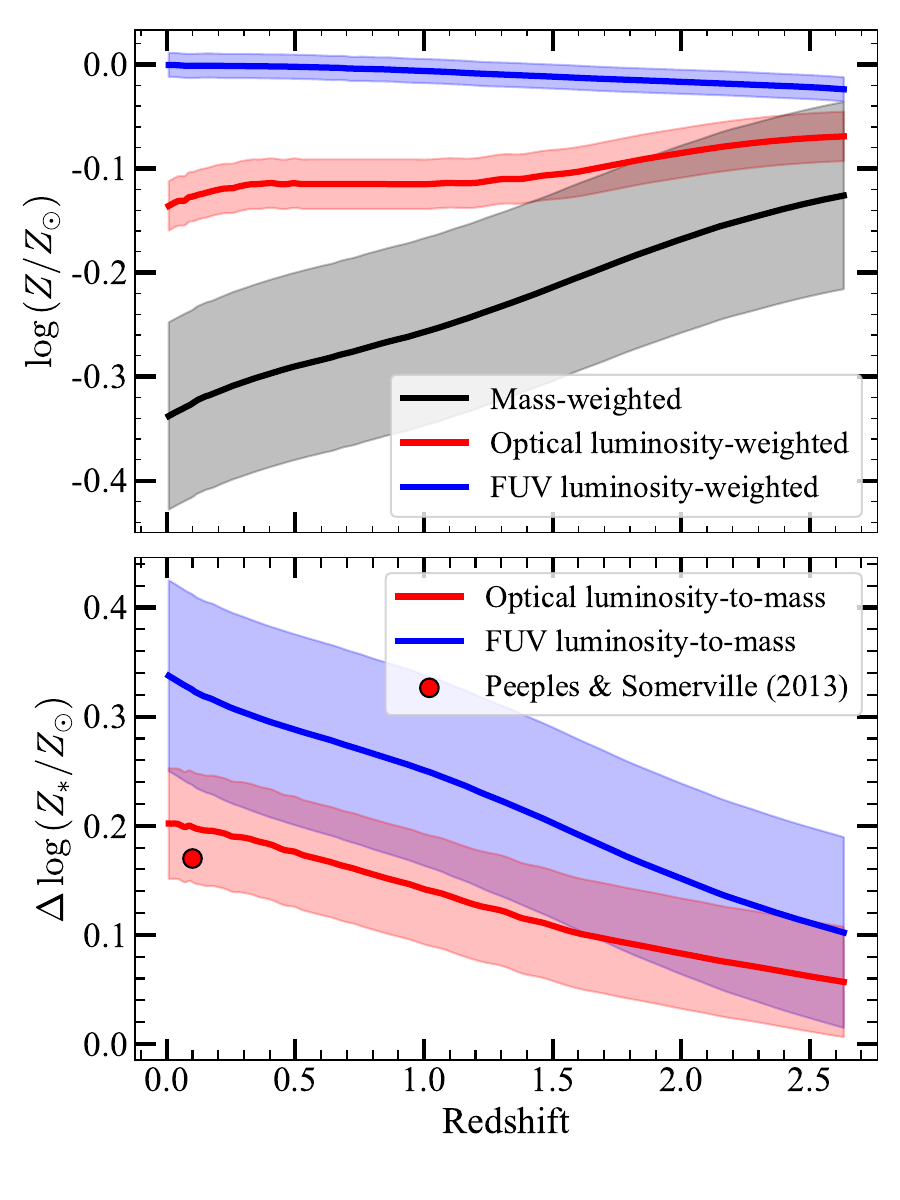}
\par\end{centering}
\begin{centering}
\caption{
{\textit{Top panel}: $Z_{*,l}$ (optical and FUV luminosity-weighted) and $Z_{*}$ (mass-weighted) metallicities as a function of redshift for galactic mass $m_*=10^{10}\,\rm{M}_{\odot}$. \textit{Bottom panel}: Correction factor to convert optical and FUV luminosity-weighted to mass-weighted metallicities, derived using simulations from \cite{kashino2022}}. Not shown here are simulations for various stellar masses, which vary very little at any redshift. To illustrate this, we plot the respective curves for $m_*=10^{8.5}\,\rm{M}_{\odot}$ and $m_*=10^{11.5}\,\rm{M}_{\odot}$. These mark the boundaries of the shaded regions along each of the curves. }
\label{kashino_corr}
\par\end{centering}
\end{figure}
The current observations are limited to $z\lesssim0.7$ and $z\sim2.5$ with a prominent gap at $z\sim 1$--2, where there is a lack of robust stellar phase metallicities for an unbiased sample containing both star-forming and quiescent galaxies. While there exist a few studies at these redshifts, they are primarily focused on quiescent galaxies \citep{estrada2019,kriek2019,carnall2022,saracco2023}. The global stellar mass density grows significantly during cosmic noon at $1\le z\le2.4$. This growth is coincident with an increase in quiescent galaxy stellar mass density consistent with a transition in population from SF galaxies to quiescent galaxies \citep{weaver2022}. At the same time, this period is characterized by a high star formation rate and, consequently, a high metal production rate. Thus, constraining the global stellar mass-metallicity relation during this transition epoch requires a comprehensive survey of stellar-phase metallicities in a sample that includes both star-forming and quiescent galaxies. This is a challenging task, primarily due to the nature of the spectra of star-forming galaxies that are dominated by younger stars and have weak absorption lines. At $z\gtrsim2$, rest-frame UV lines in composite spectra are used to measure stellar metallicities, which are more sensitive to metals in the younger stellar population \citep{cullen2019,kashino2022,chartab2023}. At $z<1$, optical absorption lines are used which better represent the mass-weighted average metallicity \citep{gallazzi2014,zahid2017}. The intermediate redshift regime $1\le z\le2$, requires high-resolution infrared spectra to constrain stellar metallicities using rest frame optical lines as well as perform dust corrections to the inferred metallicities. Even when these data are available, absorption lines are often too weak to be useful for stellar metallicity measurements for highly star-forming galaxies.

The stellar mass-metallicity relations we adopt \citep{gallazzi2014,chartab2023} are based on luminosity-weighted metallicities. They are thus biased towards the younger, brighter population of stars and do not represent the contributions from older stars, which bear the largest fraction of mass. Depending on the star formation history of a galaxy this may bias the metallicity, especially to higher metallicities, if the youngest stars are not representative of the entire population. Assessing $\Omega_{\rm{met,*}}$ as in Equation~\ref{eqn:stellar_met} requires mass-weighted metallicities. We correct light-weighted metallicities $Z_{*,l}$ to mass-weighted metallicities $Z_{*}$ using the results derived by \cite{kashino2022} (see their Appendix A and Figure 20). \cite{kashino2022} derive stellar metallicities by fitting composite spectra of star-forming galaxies to BPASSv2.2.1 templates. They describe metallicity evolution in galaxies using the ``flow-through" gas regulator model, optimizing its parameters (mass loading factor and star formation efficiency) to match observed gas phase and stellar phase mass-metallicity relations. With these best-fit parameters, they establish evolutionary tracks for SFR and metallicities, predicting the evolution of mass-weighted and luminosity-weighted metallicities. We compare these evolutionary tracks for a characteristic mass of $m_*=10^{10}$ M$_\odot$, which is roughly the mass at which $\Phi_z\left(m_*\right)m_*$ and $\Phi_z\left(m_*\right)m_*Z_*$ peak. The correction as a function of $z$ is shown in Figure~\ref{kashino_corr}; it does not depend strongly on the stellar mass in the \cite{kashino2022} models. To represent the potential variation with stellar mass, we plot the corrections for two boundary stellar mass values ($m_*=10^{8.5}$ M$_\odot$ and $10^{11.5}$ M$_\odot$) as shaded regions along each of the redshift-dependent curves in Figure~\ref{kashino_corr}. These shaded regions illustrate the range of corrections or adjustments for the lower and higher stellar mass limits. We also plot, in the lower panel of Figure~\ref{kashino_corr}, the conversion factor from optical luminosity-weighted to mass-weighted metallicity derived by \cite{peeples2013} using the stellar population synthesis models of \cite{bruzual2003} and find it consistent with \cite{kashino2022}.

\begin{deluxetable*}{lCCCCC}
\label{table:stellar}
\tablecaption{Stellar Phase Metal Densities}
\tablehead{\colhead{$\langle z\rangle$}
& \colhead{$\Delta z$} & \colhead{$\log \Omega_{\rm{met,*}}$} & \colhead{${\Omega_{\rm{met,*}}}/{\Omega_{\rm{met}}^{\rm{exp}}}$\tablenotemark{a}} & \colhead{$Z(m_*)$ reference}}
\startdata
 $0.1$ & $0.005\text{--}0.22$ & $-4.67^{+0.21}_{-0.13}$ & $0.31^{+0.20}_{-0.08}$ & \text{\cite{gallazzi2014}}\\
 $0.7$ & $0.65\text{--}0.75$ &  $-4.94^{+0.29}_{-0.20}$& $0.21^{+0.20}_{-0.08}$ & \text{\cite{gallazzi2014}}\\
 $2.5$ & $2.00\text{--}3.00$ & $-6.07^{+0.18}_{-0.17}$ & $0.08^{+0.04}_{-0.03}$ & \text{\cite{chartab2023}}\\ 
\enddata
\tablecomments{At all redshifts, we use galactic stellar mass functions from \cite{wright2018}. The uncertainties represent a 68\% confidence interval.}
\tablenotetext{a}{The uncertainties here do not incorporate the uncertainty in the expected metal density.}
\end{deluxetable*}

\subsection{Stellar Metal Mass Density}\label{subsec:stellarmassdensity}

The estimation of the global stellar metal mass density involves several key assumptions. We will now examine these nuances and potential sources of uncertainties.

We typically extrapolate the galaxy stellar mass functions towards the lower end of stellar mass at all $z$ due to the lack of data in this regime. As a result of the fast decline of $\Phi_z$ with increasing $m_*$, the upper mass limit does not affect our calculations, as the integration converges for upper mass limits in the range $m_*^{\rm{max}}=\left[10^{11},10^{12}\right]\text{M}_{\odot}$. This is not true for the lower limit of integration since $\Phi_z$ grows with decreasing mass. That being said, the contribution of low-mass or dwarf galaxies to the total stellar metal mass density is very small, particularly at low $z$. This is corroborated by studies of dwarf galaxies in the local Universe. \cite{kirby2013} evaluated the Universal stellar mass-metallicity relation for dwarf galaxies. Their $Z_*(m_*)$ is consistent with the extrapolation of the \cite{gallazzi2014} low redshift $Z_*(m_*)$ to lower stellar masses.  On integrating the mass-metallicity relationship derived by \cite{kirby2013}, in the range $m_*\in\left[10^{3},10^{9}\right]\text{M}_\odot$, we find the resulting $\rho_{\rm{met,*}}$ to be 3 orders of magnitude smaller than that obtained by integrating the \cite{gallazzi2014} $Z_*(m_*)$ between $\left[10^{9},10^{12}\right]\text{M}_\odot$. Thus, at least at low $z$, the contribution of dwarf galaxies to the total stellar phase metal density is negligible. 

We determine integration limits by assessing the contribution of galaxies across mass ranges to total stellar mass and metal densities. To ensure convergence, we calculate the total mass and metal densities by setting the lower limit to $10^5\,\rm{M}_{\odot}$. We find that the stellar mass density converges at $10^8\,\rm{M}_{\odot}$ (within a precision of $10\%$), while metal mass density converges at $10^9\,\rm{M}_{\odot}$ (within a precision of $<5\%$). That is, at all $z$, lower-mass galaxies $\left(m_*\lesssim10^9\,\rm{M}_{\odot}\right)$ contribute $<5\%$ to the total stellar metal mass density. This is due to the nature of the stellar mass-metallicity relations. Lower-mass galaxies have reduced star formation efficiency and higher metal loss through winds due to their shallow gravitational potentials, resulting in a stellar metallicity that decreases with galaxy mass. Based on these observations, we set $m_*^{\rm{min}}=10^{9}\text{M}_\odot$ and $m_*^{\rm max}=10^{12}\text{M}_\odot$ in our calculations.

We pay particular attention throughout our work to the uniform calculation of uncertainties. We use Monte Carlo techniques to estimate the uncertainties of our integrated metal densities. For the stellar metal mass density, we perform the calculation 10,000 times for each redshift. For every integration the parameter spaces of $\phi_1$, $\phi_2$, $\alpha_1$, $\alpha_2$, $M^*$ that specify the galaxy stellar mass function \citep{wright2018}, as well as the parameters specifying the mass-metallicity relations---$\gamma$ and $\bar{M}$ at low redshift (Equation~\ref{eq:gallazzi}), the slope and the intercept for the high-redshift $Z_*(m_*)$ (Equation~\ref{eqn:chartab}), are randomly sampled within their quoted $68\%$ confidence interval. Many of these parameters have asymmetric distributions. To approximate the non-Gaussian distributions, we assume that the likelihoods of the parameters follow skew-normal distributions and fit the available percentiles to the distribution function using the \texttt{scipy.stats.skewnorm} function in Python \citep{azzalini2009}. We solve for the parameters characterizing the mean, the standard deviation, and the skewness of the distribution. The parameters are optimized by minimizing the quantity $\left[\Phi_{50}-\phi_{50}\right]^2+\left[\Phi_{16}-\phi_{16}\right]^2+\left[\Phi_{84}-\phi_{84}\right]^2$, where $\Phi_{\rm{k}}$'s are the percentiles evaluated from the skew-normal distribution model, and $\phi_{\rm{k}}$'s are the corresponding values from literature \citep{wright2018}. Although this method does not guarantee a precise reconstruction of the likelihood distribution for each parameter, it is a major improvement on assuming a flat or normal distribution.

We evaluate the integral in Equation~\ref{eqn:stellar_met} 10,000 times, drawing the parameters from their distributions to generate a distribution of stellar metal density $\rho_{\rm{met,*}}$. We report our results in Table~\ref{table:stellar}, in terms of the dimensionless stellar metal density parameter $\Omega_{\rm{met,*}}=\rho_{\rm{met,*}}/\rho_c$. The trend of increasing metal density with decreasing $z$ agrees with the expectation that more metals are locked in stars over time. Table~\ref{table:stellar} also shows the contribution of stars to the total metal density of the Universe at each redshift. Stars contain a large fraction of metals at low redshift, with a notable increase in their contribution to the global metal density from 8\% at $z\sim2.5$ to 21\% at $z\sim0.7$ and 31\% at $z\sim0.1$. As discussed above, there is little information on stellar metal mass densities at intermediate redshifts ($z\sim$1--2). 

It is important to consider systematic uncertainties in our estimates of $\Omega_{\rm{met,*}}$. The use of different metallicity indicators at different redshifts and for different galaxy populations constitutes a major source of uncertainty \citep{maiolino2019}. Each diagnostic of stellar metallicity has its caveats. For instance, at low redshift \cite{gallazzi2014} use Lick indices to estimate metallicities. While the use of these indices is very common, there are known degeneracies in the age, metallicity, and dust extinction of stellar populations derived from this method \citep{gallazzi2005}. Some indices are more sensitive to the properties of the younger brighter stars that dominate the galaxy luminosity than they are to the older stars that dominate the mass of the galaxy, potentially biasing our estimates of the mass-weighted metallicity. At high redshift, \cite{chartab2023} fit the full composite spectrum of the galaxy population to stellar population synthesis models to estimate the metallicities, which is also susceptible to the age-stellar mass degeneracy mentioned above. Other sources of uncertainties arise from the assumed IMF of the stellar population synthesis models \citep{fontanot2017} and the treatment of stellar rotation in the model spectra of galaxies \citep{choi2017}. 

\section{Metal Mass Density of Hot, Virialized Gas ($T\gtrsim10^6$ K)}
\label{subsec:icmigrm}

A significant fraction of the baryons in the Universe resides in hot gas ($T \geq 10^6\ {\rm K}$) that is best probed with X-ray observations. Because X-ray observations of low-density gas are exceedingly difficult, the information available for the metal content of the hot gaseous atmospheres of individual galaxies and the WHIM is limited, a point we discuss in Section~\ref{subsec:whim}.

In group and cluster environments, ram pressure stripping and outflows cause galaxies to lose substantial amounts of enriched gas, which ultimately ends up in the intragroup medium (IGrM) and the intracluster medium (ICM). This gas phase is characterized by very high temperatures that are typically dictated by the gravitational potential of the group/cluster halos. This hot, (usually) virialized gas is traced by X-ray emission \citep{werner2008,molendi2016,gastaldello2021}. The emission spectra are fitted to models to determine the overall metallicity and baryonic mass of clusters and groups \citep{mantz2017}.

For our census, we adopt the metal densities calculated by \citetalias{peroux2020}. The combined global ICM+IGrM metal density can be derived by performing a mass integral of the product of the halo mass function \citep[adopted from][]{bocquet2016}, the hot gas baryonic fraction with mass \citep[][]{chiu2018}, and the (potentially mass-dependent) metallicity of the gas. Based on values reported in several studies \citep{mantz2017,yates2017,liu2020,flores2021,ghizzardi2021,blackwell2022,sarkar2022, molendi2024}, we assume the average ICM/IGrM metallicity to be $ Z_{\rm{ICM+IGrM}}=0.3\,\rm{Z}_\odot$ for $z\lesssim1.3$. This almost uniform large-scale metallicity is attributed to the early enrichment of the ICM/IGrM (before or shortly after cluster formation, although see \citealt{molendi2024}) and shows no evolution with redshift up to $z\sim1.3$ \citep{mantz2017,yates2017}. Significant later enrichment takes place only in cluster centers as the stellar populations of the galaxies evolve. However, since these metals constitute only a few percent by mass of all metals in the ICM \citep{liu2020}, we assume the global mass-weighted metallicity $Z_{\rm{ICM+IGrM}}$ is constant with halo mass and redshift. 

The metal density of the ICM+IGrM component increases with the age of the Universe and has almost doubled from $\Omega_{\rm met}=5.25^{+2.16}_{-1.53}\times 10^{-6}$ at $z=0.7$ to $\Omega_{\rm met}=1^{+0.35}_{-0.26}\times 10^{-5}$ at $z=0.1$. This component accounts for $\sim(15\pm4)\%$ and $\sim(10\pm3)\%$ of the total metals at $z=0.1$ and $z=0.7$, respectively. Since the overall metallicity of the ICM+IGrM remains constant over a large redshift range ($z\lesssim1.3$), the increased global metal density is due to the increasing baryonic mass density in the ICM resulting from the growth of galaxy clusters and large-scale structure formation \citep{chiu2018}.

\section{Metal Mass Density of Cool Gas ($T\lesssim10^5$ K)}
\label{sec:coldgas}

The cool gas surrounding galaxies and constituting the CGM and the cool IGM are probed using UV absorption lines in quasar spectra. We follow \cite{lehner2018a} in categorizing these intervening \HI-selected gaseous absorbers based on their neutral hydrogen column densities $N_{\rm{HI}}$\footnote{These labels are observationally motivated and have been historically used to classify absorbers based on their neutral hydrogen column density. For instance, LLSs were defined to have an optical depth $\tau\geq1$ at the Lyman limit, corresponding to $\log\,N_{\rm HI}\geq17.2$. DLAs were originally defined by \citet{wolfe1986} as absorbers with rest equivalent width $>5$ \AA, which selects absorbers having lines broadened by radiation damping and corresponds to $\log\,N_{\rm HI}\gtrsim 20.3$, a column density typically observed in disks of galaxies in contemporary 21-cm observations. We include these definitions to connect our results to the literature. These classifications serve as useful indicators of baryonic overdensities and environments while also reflecting observational techniques and limitations.} (in units of \column): (i) Ly$\alpha$ forest absorbers (LAF, $\logNHI\leq 14.5$), (ii) strong Ly$\alpha$ forest systems (SLFSs, $14.5\le \logNHI<16.2$)\footnote{For the low-$z$ sample this threshold appears at $\logNHI\gtrsim15$. This is because the UV HST/COS data used for the survey is not sensitive enough to measure low metallicities ($\left[\rm{X/H}\right]<-1$) for $\logNHI<15$ absorbers.}, (iii) partial Lyman systems (pLLSs, $16.2\le\logNHI<17.2$), (iv) Lyman limit systems (LLSs, $17.2\le\logNHI<19.0$), (v) super-LLSs or sub-DLAs (SLLSs, $19.0\le\logNHI<20.3$) and (vi) DLAs ($\logNHI\geq20.3$). The typical overdensity of gas increases with $N_{\rm{HI}}$ \citep{schaye2001}. DLAs are cool, neutral, and dense gas often associated with the inner regions of galaxies \citep{wolfe2005b,chen2017} and represent the largest reservoir of cool gas for fuelling star formation (\citealt{peroux2003}; \citealt{wolfe2005a}; \citetalias{peroux2020}). The pLLSs, LLSs, and SLFSs are often associated with the CGM or the denser IGM with metallicity as a principal indicator of their environments \citep{berg2023}. LAF absorbers trace the larger-scale diffuse intergalactic gas in the dark matter filaments threading the Universe (\citealt{mcquinn2016}; \citetalias{peroux2020}). They may also trace the extended diffuse CGM of galaxies. Thus, by studying the mass of metals in each column density regime as a function of cosmic time, we can trace the changing distribution of star formation-produced metals in cosmic gas spanning a wide range of physical scales. 

We outline, in this section, our calculations of the metal mass associated with this cool gas. To do so, we start with the total mass density of gas at any given redshift. The cosmological mass density of cool ($T\lesssim 10^5$ K), atomic absorption-line selected gas is given by:
\begin{equation}\label{om_gas}
{\rho_{\rm{gas}}}=\frac{\mu\, m_{\text{p}}\,H_0}{c}\int\frac{N_{\rm{HI}}\,f(N_{\rm{HI}})}{X({\rm{H^0}})}\,dN_{\rm{HI}},
\end{equation}

\noindent
where $m_{\text{p}}$ is the mass of a proton and $\mu=1/0.76$ accounts for the mass of helium, assuming its primordial mass fraction \citep{cyburt2016}; $X({\rm{H^0}})$ is the hydrogen neutral fraction, and $f(N_{\rm{HI}})$ is the column density distribution function (CDDF)---the number of absorbers with neutral hydrogen column densities between $N_{\rm{HI}}$ and $N_{\rm{HI}}+dN_{\rm{HI}}$ in the redshift bin ($z$, $z$+d$z$). We perform the integration in Equation~\ref{om_gas} using a piecemeal approach, integrating over column density ranges within each class of absorber individually as surveys for metallicity; even $f(N_{\rm HI})$ is often broken down along these lines. The metal mass density is then written as:
\begin{equation}
    \label{rho_met}
    \rho_{\rm{met}}  = \,\langle Z\rangle_{N_{\rm{H}}}\,\rho_{\rm{gas}}.
\end{equation}

The quantity $\langle Z\rangle_{N_{\rm{H}}}$ is the $N_{\rm{H}}$-weighted average of metallicities for each class of absorbers:
\begin{equation}
    \langle Z\rangle_{N_{\rm{H}}} = \frac{\sum_i\,Z_i\,N_{\rm{H},i}}{\sum_i\,N_{\rm{H},i}} = \frac{\sum_i\,10^{\left[\frac{\rm{X}}{\rm{H}}\right]_i}\,N_{\rm{H},i}}{\sum_i\,N_{\rm{H},i}}\,Z_{\odot}. \label{meanZ}
\end{equation}

For compiling the global metal budget, we require the total mass-weighted metal content of gas. To this end, we weight the metallicities of the gaseous absorbers by $N_{\rm{H}}$, the total hydrogen column density accounting for both neutral and ionized cool gas. For the calculation of $\rho_{\rm{gas}}$, we integrate over $N_{\rm{HI}}$ and apply an ionization correction,  $X({\rm{H^0}})^{-1}$, since we measure $f\left(N_{\rm{HI}}\right)$ and not $f\left(N_{\rm{H}}\right)$. 
The metal density parameter is calculated as:
\begin{equation}
    \Omega_{\rm{met}} = \langle Z\rangle_{N_{\rm{H}}}\,\Omega_{\rm{gas}} = \frac{\langle Z\rangle_{N_{\rm{H}}}\,\rho_{\rm{gas}}}{\rho_c}. 
\end{equation} 
Similar to the case of stars, the functions needed to carry out the above calculations do not have a standard form. They are estimated empirically at various redshifts using different functional forms, with the uncertainties in unique fitting parameters contributing to the uncertainties on the final result $\Omega_{\rm{met}}$. In the following subsections, we explore the different column density regimes and describe the methods used to represent the relevant functions for each. We note that the cool gas described above does not include molecular $\rm{H}_2$ gas \citep{tacconi2020, walter2020}. We also exclude the cool IGM traced by the LAF \citep{mcquinn2016}. While both phases have significant contributions to the total gas and even metal densities of the Universe, observational limitations prevent us from including them in our metal census. We describe these in Section~\ref{subsec:others}. 

\begin{deluxetable*}{lCCCCCC}[t]
\tablecaption{Global Metal Densities of Neutral Gas}\label{table:dlas}
\tablehead{\colhead{$\langle z\rangle$}
& \colhead{$z_{\rm{min}}$} & \colhead{$z_{\rm{max}}$} & \colhead{Number} & \colhead{$\log \Omega^{\rm{neut}}_{\rm{gas}}$} & \colhead{$\log \Omega^{\rm{neut}}_{\rm{met}}$} & \colhead{$\Omega^{\rm{neut}}_{\rm{met}}/\Omega_{\rm{met}}^{\rm{exp}}$\tablenotemark{a}}}
\startdata
 $0.54$ & $0$ & $1.0$ & $14$ & $-3.23^{+0.01}_{-0.01}$ & $-5.37^{+0.07}_{-0.08}$ & $0.07^{+0.01}_{-0.01}$ \\ 
 $1.66$ & $1.0$ & $2.0$ & $39$ & $-3.09^{+0.01}_{-0.01}$ & $-5.39^{+0.10}_{-0.09}$ & $0.16^{+0.05}_{-0.03}$\\ 
 $2.19$ & $2.0$ & $2.4$ & $44$  & $-3.04^{+0.01}_{-0.01}$ & $-5.46^{+0.11}_{-0.12}$ & $0.24^{+0.07}_{-0.06}$\\
 $2.84$ & $2.4$ & $3.5$ & $100$  & $-3.00^{+0.02}_{-0.01}$ & $-5.62^{+0.08}_{-0.08}$ & $0.32_{-0.05}^{+0.06}$\\
 $3.96$ & $3.5$ & $4.25$ & $19$  & $-2.94^{+0.02}_{-0.02}$ & $-5.60^{+0.13}_{-0.23}$ & $0.97_{-0.40}^{+0.35}$\\
 $4.52$ & $4.25$ & $5.25$ & $15$  & $-2.91^{+0.02}_{-0.02}$ & $-5.81^{+0.10}_{-0.13}$ & $0.95_{-0.24}^{+0.25}$\\
\enddata
\tablecomments{The fourth column gives the number of absorbers in each redshift bin $\left[z_{\rm{min}},z_{\rm{max}}\right]$. $\Omega^{\rm{neut}}_{\rm{gas}}$ is the mass density of neutral gas, and $\Omega^{\rm{neut}}_{\rm{met}}$ is the dust-corrected metal density of neutral gas. The metal densities are derived using DLA metallicities from \citetalias{peroux2020} with updated redshifts bins and excluding absorbers with $\sigma_{\rm{Z}}>0.3\,\,\rm{dex}$. The uncertainties represent a 68\% confidence interval.}
\tablenotetext{a}{The uncertainties here do not incorporate the uncertainty in the expected metal density.}
\end{deluxetable*}

\subsection{Neutral Gas}
\label{subsubsec:neutral_gas}

Cool neutral atomic hydrogen gas is a crucial component of the Universe and plays a vital role in galaxy formation. To understand the evolution of the neutral gas mass density, we follow the work of \citetalias{peroux2020}. At low redshift ($z\lesssim0.4$), measurements of the neutral gas mass are derived from 21-cm emission surveys. These determinations are most often done by assessing the \HI\ mass function of galaxies, either in blind or galaxy-targeted surveys, and often rely on spectral stacking at higher redshifts \citep[e.g.,][; see also references in \citetalias{peroux2020}]{zwaan2005, jones2018, chowdhury2024}. Beyond $z\sim0.4$ we are just beginning to be able to probe average \HI\, 21-cm emission properties with the next generation of radio telescopes. At $z>0.4$, \citetalias{peroux2020} used neutral gas mass densities derived from absorption line studies of DLAs. Like the ISM of all galaxies, the DLAs contain a mixture of phases, and the highly-ionized gas probed by \ion{C}{4}, \ion{N}{5}, and \ion{O}{6}, in particular, can contain a substantial amount of mass \citep[e.g.,][]{fox2007,lehner2014}. However, the gas probed by \HI\ absorption in DLAs is predominantly neutral. It is straightforward to calculate the neutral gas density $\Omega^{\rm{neut}}_{\rm{gas}}=\rho^{\rm{neut}}_{\rm{gas}}/\rho_c$ using Equation~\ref{om_gas}. 

\citetalias{peroux2020} fit the cosmological evolution of cool neutral gas density estimates from literature with a power law in redshift: $\Omega^{\rm{neut}}_{\rm{gas}}=\left[(4.6\pm0.2)\times10^{-4}\right]\left(1+z\right)^{0.57\pm0.04}$. This is consistent with more recent studies investigating the evolution of neutral cool gas with redshift \citep{walter2020,rhee2023}.\footnote{We adopt the fit from \citetalias{peroux2020}. We note that incorporating these latest 21 cm measurements from, e.g., \cite{rhee2023} at low redshift produces fits that are unchanged from that result.} We note that the extrapolation of this result to $z\gtrsim5$ is particularly uncertain, as the surveys for DLA absorption become difficult against the increased opacity of the LAF IGM absorption. 

\citetalias{peroux2020} calculated $\Omega_{\rm{met}}^{\rm{neut}}$ by combining $\Omega^{\rm{neut}}_{\rm{gas}}$ with dust-corrected metallicities from \cite{decia2018} and supplementary data (also dust-corrected, \citealt{berg2016,poudel2018} and \citealt{oliveira2014}).  We recalculate $\Omega_{\rm{met}}^{\rm{neut}}$ using the extended data from \citetalias{peroux2020}, making slight adjustments to the redshift bins (for convenient comparison with other reservoirs). There are several systems in that sample for which the dust depletion corrections are highly uncertain (and in some cases extremely large) given the limited range of elements accessible. In a few of the redshift bins, some of these systems have unrealistically large dust-corrected abundances ([Si/H]$\gtrsim0.5$ at $z>2$) with very large uncertainties due to the uncertain dust correction. Thus, for this work, we exclude absorbers in the \citetalias{peroux2020} sample with very large uncertainties in the metallicity, $\sigma_{\rm{Z}}>0.3\,\,\rm{dex}$. The excluded systems have a median metallicity consistent with the rest of the sample if one excludes the two systems with ${\rm [Si/H]} > +0.7$ (both with $\sigma_{\rm{[Si/H]}}>0.44$ dex formal uncertainties and dust corrections $\ga1$ dex). 

We summarize our results for the global dust-corrected metal density of neutral gas, $\Omega_{\rm{met}}^{\rm{neut}}$, in Table~\ref{table:dlas}. Dust contributes 20\%--30\% of the total metal density of the neutral gas in the Universe (see Section~\ref{subsec:dust}). We find that while the global metal density of cool neutral gas increases with decreasing redshift, its relative contribution to the expected metal density decreases. At high redshift ($z\sim4$), almost all metals expected to be produced by stars (95\%--97\% of the total expected metals) reside in this gas phase. At low redshift $z<1$, the contribution drops to $\sim10$\% of the total.

\subsection{Ionized and Partially-Ionized Gas}
\label{subsubsec:cgm_ion}

\begin{figure*}
\begin{centering}
\includegraphics[trim = 6 0 0 12, clip,scale=0.65]{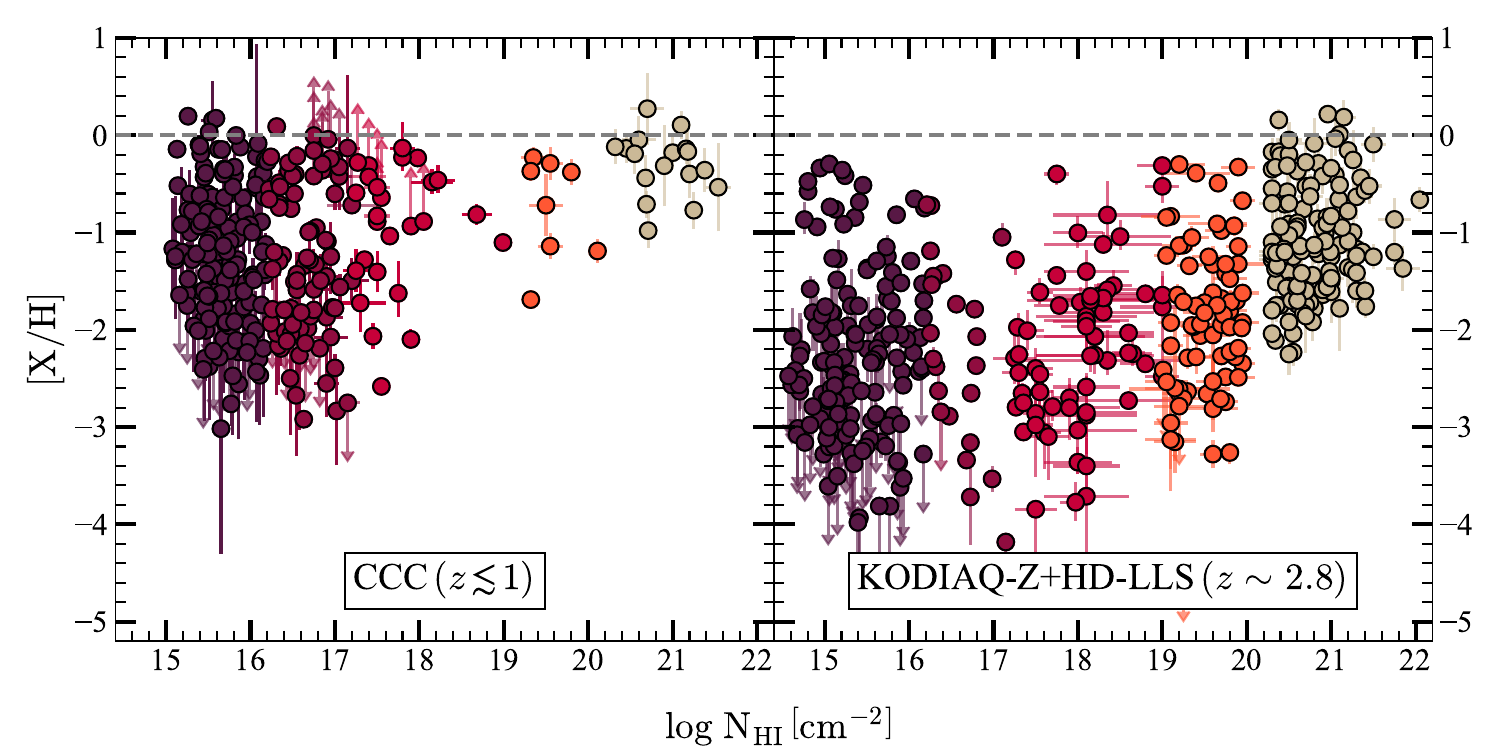}
\par\end{centering}
\begin{centering}
\caption{
{Metallicities of SLFSs, pLLSs, LLSs, SLLSs and DLAs from \citetalias{peroux2020}, the low-$z$ CCC sample (\textit{left}) covering $0.2\lesssim z\lesssim0.9$ and the high-$z$ KODIAQ-Z+HD-LLS sample (\textit{right}) covering $2.2\lesssim z\lesssim3.6$. The metallicities for the DLAs are drawn from literature \citep{decia2016,decia2018,berg2016,poudel2018,oliveira2014}. The median values for absorbers are adopted with a 68\% confidence interval, except for those with lower and upper limits, which are marked with appropriate arrows. The dashed grey line represents solar metallicity. The lack of data between $\logNHI$=20--20.3 is artificial and results from the photoionization grids stopping at \logNHI=20 \citep{lehner2022}.}}
\label{combined}
\par\end{centering}
\end{figure*}
 Unlike DLAs, whose cool gas is predominantly neutral due to self-shielding from the UV background, lower $N_{\rm{HI}}$ absorbers generally have lower densities and correspondingly lower neutral fractions/higher ionization fractions. For these absorbers, we adopt a neutral fraction function, $X\left(\rm{H}^0\right)$, derived from photoionization modeling of absorbers, to calculate $\rho_{\rm{gas}}$. The product of the gas density and the mean $N_{\rm{H}}$-weighted metallicity gives the global metal density, $\rho_{\rm{met}}$. At low redshift ($z<1$), we derive the metal densities based on a sample of 261 absorbers from the COS Circumgalactic Compendium (CCC) \citep{lehner2018a,wotta2019,lehner2019} with $15<\logNHI<19$ (see Figure~\ref{combined}). This includes 33 absorbers from \cite{wotta2016} based on low-resolution spectra with only $N_{\rm{HI}}$ and $N_{\rm{MgII}}$ measurements. Ionization modeling requires multiple ions to constrain the ionization parameter $U$ and the metallicity. For these absorbers, a Gaussian prior on $\log{U}$ was assumed, with parameters derived from the probability distribution functions of absorbers with well-constrained $\log{U}$ \citep{wotta2016}. For 17 of these absorbers, the \MgII\/ lines were saturated, resulting in lower limits on the metallicity. At the low $N_{\rm{HI}}$ end, metal line non-detections yield only upper limits on the metallicities of 41 absorbers. These lower and upper limits on metallicity estimates need to be treated carefully while calculating the $N_{\rm{H}}$-weighted mean metallicity.  
 
 At high redshift, $z\sim 2.5$--3.5, we assess the metal density of cool gas using a sample of 321 absorbers assembled by \cite{lehner2016,lehner2022}. Their sample includes targets chosen from the Keck Database of Ionized Absorbers toward Quasars (KODIAQ-Z) survey \citep{lehner2014,omeara2015,omeara2017} in addition to 157 absorbers from the HD-LLS survey \citep{prochaska2014,fumagalli2016} and 77 absorbers from literature (see references in \citealt{fumagalli2016}). The whole sample of high-$z$ absorbers spans a wide range in $N_{\rm{HI}}$, $14.6\le\logNHI<20$, and their metallicities were estimated consistently by \cite{lehner2016}. 
 The absorbers comprising both high-$z$ KODIAQ-Z+HD-LLS and the low-$z$ CCC samples are \HI-selected. This selection method has the advantage of giving a relatively unbiased census of the metallicity since it is sensitive to low as well as high metallicities. At both redshifts, the authors used the spectral synthesis code $\rm{Cloudy}$ \citep{ferland2013} to construct grids of ionization models that constrain the posterior probability distributions of (among other parameters) the volume density and metallicity of absorbers (following \citealt{fumagalli2016}). For both the low and high $z$ samples, the HM05 EUV background (EUVB) radiation field \citep{haardt1996} was adopted.\footnote{\cite{fumagalli2016} used the HM12 EUVB \citep{haardt2012} field to determine metallicities for the HD-LLS sample. \cite{lehner2022} repeated the ionization modeling with the HM05 EUVB in their work to eliminate this systematic.} Since the same EUV background models are adopted at $z<1$ and $z\sim2.8$, the systematic effect of the assumed radiation field on the inferred metallicity is mitigated. Thus, any observed evolution in the metallicity between these two redshifts is a real effect. For ionized gas, systematic uncertainties in the derived metallicities resulting from the choice of the UV radiation field can be a factor of 1.5--2.5 \citep{shull2014,wotta2016,wotta2019,chen2017a,gibson2022}. For instance, adopting the harder HM12 field \citep{haardt2012} instead of the softer HM05 field produces a systematic increase of the metallicities by $+0.4$ dex on average \citep{wotta2019,gibson2022}. The ionization corrections derived from the photoionization models are another significant source of uncertainty and are propagated through the neutral fraction function as well as the derived metallicities (detailed analyses of the dependence of the derived ionization corrections for specific ions on the assumed EUVB can be found, e.g., in \citealt{howk2009,shull2014,chen2017a}).
 
Figure~\ref{combined} shows the metallicity as a function of $N_{\rm{HI}}$ for the two redshift regimes. Both indicate an increase of metallicity with increasing $N_{\rm{HI}}$. As seen in Figure~\ref{combined}, a major difference between the two samples is the weighting of the \HI\ column densities covered in each study as a result of observational limitations. While it is possible to use the Lyman decrement to determine $N_{\rm{HI}}$ for the low-$z$ CCC sample, the same cannot be done for the high-$z$ KODIAQ-Z sample. This is because the HIRES spectra used for the high-$z$ sample are flux-normalized before coaddition, making determination of flux decrements unreliable \citep{lehner2022}. Thus, only Lyman series lines can be used to determine $N_{\rm{HI}}$, even for LLSs. For these column densities, most of the absorbers are taken from the HD-LLS survey \citep{prochaska2015}. At higher column densities $\log{N_{\rm{HI}}}\gtrsim17$, the Lyman series lines are all saturated, considerably limiting the sample size until the column density is high enough for damping wings to appear in Ly$\alpha$ absorption ($\log{N_{\rm{HI}}}>18.5$). The low occurrence rate of SLLSs at low $z$ in HST/COS spectra limits the sample to 8 \HI-selected absorbers. We do not include metal line-selected absorbers to avoid any bias when deriving the metal density \citep{meiring2009,lehner2013,som2015}.

In addition, the sensitivity to $\left[\rm{X/H}\right]$ with $N_{\rm{HI}}$ is not constant between low- and high-redshift samples. The KODIAQ-Z sample with higher signal-to-noise ratio (SNR) spectra and higher resolution affordable by Keck/HIRES is sensitive to very low-metallicity absorbers down to $\log{N_{\rm{HI}}}\sim14.5$. The CCC sample observed with HST/COS is limited to metallicity determinations for absorbers with $\log{N_{\rm{HI}}}\geq15.3$ (see Figure~\ref{combined}). The advantage of the two datasets is that they probe the metal densities of absorbers at two very different epochs. They complement each other in redshift (and thus cosmic time), allowing us to study the change in the distribution of metals with cosmic overdensity over cosmic time. Below, we detail the calculation of metal densities for the two samples. We address the nuances, the limitations, and the statistical methods needed to assess $\Omega_{\rm{met}}$ in each class and at each redshift.
\begin{figure}
\begin{centering}
\includegraphics[trim = 6 0 0 0, clip,scale=0.423]{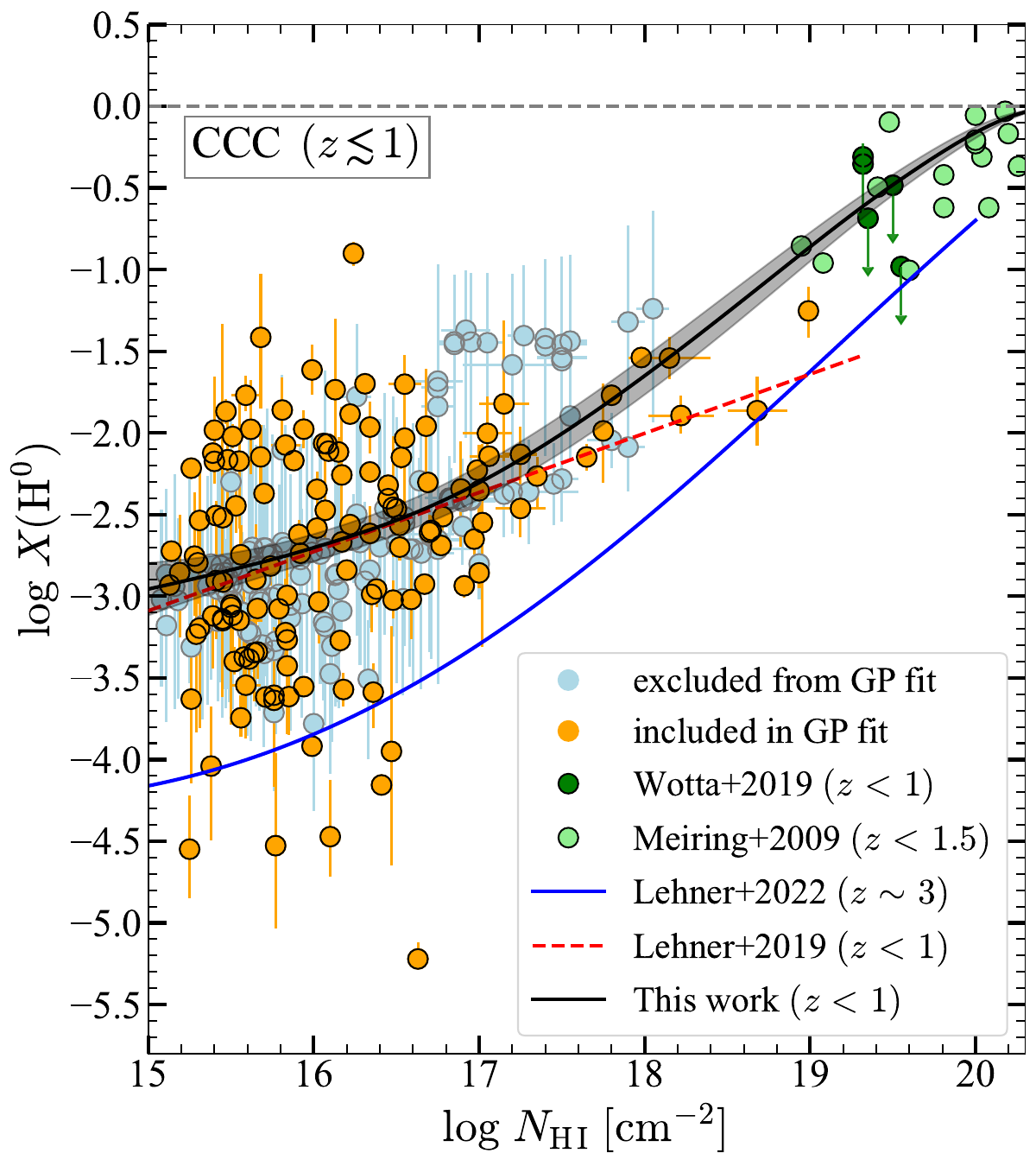}
\par\end{centering}
\begin{centering}
\caption{
Neutral fractions of absorbers $X_{\rm{H}^0}$ as a function of $N_{\rm{HI}}$ for the low-$z$ CCC sample. The black curve is the resulting GP model, and the shaded area around it represents the standard deviation predicted by the fit. Orange circles represent absorbers with a flat prior on $\log\,{U}$, whereas blue circles are absorbers with a Gaussian prior on $\log\,{U}$ and are thus excluded from the Gaussian process fit.}
\label{gp_ccc}
\par\end{centering}
\end{figure}
\subsubsection{Low Redshift Sample $\left(z\lesssim1\right)$:}

To estimate $\Omega_{\rm{gas}}$, and subsequently $\Omega_{\rm{met}}$, for absorption-line selected ionized gas we need the column density distribution function $f(N_{\rm{HI}})$, the neutral fraction $X(\rm{H}^0)$ as a function of $N_{\rm{HI}}$, and the $N_{\rm{H}}$-weighted mean metallicity for each class of absorbers \citepalias{peroux2020}. At low $z$, we adopt the power law CDDF derived by \cite{shull2017} for $0.24\leq z\leq0.84$ \HI-selected absorbers with $15\leq\logNHI\leq17.5$. They fit their results as a power law in $N_{\rm{HI}}$ and $(1+z)$, $f(N_{\rm{HI}},z)=C_0N_{\rm{HI}}^{-\beta}(1+z)^\gamma$, with $C_0=4\times10^7$, $\beta=1.48\pm0.05$, and $\gamma=1.14^{+0.88}_{-0.89}$. The redshift evolution for the CDDF is modest and poorly constrained. Thus, we assume a fixed value of $(1+z)^\gamma$ setting $z\sim0.45$, the $N_{\rm{HI}}$-weighted mean redshift for the CCC sample. \cite{shull2017} derived their CDDF over redshifts consistent with the redshift for which metallicities are available from the CCC sample ($0.2\le z\le0.9$). For our purposes, we adjust the power-law expression by introducing a pivot in $N_{\rm{HI}}$: $f(N_{\rm{HI}},z)=1.53\left[C_010^{-15\beta_0}\right]{\left[{N_{\rm{HI}}}/{10^{15}\,\rm{cm}^{-2}}\right]}^{-\beta}$, where $\beta_0=1.48$. This adjustment is crucial to our calculation of $\Omega_{\rm{gas}}$ as in the absence of a pivot, even slight changes in $\beta$ can lead to significant variations in $N_{\rm{HI}}^{-\beta}$. 

Assessing the total mass density of ionized gas requires correcting the observed \HI\ to the total hydrogen $\rm{H}$ by dividing it by the neutral fraction, $X\left(\rm{H}^0\right)$ (Equation~\ref{om_gas}). We approach this by fitting the typical neutral fractions as a function of $N_{\rm{HI}}$ using the $\rm{Cloudy}$ models generated by \cite{lehner2019}. Following the treatment in \citet{lehner2022}, we use a Gaussian process (GP) model for non-parametric regression to describe the mean $X\left(\rm{H}^0\right)$ as a function of $N_{\rm{HI}}$. This model allows for probabilistic fitting of the data without specifying a functional form. The predictions generated by the GP model can be fit to simpler polynomial forms while retaining information about the posterior distribution. We use the empirical confidence intervals generated by the model to assess uncertainties in the fit. We use the Python $\textsc{gaussian process regression}$ package within $\textsc{scikit-learn}$ \citep{scikit-learn,sklearn_api} to fit the form of $X\left(\rm{H}^0\right)$ versus $N_{\rm{HI}}$. We adopt a Matern kernel in the fit, specifying the variance, the length scale of influence, and the scatter of the fit, approximated by the mean standard deviation of the distribution of absorbers in the neutral fraction space. 

We restricted our dataset to only those absorbers for which robust, multi-ion models were calculated. To anchor the GP fit at large $N_{\rm HI}$, we assume the neutral fraction for DLAs is $X_{\rm{DLA}}(\rm{H}^0)=0.96\pm0.04$. This is consistent with our treatment of neutral gas in section~\ref{subsubsec:neutral_gas}. We include estimates of the neutral fraction of SLLSs from \cite{wotta2019} and \cite{meiring2009}.
These estimates are consistent with the GP fit even when they are excluded from the fit, but their inclusion reduces the uncertainties significantly. The \cite{meiring2009} absorbers are \MgII-selected, so we do not use their metallicity measurements in our global metal density calculations. However, this potential metal bias should not significantly affect the ionization state.

\begin{deluxetable*}{lCCCCC}[t] \label{table:ccc}
\tablecaption{Metal Densities of Gaseous Absorbers}
\tablehead{\colhead{Label}
& \colhead{$\log N_{\rm{HI}}$ $\left[\rm{cm}^{-2}\right]$}
& \colhead{${\Omega_{\rm{gas}}}/{10^{-3}}$}
& \colhead{$\Omega_{\rm{met}}/{10^{-7}}$} & \colhead{$\Omega_{\rm{met}}/\Omega_{\rm{met}}^{\rm{exp}}$\tablenotemark{a}\protect}}
\startdata
\multicolumn{4}{l}{\emph{Low redshift $z\lesssim1$}}\\
 SLFS & $(14.5,16.2]$\tablenotemark{b}\protect & $0.33^{+0.07}_{-0.08}$ & $6.39^{+1.28}_{-1.65}$ & $0.010_{-0.002}^{+0.003}$\\ 
 pLLS & $(16.2,17.2]$ & $0.35^{+0.08}_{-0.11}$ & $1.34^{+0.32}_{-0.42}$ & $0.002_{-0.001}^{+0.001}$\\
 LLS & $(17.2,19.0]$ & $0.41^{+0.13}_{-0.19}$ & $17.86^{+5.78}_{-8.44}$ & $0.029_{-0.014}^{+0.009}$\\ 
 SLLS/sub-DLA & $(19.0,20.3]$ & $0.16^{+0.07}_{-0.11}$ & $7.40^{+3.12}_{-5.24}$ & $0.012_{-0.008}^{+0.005}$\\ 
 DLA & $\geq20.3$ & $0.59^{+0.02}_{-0.01}$ & $42.50^{+8.00}_{-6.90}$ & $0.069_{-0.011}^{+0.015}$\\
 \\
\multicolumn{4}{l}{\emph{High redshift $2.2\lesssim z\lesssim3.6$}}\\ 
 SLFS & $(14.5,16.2]$ & $3.82^{+0.82}_{-0.68}$ & $7.99^{+1.84}_{-1.51}$ & $0.106_{-0.024}^{+0.020}$\\ 
 pLLS & $(16.2,17.2]$ & $1.21^{+0.49}_{-0.31}$ & $0.98^{+0.38}_{-0.27}$ & $0.013_{-0.004}^{+0.005}$\\
 LLS & $(17.2,19.0]$ & $1.17^{+0.41}_{-0.30}$ & $1.47^{+0.56}_{-0.38}$ & $0.020_{-0.005}^{+0.007}$\\ 
 SLLS/sub-DLA & $(19.0,20.3]$ & $1.33^{+0.33}_{-0.27}$ & $5.85^{+1.60}_{-1.23}$ & $0.078_{-0.016}^{+0.021}$\\ 
 DLA & $\geq20.3$ & $0.99^{+0.04}_{-0.03}$ & $23.90^{+5.00}_{-4.10}$ & $0.318_{-0.055}^{+0.067}$\\
\enddata
\tablecomments{ The uncertainties represent 68\% confidence interval. For DLAs, we adopt results from \citetalias{peroux2020}.}
\tablenotetext{a}{The uncertainties here do not incorporate the uncertainty in the expected metal density.}
\tablenotetext{b}{The observed $N_{\rm{HI}}$ of SLFSs in the low-$z$ sample is bounded by $\log\,N_{\rm{HI}}\geq15.3$ below which the UV data is not sensitive to metallicities $\left[\rm{X/H}\right]<-1$. We extrapolate the neutral fraction fit to $\log\,N_{\rm{HI}}=14.5$ for calculating $\Omega^{\rm{SLFS}}_{\rm{gas}}$.}
\end{deluxetable*}

Figure~\ref{gp_ccc} shows the individual absorber neutral fractions, our GP model, as well as the linear fit performed by \cite{lehner2019}. The non-parametric GP fit captures the non-linear relation between the hydrogen neutral fraction and \HI\, column density while also providing a probabilistic framework to constrain the statistical uncertainties. The GP fits at $z<1$ and $z\sim3$ \citep{lehner2019} have similar shapes, with the higher redshift fit being slightly steeper at lower $N_{\rm{HI}}$. At both redshifts, the slope of the GP fit is shallower for lower $N_{\rm{HI}}$ ($\logNHI<16$) and steeper for higher $N_{\rm{HI}}$ ($16\le\logNHI<18$). Absorbers in the low-$z$ CCC sample have higher neutral fractions compared to their high-$z$ counterparts at the same $N_{\rm{HI}}$. This is expected given the lower intensity and softer spectral shape of the low-redshift UV background.
\begin{figure}
\begin{centering}
\includegraphics[trim = 7 35 65 80, clip,width=\columnwidth]{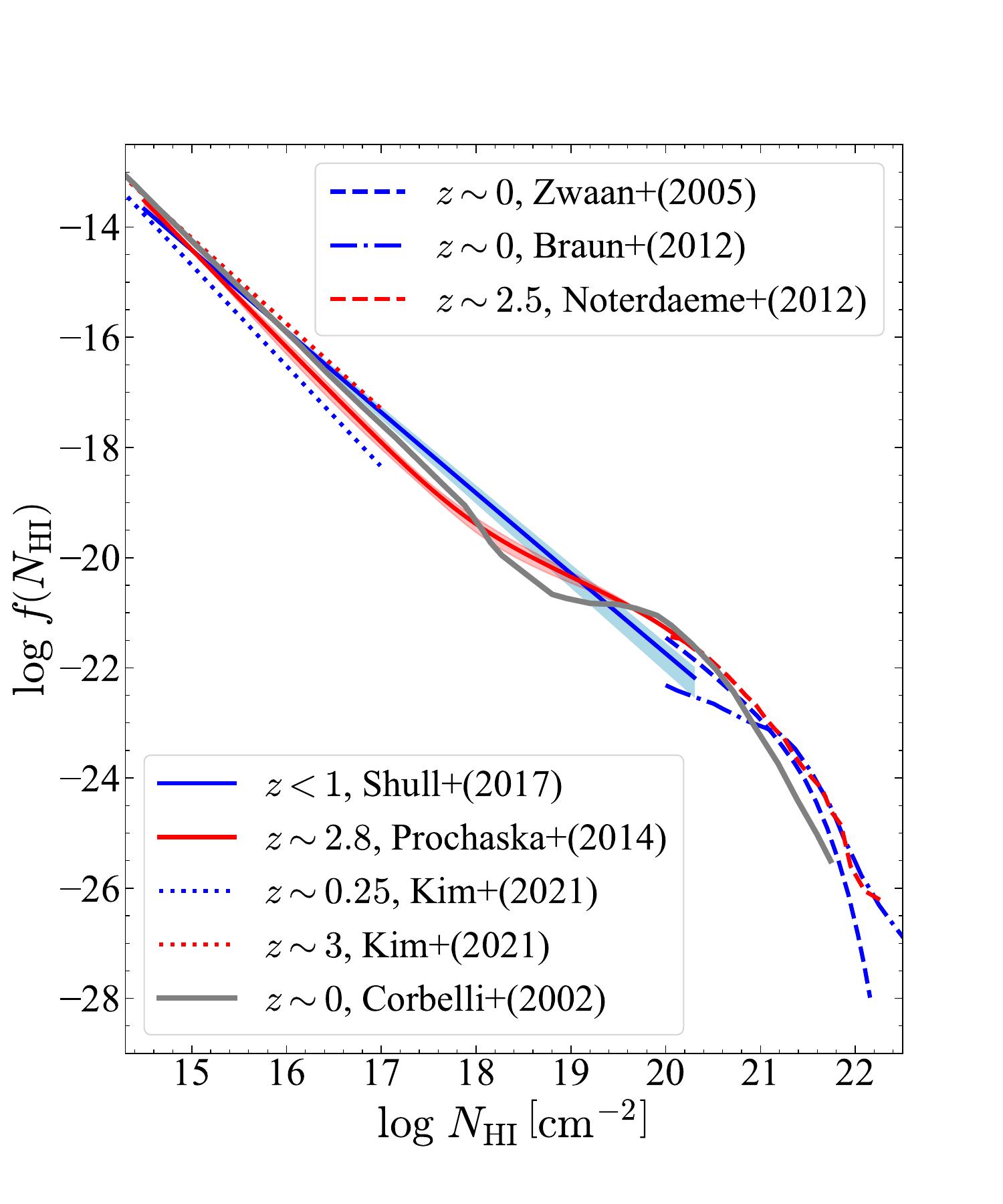}
\par\end{centering}
\begin{centering}
\caption{\HI\ column density distribution functions (CDDFs) at $z<1$ and $z\sim2.8$ with references in the legend. We adopt the \cite{shull2017} and \cite{prochaska2014} CDDFs for our calculations.}

\label{fig:cddf}
\par\end{centering}
\end{figure}
To assess the effect of sample variance and uncertainties in the parameters of $f(N_{\rm{HI}})$ and $X(\rm{H}_0)$ we use Monte Carlo sampling techniques. For the CDDF, we perform a Monte Carlo sampling of the power law index $\beta=1.48\pm0.05$, assuming a normal distribution (using 10,000 samples).  
To account for uncertainties in the neutral fraction $X(\rm{H}^0)$, we generate 10,000 realizations from the posterior distribution of the GP model. Figure~\ref{fig:cddf} shows realizations of the CDDF plotted within a 68\% confidence interval. We also plot CDDFs from different analyses at the two redshifts $z<1$ \citep{zwaan2005,corbelli2002,braun2012,shull2017,kim2021} and $z\sim2.8$ \citep{noterdaeme2012,prochaska2014,kim2021}.

We integrate Equation~\ref{om_gas} using the multiple realizations of the functions $f(N_{\rm{HI}})$ and $X(\rm{H}^0)$ to obtain a posterior distribution for the gas density, $\Omega_{\rm{gas}}$, for each column density class. For determining $\Omega_{\rm{met}}$, we calculate $\langle Z\rangle_{N_{\rm{H}}}$ separately for each of the \HI\ column density regimes and derive their piecewise contributions to $\Omega_{\rm{met}}$. Saturation, contamination, and non-detection of metal lines result in both lower and upper limits for derived metallicities (see  Figure~\ref{combined} and \citealt{lehner2022}). To find the sample average in the presence of left-censored (upper limits) or right-censored (lower limits) data points, we use survival analysis techniques. Specifically, we use the non-parametric Kaplan-Meier estimator from the Python package $\textsc{sksurv}$ \citep{sksurv} in $\textsc{scikit-learn}$ \citep{scikit-learn,sklearn_api}. The Kaplan-Meier distribution accurately represents the true sample distribution when two conditions are satisfied---the upper/lower limits are statistically independent of one another, and the probability of their occurrence does not depend on sample selection. Since the metallicities were derived using resolved metal lines without significant contamination from other absorbers, and all the absorbers used in this work are \HI--selected, the samples satisfy both conditions.\footnote{See section 3.1 in \cite{lehner2022} for a detailed description on the treatment of contaminations.} 
 
We perform left-censored survival analysis \citep{feigelson1985} for SLFSs that have some non-detections, and right-censored survival analysis for LLSs that have some saturated lines and lower limits for metallicities. The pLLSs subset in the CCC sample has both upper and lower limits for metallicities. In this case, we analyze the effects of excluding the right-censored data with lower limits and the left-censored data with upper limits, one at a time. We find that the presence of upper limits marginally affects the mean metallicity, whereas $\langle Z \rangle_{\rm{N_H}}$ is very sensitive to the lower limits. This is expected because upper limits are bounded within the detection limits, whereas lower limits are unbound and can be very large in magnitude. Thus, we fix the upper limits and perform the standard right-censored survival analysis for the pLLSs. For each of the column density bins, we use 10,000 bootstrap realizations of the sample to compute the posterior distribution of $\langle Z \rangle_{\rm{N_H}}$. We multiply these with the $\Omega_{\rm{gas}}$ samples to obtain posterior distributions of $\Omega_{\rm{met}}$. We summarise our results as the median value of the $\Omega_{\rm{met}}$ PDF with $68\%$ confidence interval for each class of absorber in Table~\ref{table:ccc}.

\subsubsection{High Redshift Sample $\left(2.2\lesssim z \lesssim 3.6\right)$:}

At high $z$, we adopt the spline fit from \cite{prochaska2014} to describe the CDDF. They report their results as a set of anchor points describing the fit. We perform Monte Carlo sampling, assuming a skew-normal distribution and generate 10,000 sets of anchor points following Section~\ref{subsec:stars}. Interpolations of each of these anchor sets yield 10,000 realizations of the CDDF. For the neutral fraction, $X(\rm{H}^0)$, we adopt the results from the GP model generated by \cite{lehner2022}. As in the case of the low-redshift data, we sample the posterior and use the predictions along with the CDDF realizations to calculate the gas density $\Omega_{\rm{gas}}$ and, ultimately, $\Omega_{\rm{met}}$ for each class of absorber. The high redshift sample does not have any lower limits. We use the left-censored survival analysis technique described in the previous paragraph for the estimation of $N_{\rm{H}}$-weighted mean metallicity. 

Table~\ref{table:ccc} summarizes our results for this section. We note that, unlike neutral gas, the metal densities of the ionized gas are not dust-corrected. Both at low and high redshifts, \cite{lehner2018a} and \cite{fumagalli2016} studied the dust depletion for $15\leq\log\,N_{\rm{HI}}\leq19$ absorbers and found it to be negligible. We discuss the contribution of dust to global metal density in the next section.

\section{Other contributors to the global metal budget}
\label{subsec:others}


Our global-scale census of metal reservoirs in the Universe is inclusive of contributions that are well-characterized, amenable to robust uncertainty estimation, free from systematic biases, and not subject to ``double counting" of metals. There are, however, contributors to the global metal budget of the Universe that we have not included in our census, as our current determinations of their metal content do not meet these criteria. These include metals associated with (1) the cool and (2) the warm-hot intergalactic medium or the hot gas in the circumgalactic medium of galaxies, (3) metals associated specifically with the molecular medium in galaxies, and (4) metals bound in dust grains beyond the ISM of galaxies. We discuss these contributors below.

\subsection{Cool Intergalactic Medium $(T<10^5\,\rm{K})$} \label{cool_igm}

The IGM represents the majority of all the baryons in the Universe, including $>90\%$ of all baryons at $z\ge1.5$ \citep{meiksin2009} and even $\approx80\%$ at $z\la 0.4$ \citep{shull2012, wang2023}. Given its large baryonic content, the IGM is potentially also a major reservoir of metals, one that is not included in our census. 

The cool diffuse IGM manifests itself observationally as the LAF. At high redshift, the LAF houses most of the baryons in the Universe \citep{rauch1998, mcquinn2016}. Even so, studies of IGM metals find its metal mass density is negligible compared to the neutral gas \citep{schaye2003,simcoe2004,aguirre2004,songaila2005,scannapieco2006,simcoe2011b,boksenberg2015,cai2017,dodorico2013,dodorico2022,davies2023}, contributing only $\approx2$--$4\%$ of the total metal budget. For example, based on \ion{C}{4} measurements by \cite{dodorico2013}, \citetalias{peroux2020} estimate $\Omega^{\rm{LAF}}_{\rm{met}}\simeq1.4\times10^{-7}$  at $z\sim3$, about $2\%$ of $\Omega_{\rm{met}}^{\rm{exp}}$ at this redshift. \cite{simcoe2011b} similarly estimated the ionization-corrected carbon abundance of the intergalactic medium using \ion{C}{4} detections in individual absorbers at $z\sim 2.4$--4.3. If we assume solar relative abundance ratios, their results translate to $\Omega^{\rm{LAF}}_{\rm{met}}(z=2.4)\simeq 2.6\times10^{-7}$ and $\Omega^{\rm{LAF}}_{\rm{met}}(z=4.3)\simeq1.5\times10^{-7}$, consistent with the \cite{dodorico2013} results and account for only 2\% and 7\% of the total metal budget at these redshifts. Other surveys spanning $z\sim$1.9--4.5 give similar results---a steady rise in $\Omega_{\rm C IV}$ and thus $\Omega^{\rm{LAF}}_{\rm{met}}$ with cosmic time \citep{songaila2001,songaila2005,aguirre2004,scannapieco2006,cooksey2013,boksenberg2015}.

We do not include these estimates in our census for several reasons. First, they often rely on an assumed baryon budget rather than one derived from the data themselves. Second, \cite{simcoe2004} (see their Figure 2) show that most of the strong \ion{C}{4} absorbers have neutral hydrogen column densities in the range $15\lesssim\log\,N_{\rm{HI}}\lesssim16$, overlapping the lowest column density absorbers already included in our census (the SLFSs). We do not include these studies in our census to avoid double counting.  


At low redshift, determining the metallicity of the LAF is challenging because: (i) detecting metals in the low $N_{\rm{HI}}$ absorbers requires higher SNR spectra than are typically available in space UV data; (ii) even when metals are detected, often not enough ions in differing ionization states are detected to allow the detailed ionization modeling required to derive metallicities of individual absorbers; and (iii) when ions such as \CIV\ or \OVI\ are detected in LAF absorbers at low redshift, there is an ambiguity in the ionization mechanism that makes it difficult to determine the total metal content (see Section~\ref{subsec:whim}), as they may be photoionized by the UV background as part of the cool LAF, collisionally ionized as part of the WHIM, or a combination of both.

While we do not include these metal reservoirs in our formal census, we mention some of these estimates here to give a sense of their potential contribution to the total \citep{cooksey2010,danforth2008,tilton2012,danforth2016}. \citet{shull2014} present a comprehensive estimate of the global metal density at $z\le0.4$ based on a survey of metal lines (which in their survey probe \HI\ column densities $12 \la \log \NHI \la 17.5$). They assume \ion{C}{3}, \ion{C}{4}, \ion{Si}{3}, and \ion{Si}{4} absorption are associated with photoionized gas, while \ion{O}{6} absorption traces collisionally-ionized gas. They estimate the combined contribution as $\Omega_{\rm{met}}=\left(1.1\pm0.6\right)\times10^{-5}$, approximately 18\% of $\Omega_{\rm{met}}^{\rm{exp}}$ at $z\le0.4$. It is important to note the assumption that these metal ions can be uniquely associated with each ionization mechanism may lead to double counting (see \citealt{tripp2008,stern2018}). 

The \citet{shull2014} estimates are derived by integrating the column density distribution functions for each of the ions as a function of the ionic column density and applying global ionization correction estimates. Given the broad range of \HI\ column densities sampled by this survey, and the propensity for the highest column density metal lines to be associated with the highest \HI\ column density absorbers, this approach will be highly-influenced by metals associated with absorbers that strongly overlap our selection, and the results do not uniquely apply to the IGM as we have defined it in our work.

\subsection{Hot CGM and IGM Gas $(T\sim10^{5-7}\,\rm{K})$}
\label{subsec:whim}

Cool gas in the CGM and IGM is included in our census of metals through the \HI -selected absorbers tracing $T\approx10^{4-5}\,\rm{K}$ gas. The hotter gas described as ``warm-hot'' ($T\approx10^{5-6}\,\rm{K}$) or ``hot'' ($T\approx10^{6-7}\,\rm{K}$) present in galaxy halos \citep[e.g.,][]{tumlinson2017,chen2024} and the IGM \citep[e.g.,][]{bertone2008,shull2012} is not included in our census. This gas is a particularly important contributor to the baryon budget of gas at $z<1$, where the high-temperature IGM (WHIM) may contain as much as $\approx 50\%$ of the all the baryons \citep{cen2006, mcquinn2016}. CGM absorption experiments find that higher-ionization gas that may trace this hotter material can account for a substantial fraction of the total metal budget within \lstar\ galaxies, especially at low redshift \citep[e.g.,][]{peeples2014}. However, the uncertain ionization state of the gas \citep[e.g.,][]{howk2009,hussain2015,werk2016,pachat2017},  uncertainties in separating the contributions from the WHIM and the warm-hot CGM (to avoid double counting), and observational limitations make a robust accounting (with reliable uncertainties) of the metal mass contribution difficult. 

To date, the largest sample of detected metals in the WHIM and warm-hot CGM come from UV absorption line studies of \OVI, \ion{Ne}{8}, and other ``highly-ionized'' metal ions \citep{savage2005,tumlinson2011,narayanan2012,savage2014,stocke2014,pachat2017,frank2018,burchett2019,tchernyshyov2022,qu2024,sameer2024}. However, even with detailed ionization modeling, the origins of the gas traced by these ions are not easy to constrain. \OVI\ may trace the collisionally-ionized hot phase or a low-density photoionized gas, whose metal contributions may already be included in studies of the cool gas, or a combination of both \citep{howk2009,werk2016,oppenheimer2016,gnat2017,pachat2017}. Even if the absorption is known to trace hot gas (which is usually the case for \ion{Ne}{8}), this does not uniquely constrain the ionization fraction needed to transform the ionic column densities accessible in UV spectra into total metal densities. Working directly from observations can yield lower limits to the total contribution to the metal density, but these may be an order of magnitude or more below the true metal content of that phase.

X-ray observations of highly ionized metals (\ion{O}{7} and \ion{O}{8}) that represent the dominant ionization states of the warm-hot gas can be used to estimate the contribution of this material to the total metal mass density. However, the limitations of current X-ray instrumentation make the detection of this gas very difficult, and compiling robust surveys is beyond our current capability. Thus, we do not include results based on these limited datasets in our census. 
Future missions such as the Line Emission Mapper \citep{patnaude2023} and the Athena X-ray observatory \citep{nandra2013} with significantly higher energy resolutions and sensitivity may allow robust surveys of weak metal lines tracing the WHIM and the warm-hot CGM \citep{nicastro2021}. However, even the currently planned missions may be limited to detecting only the most metal-enriched portions of the WHIM, those closest to galaxies \citep{tuominen2023}.

\subsection{Molecular Gas: $\rm{H}_2$}
\label{subsection:molecular}

We have not included in our census the metal contribution from molecular hydrogen gas \citep{klitsch2019,tacconi2020,walter2020,guo2023,Hamanowicz2023,bollo2025}. This cold phase ($T\sim100\,\rm{K}$) hosts star formation in galaxies and draws its fuel from the neutral atomic hydrogen reservoirs. The molecular phase represents a significant amount of mass: the global mass density of molecular gas reaches as high as $\sim40\%$ the neutral gas density at redshift $z\sim 1.5$--2 (\citetalias{peroux2020}; \citealt{bollo2025}), which corresponds to the peak in the cosmic star formation rate density. At lower redshifts ($z<1.5$), molecular gas mass can be $\sim25\%$  as high as the neutral gas mass.  At higher redshifts ($z>4$), most of the baryonic gas mass resides in neutral gas, and the mass density of $\rm{H}_2$ gas represents $<8\%$ of neutral gas density at $z\sim4.5$ \citep{guo2023,aravena2024,casavecchia2024,bollo2025}. It is difficult to assess the metallicity distribution of the molecular gas directly. Molecular gas is typically studied by observing the integrated flux of CO rotational lines, which is used to estimate the molecular gas mass by applying a conversion factor \citep{decarli2019,tacconi2020,decarli2020,walter2020,Boogaard2023}. There exist studies investigating the correlation between molecular gas mass and gas phase metallicities of galaxies as measured from \ion{H}{2} region metallicities \citep{boogaard2020,sanders2023}. However, the lack of an appropriate statistical sample and a consistent treatment of metallicities prevents us from deriving a reliable $\Omega_{\rm{met}}$ for molecular gas. Moreover, the uncertainties in the $\rm{CO}$ to $\rm{H}_2$ conversion factor also limit the precision of the estimates of the molecular gas density and its uncertainties. 

While we do not include an estimate of the metals found in the cold molecular phase in our census, we can roughly estimate its contribution if we assume the molecular gas has the same metallicity distribution as the cool neutral atomic gas traced by DLAs. For the molecular gas mass density, we adopt results from \cite{bollo2025}. Using the ALMACAL-22 survey of CO-selected galaxies, as well as previous estimates from the literature, \citeauthor{bollo2025} constrained the molecular gas density across $z\sim0$--6 while addressing cosmic variance effects. Their updated estimates are lower than those reported in \citetalias{peroux2020}. Combining their mass distribution with the metallicity estimates for the neutral atomic gas, we find that the metal content of the molecular gas peaks at $z\sim2$, with a metal mass $\sim40\%$ of that in the neutral gas phase or $\sim7\%$ of the expected total metal density. Thus, the molecular gas phase is never expected to be a dominant contributor to the metal mass density of the Universe, but it does make a modest contribution to the whole.

\subsection{Dust}
\label{subsec:dust}

In the ISM of galaxies, a significant fraction of the metals are locked into solid-phase material that is not directly measured in our accounting of metal reservoirs based on absorption line studies (e.g., \citealt{jenkins2009}, \citealt{jenkins2017}, \citetalias{peroux2020}, \citealt{roman-duval2022}, \citealt{konstantopoulou2024}). For example, in the Milky Way, roughly $45\%$ of the total metal mass is locked into solid-phase dust grains \citep{draine2007, draine2014}. We do not separately catalog the contribution of dust to the total metal budget of the Universe. However, the neutral gas (DLA) metal budget is inclusive of the metals locked into dust (\citealt{peroux2023}, \citetalias{peroux2020}). The individual DLA metallicity measurements follow the approach of \citet{decia2018} to correct the measured gas-phase abundances to total (gas+dust) metal abundances using relative elemental abundances and a characterization of the differential incorporation of those elements into dust grains. The total dust contribution to $\Omega^{\rm{neut}}_{\rm{met}}$ ranges from $\sim20\%$ at $z\ga3$ to $\sim30\%$ at $z\la2.5$ (using data from \citetalias{peroux2020}). Thus, while dust is not separately included in our census, its impact is accounted for in the neutral gas metal budget.

Determining metallicities for low column density systems requires complex ionization models. These models must account for the interplay between ionization, non-solar abundances, and depletion when analyzing mixed refractory and volatile elements \citep{fumagalli2016,quiret2016,wotta2019}. However, lower (column) density absorbers typically trace more diffuse, CGM/IGM-like gas. In this case, the gas is typically more metal-poor, and hence intrinsically more likely to have lower dust content \citepalias{peroux2020}), and any dust that does exist is often exposed to harsher conditions, leading to more rapid destruction (and hence the return of the metals to the gas phase). That being said, the ions used to derive metallicities in these systems (typically using $\alpha$-elements such as Si, Mg, or even O) are not as prone to dust depletion as the refractory elements, and even the refractories seem to be only lightly depleted, if at all, in these systems. \cite{lehner2018a} and \cite{fumagalli2016}, both find the impact of dust depletion on the metallicities to be negligible. Thus, we have not performed dust correction for these systems.

\begin{figure*}
\begin{centering}
\includegraphics[trim = 0 40 0 46, clip, width=0.75\linewidth]{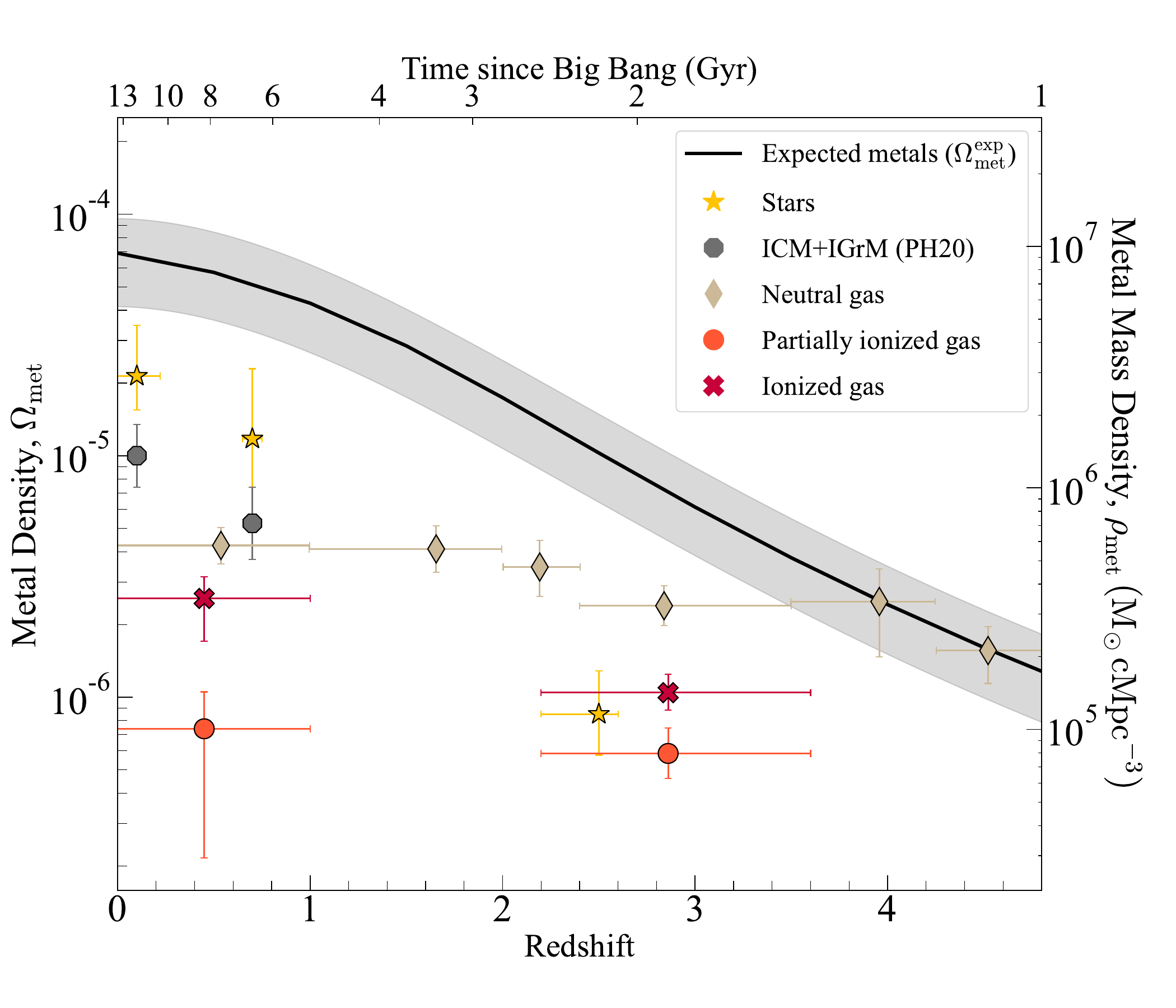}
\par\end{centering}
\begin{centering}
\caption{
{A global census of the metals in the Universe traced by the evolution of the global metal densities with cosmic time compared with the expected amount of total metals ever output $\Omega_{\rm{met}}^{\rm{exp}}$. Here, ionized gas metal density is a sum of the global metal densities of SLFSs, pLLSs, and LLSs. The partially ionized gas is traced by SLLSs.}}
\label{omega_met}
\par\end{centering}
\end{figure*}

\section{Evolution of Cosmic Metal Budget}
\label{sec:budget}
\noindent

\subsection{Global Metal Densities of Reservoirs}

We plot in Figure~\ref{omega_met} the global metal densities as a function of redshift for the reservoirs cataloged in our census. We plot the cosmic metal densities of stars, ionized gas (SLFS, pLLS, and LLS in Table~\ref{table:ccc}), partially ionized gas (sub-DLAs/SLLSs), and neutral gas (traced by 21 cm studies at low $z$ and DLAs at higher redshift). We also plot the expected total metal density derived using Equation~\ref{eq:expmet} for an integrated metal yield $y=0.033\pm0.010$ \citep{peeples2014}. The grey region around the curve represents the uncertainty in $\Omega^{\rm{exp}}_{\rm{met}}$ due to the uncertainty in the yield $y$ derived from chemical evolution models. 

The global metal density of the neutral gas increases with time. However, its relative contribution to the total metal density decreases drastically (see Table~\ref{table:dlas}) as the expected metal density rapidly increases and the metal reservoirs become more diverse at low redshifts ($z\sim0.5$). The steady increase in the metal density with decreasing redshift in neutral gas suggests metals are deposited in the condensed gas clouds traced by DLAs. This may indicate the recycling of cool enriched gas, producing the next generation of stars. The diverse reservoirs at low redshift include cool, ionized, and partially ionized gas, such as that observed in the CGM or the denser regions of the IGM. These absorbers have baryon overdensities ($\delta=\rho_b/\bar{\rho}_b$) of the order of $\delta=10^1$--$10^3$ \citep{schaye2001}. The higher overdensities ($\delta>200$) most likely probe CGM-like gas (SLLSs, LLSs); the lower \HI-column density gas (pLLSs, SLFSs) have lower overdensities ($\delta<200$) and may also be associated with the denser IGM. Metallicity distribution studies offer another indirect method to link \HI-selected absorbers to CGM or IGM structures \citep{battisti2012,bouche2012,johnson2013,lundgren2021,hamanowicz2020,weng2023,berg2023,weng2024}. Notably, \cite{wotta2019} found that the metallicity distributions of pLLSs and LLSs in the low-$z$ CCC dataset exhibit bimodality. \cite{berg2023} showed that this bimodality in the metal distribution of pLLSs and LLSs can be related to their association (or lack thereof) with galaxies. They demonstrated that we are more likely to find high-metallicity absorbers tracing the CGM of galaxies ($\delta>10^2$), whereas a large fraction of the low-metallicity absorbers may instead characterize overdense regions ($\delta\sim10^1$--$10^2$) of the Universe that may be more representative of the IGM. We find the global metal density of LLSs increases by a factor of $>10$ from $z\sim2.8$ to $z\lesssim1$, whereas the metal densities of pLLSs and SLFSs remain nearly constant (see Table~\ref{table:ccc}). This suggests that metals are more readily deposited in the denser CGM (resulting in the large increase in the global metal density of the LLSs), and the pollution of lower-density clouds tracing the IGM (SLFSs and pLLSs) is a slower process. The timescales of such enrichment can be better constrained by studying the distribution of metals in the CGM and the IGM at intermediate redshifts. This will require a comprehensive analysis of CGM-like ($\delta\sim10^2$--$10^4$) and IGM-like gaseous absorbers ($\delta<10^2$) at $z\sim 1$--2.

\begin{figure}
\begin{centering}
\includegraphics[trim = 7 35 65 80, clip,width=\columnwidth]{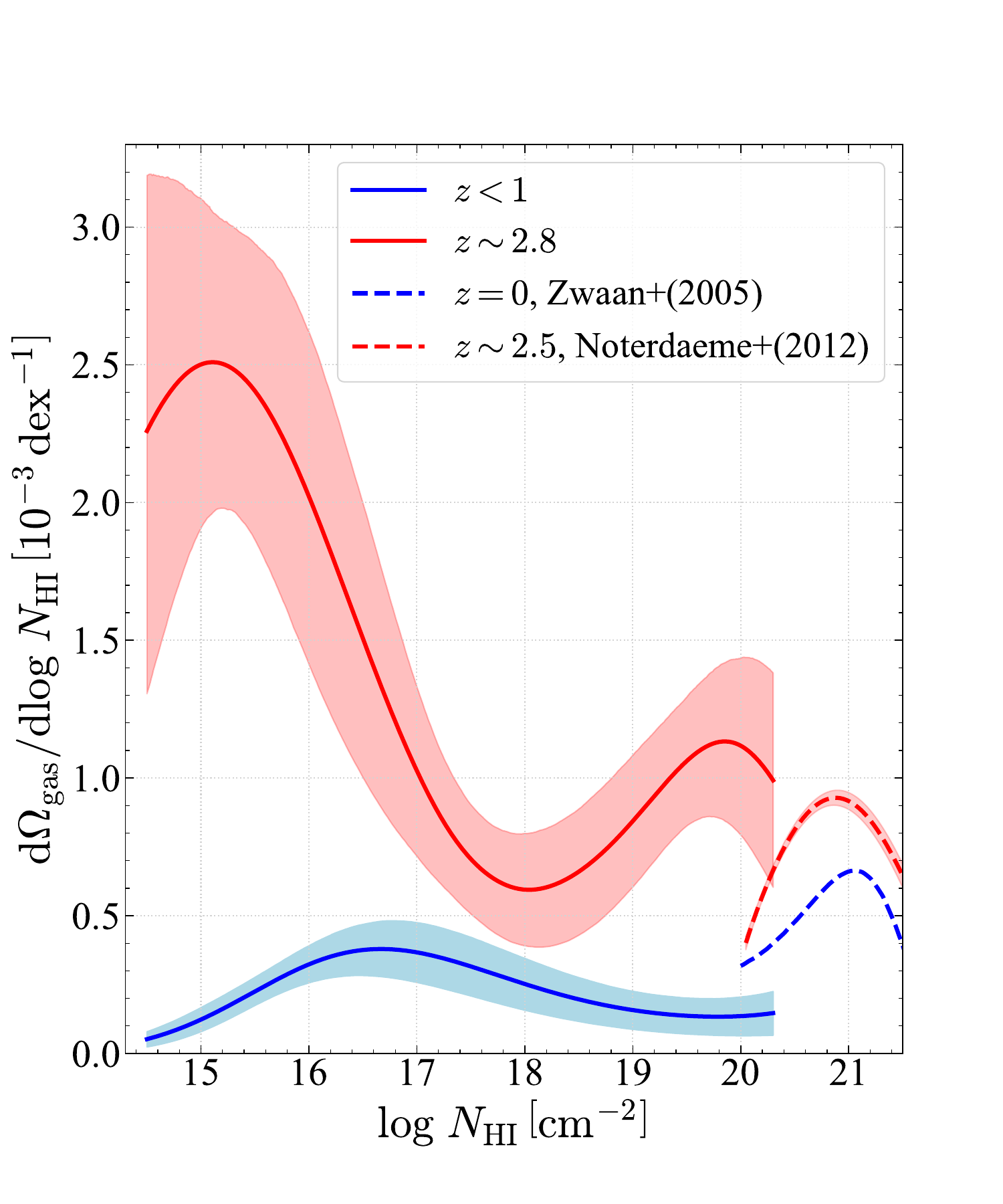}
\par\end{centering}
\begin{centering}
\caption{Differential contributions to the cool gas mass density, $\Omega_{\rm gas}$, with $\log\,N_{\rm HI}$ shown as ${\rm{d}}\Omega_{\rm{gas}}/{\rm{d}}(\log N_{\rm{HI}})$. The area under the curve over any column density interval gives the total gas density (\HI\, and ionized) $\Omega_{\rm gas}$ of absorbers in that regime. For the DLAs, we adopt the \HI\,mass density distribution functions from \cite{noterdaeme2012} and \cite{zwaan2005} and neutral fraction estimates from \cite{vladilo2001}. The lower column density results are from \cite{shull2017} at $z<1$ and \cite{prochaska2014} at $z\sim2.8$.}
\label{fig:CDDF_OmG}
\par\end{centering}
\end{figure}

To assess the contributions of the different column density regimes to the total metal density of cool gas ($T\lesssim 10^5\,\rm{K}$), we show their contributions to the total baryon budget by plotting gas density distribution, ${\rm{d}}\Omega_{\rm{gas}}/{\rm{d}}(\log N_{\rm{HI}})$ in Figure~\ref{fig:CDDF_OmG}.\footnote{The gas density distribution is estimated using: \begin{equation*}
    \frac{1} {\ln{10}}\frac{{\rm{d}}\Omega_{\rm{gas}}}{{\rm{d}}(\log N_{\rm{HI}})}=\frac{\mu m_{\rm{p}}H_0}{c\rho_c}\frac{f(N_{\rm{HI}})N_{\rm{HI}}^2}{X(\rm{H}^0)}.
\end{equation*}}. Integrating this function over bins of $\log N_{\rm{HI}}$ gives us the total gas density contributed by absorbers whose \HI\ column densities lie within those bins. We use this to assess how the different overdensity regimes, characterized by their column densities, are weighted in terms of their contribution to the total $\Omega_{\rm{met}}$. At low redshift, most of the total mass density of cold gas comes from higher \HI\ column density gas, more specifically absorbers with $\logNHI>16$. This corresponds to pLLSs, LLSs, and SLLSs, which are often associated with the CGM and denser IGM. On the other hand, at high redshift, most of the mass of cold gas is found in lower $N_{\rm{HI}}$ absorbers. For all column densities, the gas densities are higher at $z\sim2.8$ than $z<1$. This decline in $\Omega_{\rm gas}$ is most pronounced in SLFSs, followed by the pLLSs and the LLSs (see Table~\ref{table:ccc}). However, the mean metallicities increase with cosmic time for all column density regimes. The combination of these two factors results in a large increase in $\Omega_{\rm met}$ of LLSs, whereas the pLLSs experience a much smaller rise. The $\Omega_{\rm{met}}$ of SLFSs, on the other hand, decreases with time. This change in the distribution of metals is driven by the changing distribution of baryonic mass among these $N_{\rm HI}$ regimes. It likely results from the cool diffuse intergalactic gas at high $z$ being heated to warm-hot intergalactic gas within cosmic filaments (not included in our census) and the denser gas at low redshift, a result of the large-scale collapse and condensation of cosmic gas with time \citep{dave2010}. The peak in the gas mass density distribution at higher column densities ($\logNHI\gtrsim20$, see Figure \ref{fig:CDDF_OmG}) indicates that a large fraction of the cool gas also resided in the dense regions traced by DLAs. This results in the large observed metal density of neutral gas, especially at high redshift ($z\gtrsim4$), where almost all metals reside in the neutral gas traced by DLAs.

\begin{figure*}
\begin{centering}
\includegraphics[trim = 40 15 47 0,clip,width=0.85\textwidth]{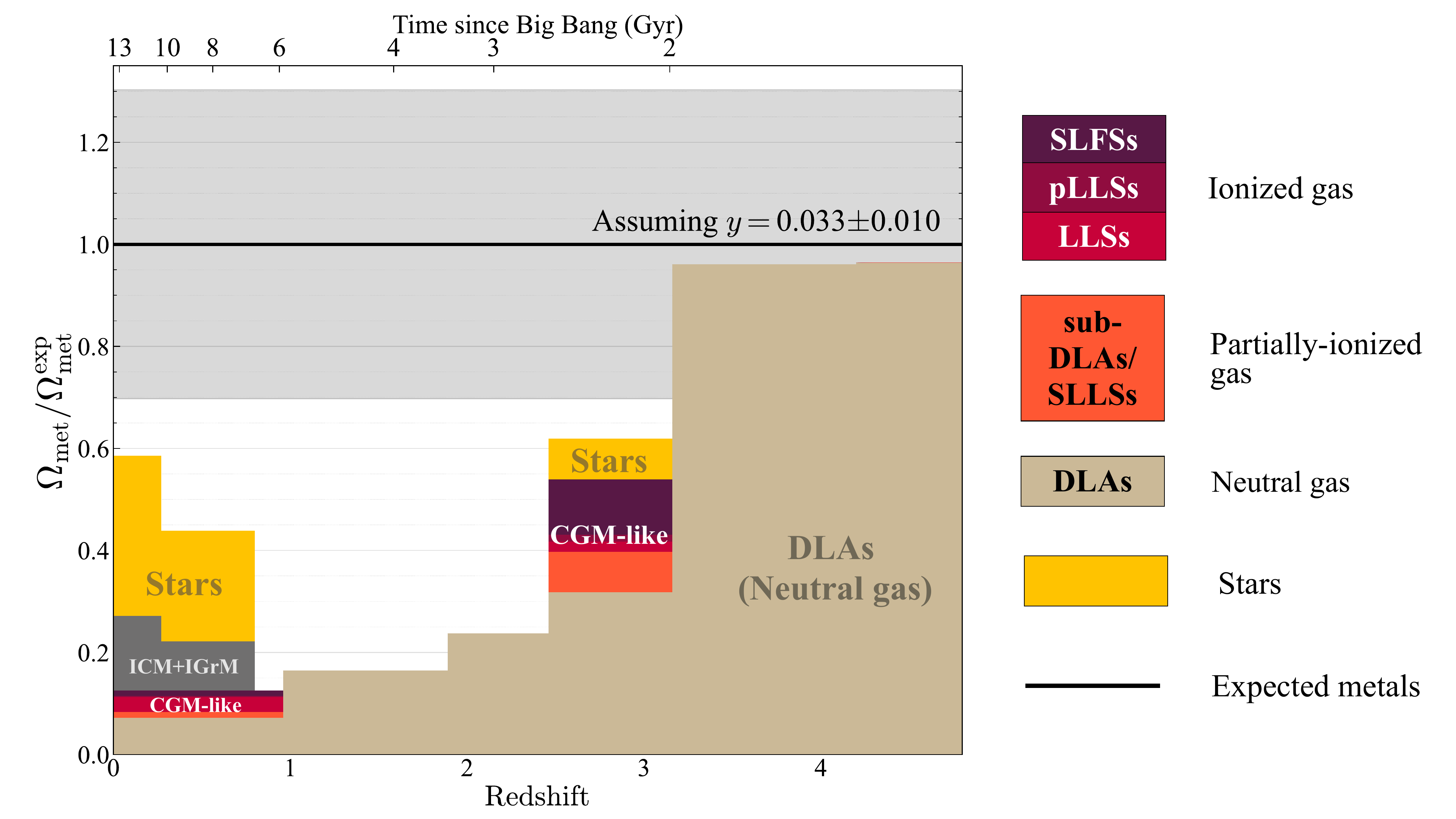}
\par\end{centering}
\begin{centering}
\caption{
A ``sand plot" of the normalized contributions of each of the reservoirs to the total expected metal density of the Universe, as a function of redshift. Here, we have shown the contributions for all the classes of \HI\ absorbers to compare their relative significance to the cosmic metal budget. The bin size along $z$ for each component is fixed to show the redshift range covered by its source survey. An empty bin indicates the absence of robust metal density estimates. The grey region around the expected metals marks the uncertainty in $\Omega^{\rm{exp}}_{\rm{met}}$.}
\label{global_stack}
\par\end{centering}
\end{figure*}

We can now compile our results into the global metal budget as a function of redshift. Figure~\ref{global_stack} shows the stacked contribution of each of the metal reservoirs normalized to the expected total metal density of the Universe as a function of redshift. As evident from Figure~\ref{global_stack}, for the assumed yield, at low redshift ($z\sim0.5$), the largest contributor to the global metal budget is the metal mass in stars. The contributions from the hot gas in the ICM, cool neutral gas, and ionized gas are all comparable, demonstrating the diversification of metal reservoirs with cosmic time.

Using a fundamentally different approach to estimating the metal budget, \cite{molendi2024} report that most of the metals today have not yet been cataloged. We note on one hand that their census focuses on $z=0$, while we are referring here to $z\le1$. Second, as the estimated amount of metals in stars strongly relies on the assumption of the yields \citep[i.e.][]{peeples2014}, the findings of \cite{molendi2024} that the fraction of metals in stars is $0.15$--0.28, are consistent with the results reported here within the errors (we find $31^{+20}_{-8}$\% of $z\sim0.1$ metals in stars). Finally, our census also accounts for metals observed in neutral and partially ionized gas detected in absorption, whose contribution ($\sim$10\%) naturally adds to the fraction of metals already accounted for at low redshifts. We conclude that our results are consistent with previous findings published in the literature within the uncertainties. 

We note that the large uncertainty in the assumed yield ($y=0.033\pm0.010$) implies significant uncertainty in the expected metal content of the Universe. This remains among the largest uncertainties in assessing whether the global metal budget can be considered ``closed.''

We may now revisit the ``missing metals" problem. Given the likely contribution of metals in the IGM and the warm-hot CGM, our results suggest our census of cosmic metals is consistent with the total expected metal density of the Universe. The overall trend of $\Omega_{\rm{met}}/\Omega^{\rm{exp}}_{\rm{met}}$ in Figure~\ref{global_stack} demonstrates the global cycling of metals as the Universe matures, with---(i) star formation and metal production in the cool dense gas at $z>3$, (ii) recycling of gas in the CGM at $z\sim$2--3, (iii) a rise in the global star formation rate (and its subsequent decline) resulting in a large increase in the stellar metal density at $z\sim$1--2, and (iv) redistribution of metals by outflows into the hot gas in galactic halos, the ICM/IGrM and the IGM at $z<1$. At all cosmic times, metals are being produced in the centers of galaxies, causing an ``inside-out" chemical enrichment of the Universe. The cool, neutral, dense gas traced by DLAs is enriched first and remains the dominant metal reservoir until $z\sim4$. This is followed by the stellar feedback-fueled dispersion of metals into the CGM, which recycles and mediates the metal transfer between the ISM of galaxies and the IGM. This is observed in the rise of the global metal density of CGM-like gas (Figure~\ref{global_stack}), which hosts $\sim20\%$ of the expected metals at $z\sim2.8$. Simultaneously, the next generation of stars is formed within the enriched ISM, while low mass stars and stellar remnants trap metals, contributing to a monotonic increase in the global stellar metal density with cosmic time ($\sim8\%$, $\sim21\%$, and $\sim31\%$ of the total expected metal density at $z\sim2.5$, $0.7$, and $0.1$, respectively). The outflows powered by supernovae and AGN (which peak at intermediate redshifts $z\sim$1--2.4) pollute the IGM, which contributes $\sim18\%$ of the total at $z\leq0.4$ and would most likely close the gap between the observed and the expected total metals of the Universe at $z<1$ (see Section \ref{cool_igm}).

We note a gap in our census across ``cosmic noon" at $z\sim 0.9$--2.4. This is a critical epoch in the evolution of the global properties of the Universe, characterized by a peak in the star formation rate density as well as increased AGN activity. The enhanced stellar and AGN feedback during this period also makes it a pivotal point in galaxy evolution \citep{maiolino2019}. For \HI-selected cool gas, this epoch has remained a challenge due to technical limitations. While there exist some \HI\, surveys at this epoch \citep{janknecht2006,omeara2013}, they lack corresponding metallicity measurements. At these redshifts, some key gas diagnostics lie in the near UV, while others lie in the optical. Identifying \HI\, absorbers and measuring their column densities at $0.9$--1.6 requires space-based UV observations in a wavelength regime at which Hubble’s instruments are not as sensitive (NUV). This is even more problematic for the weak metal lines of IGM-like absorbers. However, many metal lines for absorbers are accessible in the far blue of the optical band. In future papers, we will couple HST observations of the Lyman series and strong metal lines with high-quality, high-resolution ground-based spectra of metals to calculate the metallicities of CGM and IGM-like gas at cosmic noon. This will allow us to study the temporal evolution of cool gas metallicity, global metal densities, and consequently the timescales of global metal enrichment in the CGM and IGM.

\subsection{Comparison with Empirically Motivated Chemical Evolution Models}

As we strive to empirically account for the global metal content in stars and cosmic gas, it is useful to compare our observational results with theoretical models of chemical evolution in galaxies. These models are grounded in physical principles, and our ultimate goal is to establish a direct connection between the underlying physics driving metal enrichment in the Universe and the inferences drawn from observations. Chemical evolution models have been around for decades, but fine-tuning their parameters and relevant physical processes requires constant comparison with observational data. 

\begin{figure}
\begin{centering}
\includegraphics[trim = 0 35 0 40, clip,width=\columnwidth]{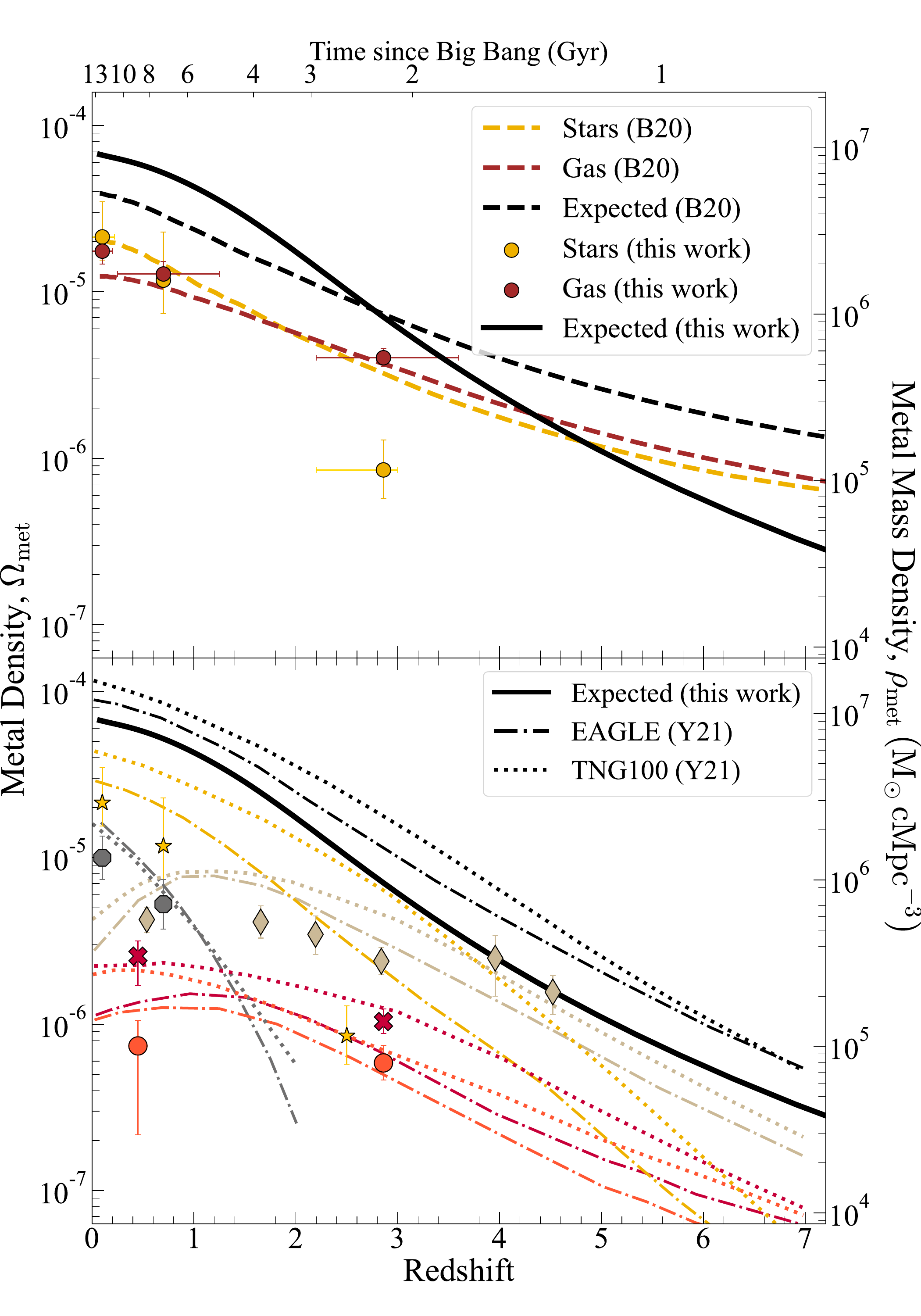}
\par\end{centering}
\begin{centering}
\caption{
{Comparison of predictions from the chemical evolution model in \cite{bellstedt2020} (B20) (\textit{Top panel}), and those from the EAGLE and TNG100 simulations in \cite{yates2021} (Y21) (\textit{Bottom panel}) to the observed global metal densities residing in stars and gas as a function of redshift. The gas phase metal densities are the sum of the available $\rho_{\rm{met}}$ of the ICM+IGrM, SLFSs, pLLSs, LLSs, SLLSs, and DLAs. In both panels, we also plot the expected metal densities from observations, the B20 model, and the simulations. The markers in the bottom panel are consistent with those used in Figure~\ref{omega_met}.}}
\label{bellstedt_comparision}
\par\end{centering}
\end{figure}

To this end, we compare our results to predictions from the chemical evolution models of \cite{bellstedt2020}, who use optimize parameters of their evolutionary models using observations. They employ the \textsc{ProSpect} \citep{robotham2020} SED-fitting code on multi-wavelength photometric data from the GAMA survey \citep{driver2011} at $z<0.06$. By allowing for evolving gas phase metallicity, they derive star formation histories for $\sim7000$ galaxies. These histories are stacked to determine the evolution of stellar mass density, as well as cosmic gas and stellar phase metal densities as a function of redshift. In the top panel of Figure~\ref{bellstedt_comparision}, we plot the total expected mass density of metals produced in stars, those that remained locked in the stellar phase, and those residing in various gaseous phases as predicted by the \cite{bellstedt2020} closed-box model. Our results for the total gas phase metal mass are consistent with their predictions at $z\sim0.5$ and $z\sim3$. At $z\sim0.1$, the observed gas phase metal density is higher than their prediction. This is most likely because we also account for the metals in the ICM+IGrM, while \cite{bellstedt2020} model the gas phase metal density only of the gas within galaxies. The stellar metal density, on the other hand, shows remarkable agreement at $z\sim0.1$ and $z\sim0.7$. However, the models are discrepant with the stellar metal densities at high redshift. Their prediction for the total metal mass produced is much higher than our assumed metal mass density at high redshift. These disparities result from the difference in the stellar mass densities estimated by \cite{bellstedt2020} and \cite{madau2014} (adopted by us) at early times (see Figure~\ref{SMD_comp}). Based on comparison with observations and other estimates from literature, \cite{bellstedt2020} report that their models overestimate the cosmic star formation rate at high redshift, which ultimately results in higher predictions for $\rho^{\rm{exp}}_{\rm{met}}$ and $\rho_{\rm{met,*}}$ (see Figure~5 in \citealt{bellstedt2020}). Figure~\ref{bellstedt_comparision} shows an overall increase in the observed metal density of the gas phase with an evolution similar to that in the stellar phase with cosmic time for the models as well as our observations. The global gas phase metal density increases by a factor of $\sim4.5$ from $z\sim3$ to $z\sim0.1$. The observed stellar phase metal density shows a much larger increase (a factor of 25) over the same redshift interval. This may indicate a period of rapid metal production and astration into stars over $z\sim$3-1. 

\subsection{Comparison with Simulations}

Semi-analytical models \citep{somerville2015,croton2016,cousin2016,hirschman2016,lagos2018} and hydrodynamical cosmological simulations \citep{somerville2015b,dave2017,peeples2019,hough2023,moser2022,weng2024} have greatly improved our understanding of galaxy evolution, and they help guide the goals and strategies for future observational missions. Although they broadly agree with observed galaxy trends, they suffer from unknown physics and have difficulty accurately reproducing key observables, such as the distribution of baryons and metals in galaxies. As simulations push for higher resolutions \citep{peeples2019,wetzel2023,saeedzadeh2023,ramesh2024,strawn2024}, a detailed accounting of metals as a function of overdensity provides empirical benchmarks for testing feedback models on physical scales comparable to those of the absorbers probing the CGM. In this section, we compare our results to cosmological simulations.

\cite{yates2021} compare the cosmic density of metals in stars and various gas phases predicted by three cosmological simulations: \textsc{L-Galaxies} 2020 \citep{henriques2020}, EAGLE \citep{schaye2015}, and TNG100 \citep{pillepich2018}. Since \textsc{L-Galaxies} 2020 focuses more on probing the ISM of galaxies, we compare our results with EAGLE and TNG100 predictions in the bottom panel of Figure~\ref{bellstedt_comparision}. Both simulations reproduce the observed trends for all reservoirs except for a notable discrepancy in stellar metal densities at high redshift. Similar to \cite{bellstedt2020}, the simulations predict much higher stellar metal densities than our empirical values. Another crucial difference between observations and simulations is the total expected metal density of the Universe. All three simulations, including \textsc{L-Galaxies}, predict higher total metal mass densities than our preferred values. Consequently, while observations suggest that almost all metals at $z \ge 3$ reside in neutral gas, the simulations propose that a large fraction of metals ($>60$\%) remain unaccounted for and may reside in hot, low-density gas, resulting from feedback-driven outflows from highly star-forming galaxies or active galactic nuclei in the early Universe. The predictions could be reconciled if there is a significant amount of obscured star formation at high redshift. At even higher redshift, measurements of the metal density of the ISM by  \cite{heintz2023}, based on the poorly constrained [\CII] luminosity function, at $z \sim 5$ and $z \sim 7$  are consistent with our conclusion that almost all metals reside in the neutral/molecular gas.

Cosmological simulations can produce vastly different results for gas and metal mass distribution across cosmic reservoirs, even with identical initial conditions, primarily due to differences in the implementation of stellar and AGN feedback processes \citep{strawn2024}. This highlights the importance of accurately modeling feedback processes inferred from observations to enhance the reliability and accuracy of our cosmological models.

\subsection{Future Directions}

There is a prominent gap in our metal census at intermediate redshifts ($z\sim 1$--2.4) during cosmic noon. While the metal density of the cool neutral gas is constrained during this period, robust estimates of the metal content of other classes of absorbers that trace the CGM and the IGM are missing. A comprehensive study of the global distribution of metals in the CGM and IGM over this epoch is needed to understand the efficiency of galactic outflows in polluting the outer regions of galaxies at these redshifts. Constraining the metal budget at cosmic noon requires a synergistic approach that combines multi-wavelength observations to probe the multi-phase CGM and IGM gas \citep{peroux2023b}. 

The epoch of cosmic noon also lacks a comprehensive survey quantifying stellar metallicities for a representative sample spanning both star-forming and quiescent galaxies.  New observations from the James Webb Space Telescope may help fill this gap in our census in the near future \citep{Li2023,slob2024,He2024}. To completely close the gap between the expected metals and the observed metals, we also need robust measurements of metallicities for the gas tracing the warm-hot CGM. The eROSITA mission has made significant strides in mapping the properties of the warm-hot CGM \citep{zhang2024a,zhang2024b}. Future studies aimed at constraining metallicities will play a crucial role in bridging the gap in the cosmic metal budget. 

Finally, while observations are essential, semi-analytical models and cosmological simulations can provide valuable insights into the processes governing metal production, transportation, and mixing. Cosmological hydrodynamical simulations that incorporate observationally calibrated models for stellar and AGN feedback processes will prove to be invaluable in translating newer metallicity observations to the underlying physics governing cosmic gas and galaxies. For instance, the FLAMINGO project \citep{schaye2023} will generate a suite of hydrodynamical simulations with subgrid prescriptions calibrated to observed galaxy stellar mass functions and cluster gas fractions at low redshift. \cite{schimd2024} employ the \textsc{Simba} suite of hydrodynamical simulations to investigate the temporal evolution of global trends in gas properties, including the density and metallicity fields for atomic and molecular hydrogen, spanning the redshift range of $z=0$ to 5. The simulated data from such studies can be leveraged to estimate global metal densities for the gaseous components, facilitating comparisons with observational constraints.

\section{Summary}\label{sec:summary}

We present an updated accounting of metals in the Universe and their evolution with cosmic time. Estimating the global metal density of a given metal reservoir requires combining its mass density distribution with the mass-weighted mean metallicity at each redshift. We utilize data from existing surveys to calculate the global metal content of condensed galactic and circumgalactic matter in the Universe and perform a careful assessment of associated uncertainties. Specifically, we adopt renormalized stellar mass functions from \cite{wright2018} and stellar mass-metallicity relations from \cite{gallazzi2008} and \cite{chartab2023} to constrain the stellar metal densities. Using archival data from the CCC \citep{lehner2018a,lehner2019} and the KODIAQ-Z survey \citep{lehner2016,lehner2022} and HD-LLS survey \citep{prochaska2015,fumagalli2016}, we calculate the global metal densities of cool ionized gas as a function of neutral hydrogen column density at $z<1$ and $z\sim2.8$, respectively. By synthesizing these results, we derive the global metal budget across cosmic time. Our main findings are summarized as follows:
\begin{enumerate}
    
    \item We estimate the global stellar metal densities at key cosmic epochs: $z = 0.1, 0.7$ and $2.5$ (spanning $\sim11$ Gyrs). The relative contribution of stars to the total metal density of the Universe has increased from $8^{+4}_{-3}\%$ at $z=2.5$ to $21^{+20}_{-8}\%$ at $z=0.7$. This indicates a remarkably rapid buildup---a factor of $13.5^{+15}_{-7}$ increase---in the global stellar metal density during cosmic noon ($z\sim1$--2.4). At low redshifts ($z\sim0.1$), stars emerge as the dominant contributor to the total metal content of the Universe, housing approximately $31^{+20}_{-8}\%$ of the total expected metal density.

    \item At $z\gtrsim4$, most metals in the Universe reside in the cool, neutral gas traced by DLAs. We observe a shift of metal distribution at $z\sim3$ from neutral to also ionized gas. Together, neutral ($31.8\pm{7}\%$) and ionized gas ($21.7\pm{3}\%$) house $\sim54\pm8\%$ of all metals in the Universe at $z\sim3$. At $z<1$, we observe further diversification of metal reservoirs: hot virialized gas accounts for {$\sim15\pm4\%$ and $\sim10\pm3\%$} of the expected metals at $z\sim0.1$ and $z\sim0.7$, respectively; cool gas accounts for $\sim12\pm2\%$ of total metals at $z<1$. Thus, the total metal content in the gas phase is comparable to that of stars at low redshift. 
    
    \item To understand the evolving contribution of cool atomic and ionized gas to the Universe's metal and baryon budgets, we analyze their contribution to the baryonic mass density as a function of $N_{\rm{HI}}$. We find that across all column density regimes, the total mass density of cool gas has decreased with cosmic time as more baryons condense into stars. At $z\sim3$, lower column density absorbers ($14.5 < \logNHI < 16$) are the dominant gas reservoirs, whereas, at $z<1$, higher column density absorbers ($\logNHI > 16$) contain more gas mass. These changes drive the increase in the global metal density of LLSs and pLLSs, while the metal density in SLFSs decreases with cosmic time.

    \item We estimate that the combined photoionized components of cool gas (SLFSs, pLLSs, and LLSs) account for approximately $4\pm2\%$ and $14\pm2\%$ of the total expected metals at $z<1$ and $z\sim2.8$, respectively. While their relative contributions have decreased with cosmic time, their total metal density has more than doubled. Globally, the higher column density absorbers (pLLSs and LLSs) typically associated with CGM-like gas are more enriched compared to the lower column density absorbers (SLFSs) that probe IGM-like gas over the same period. This reflects the preferential enrichment of gas closer to galaxies by feedback-driven outflows. The substantial increase in metal content of both the cool ionized phase tracing CGM and IGM-like gas and the hot ICM/IGrM suggests enhanced stellar and AGN feedback during intermediate redshifts ($z\sim 1$--2.4). 

    \item At $z>3$ we find that all metals can be plausibly accounted for in cool neutral gas traced by DLAs. We find little evidence for a ``missing metals problem'' at these redshifts. At $z<3$, our cataloged reservoirs do not completely close the metal budget. However, we have identified significant metal reservoirs (cool IGM traced by the Ly$\alpha$ forest, warm-hot CGM, WHIM, and molecular gas) that could plausibly complete the budget. We have not included these in our census due to a lack of robust uncertainty estimates, poor probes of their metal content, or limited statistical samples. In the most optimistic case in which we assume existing estimates of $\Omega_{\rm met}$ for cool IGM ($\sim18\%$) and molecular gas ($\sim8\%$) to be close to the true value, we can account for $\approx90\%$ of the expected metals at $z<0.4$. We expect that if the contributions from the hot CGM and WHIM were included, the census would be complete at low redshifts too.

    \item The largest source of uncertainty while compiling the global metal budget remains in the assumed integrated stellar yield and initial mass function, that is, $y\times(1-R)$. This term introduces a $\sim30\%$ uncertainty on the total expected metal density, which makes the fraction of unaccounted metals highly uncertain.
    
    \item A notable gap in our census spans $z\sim0.9$--2.4. This epoch currently lacks metal studies for both the CGM and IGM, as well as a survey of stellar metals for a representative sample of galaxies. A detailed study of the metallicity distribution in CGM- and IGM-like cool, \HI-selected gas will be presented in forthcoming papers.

    \item Our results are mostly consistent with the chemical evolution models derived by \cite{bellstedt2020}. However, our estimates of the global stellar metal densities are low compared to the model predictions. This discrepancy may be due to the assumptions of closed-box evolution, which does not allow for the removal of metals from halos. We observe a good agreement between observed global metal densities and the EAGLE and TNG100 simulations at low redshift. At high redshift, simulations predict a higher amount of expected metals compared to our estimates, which would imply a larger fraction of unaccounted metals than our estimates. 

\end{enumerate}

Future JWST surveys will help quantify the metal content of stars at cosmic noon. The synthesis of global metal densities in the CGM and IGM at cosmic noon, along with compiling a comprehensive metal census, requires multi-wavelength efforts. Results from our forthcoming surveys on the CGM and IGM metals will help fill the current gaps in our census. Next-generation X-ray telescopes and upcoming metallicity surveys will significantly advance our understanding, helping to complete the current picture of the cosmic metal cycle of the Universe. Global cosmological simulations will also guide future missions, mitigate known systematics in calculating global quantities, and, with more precise feedback implementations, aid in interpreting empirical results and understanding the underlying physics driving galaxy evolution.

\section*{Acknowledgements}

We thank the anonymous referee for a thorough review and insightful comments that helped improve the overall quality of our manuscript. We greatly appreciate the discussions with Prof. Evan Kirby regarding the current state of our knowledge of stellar metals and the future scope of work in this area. 
Support for this research was provided by NSF grants AST-1910255 and NSF AST-2206854. 
A part of this work was supported by NASA through grant 80NSSC21K0648 and grant HST-AR-15634 from the Space Telescope Science Institute (STScI), which is operated by the Association of Universities for Research in Astronomy, Incorporated, under NASA contract NAS5-26555. This research has made use of NASA's Astrophysics Data System and {\tt adstex} (\url{https://github.com/yymao/adstex}).

\software{Astropy \citep{astropy2018},
  Matplotlib \citep{hunter2007}}, $\textsc{scikit-learn}$ \citep{scikit-learn,sklearn_api}, $\textsc{sksurv}$ \citep{sksurv}. 

\bibliography{references}

\begin{thebibliography}{}
\expandafter\ifx\csname natexlab\endcsname\relax\def\natexlab#1{#1}\fi
\providecommand{\url}[1]{\href{#1}{#1}}

\bibitem[{{Aguirre} {et~al.}(2004){Aguirre}, {Schaye}, {Kim}, {Theuns},
  {Rauch}, \& {Sargent}}]{aguirre2004}
{Aguirre}, A., {Schaye}, J., {Kim}, T.-S., {et~al.} 2004, \apj, 602, 38

\bibitem[{{Anderson} {et~al.}(2013){Anderson}, {Bregman}, \&
  {Dai}}]{anderson2013}
{Anderson}, M.~E., {Bregman}, J.~N., \& {Dai}, X. 2013, \apj, 762, 106

\bibitem[{{Andrews} {et~al.}(2017){Andrews}, {Driver}, {Davies}, {Kafle},
  {Robotham}, {Vinsen}, {Wright}, {Bland-Hawthorn}, {Bourne}, {Bremer}, {da
  Cunha}, {Drinkwater}, {Holwerda}, {Hopkins}, {Kelvin}, {Loveday},
  {Phillipps}, \& {Wilkins}}]{andrews2017}
{Andrews}, S.~K., {Driver}, S.~P., {Davies}, L.~J.~M., {et~al.} 2017, \mnras,
  470, 1342

\bibitem[{{Angl{\'e}s-Alc{\'a}zar} {et~al.}(2017){Angl{\'e}s-Alc{\'a}zar},
  {Faucher-Gigu{\`e}re}, {Kere{\v{s}}}, {Hopkins}, {Quataert}, \&
  {Murray}}]{angles-alcazar2017}
{Angl{\'e}s-Alc{\'a}zar}, D., {Faucher-Gigu{\`e}re}, C.-A., {Kere{\v{s}}}, D.,
  {et~al.} 2017, \mnras, 470, 4698

\bibitem[{{Aravena} {et~al.}(2024){Aravena}, {Heintz}, {Dessauges-Zavadsky},
  {Oesch}, {Algera}, {Bouwens}, {da Cunha}, {Dayal}, {De Looze}, {Ferrara},
  {Fudamoto}, {Gonzalez}, {Graziani}, {Hygate}, {Inami}, {Pallottini},
  {Schneider}, {Schouws}, {Sommovigo}, {Topping}, {van der Werf}, \&
  {Palla}}]{aravena2024}
{Aravena}, M., {Heintz}, K., {Dessauges-Zavadsky}, M., {et~al.} 2024, \aap,
  682, A24

\bibitem[{{Asplund} {et~al.}(2021){Asplund}, {Amarsi}, \&
  {Grevesse}}]{asplund2021}
{Asplund}, M., {Amarsi}, A.~M., \& {Grevesse}, N. 2021, \aap, 653, A141

\bibitem[{{Azzalini} \& {Capitanio}(2009)}]{azzalini2009}
{Azzalini}, A., \& {Capitanio}, A. 2009, arXiv e-prints, arXiv:0911.2093

\bibitem[{{Balashev} \& {Noterdaeme}(2018)}]{balashev2018}
{Balashev}, S.~A., \& {Noterdaeme}, P. 2018, \mnras, 478, L7

\bibitem[{{Balashev} {et~al.}(2019){Balashev}, {Klimenko}, {Noterdaeme},
  {Krogager}, {Varshalovich}, {Ivanchik}, {Petitjean}, {Srianand}, \&
  {Ledoux}}]{balashev2019}
{Balashev}, S.~A., {Klimenko}, V.~V., {Noterdaeme}, P., {et~al.} 2019, \mnras,
  490, 2668

\bibitem[{{Barrufet} {et~al.}(2024){Barrufet}, {Oesch}, {Marques-Chaves},
  {Arellano-Cordova}, {Baggen}, {Carnall}, {Cullen}, {Dunlop}, {Gottumukkala},
  {Fudamoto}, {Illingworth}, {Magee}, {McLure}, {McLeod}, {Micha{\l}owski},
  {Stefanon}, {van Dokkum}, \& {Weibel}}]{Barrufet2024}
{Barrufet}, L., {Oesch}, P., {Marques-Chaves}, R., {et~al.} 2024, arXiv
  e-prints, arXiv:2404.08052

\bibitem[{{Battisti} {et~al.}(2012){Battisti}, {Meiring}, {Tripp}, {Prochaska},
  {Werk}, {Jenkins}, {Lehner}, {Tumlinson}, \& {Thom}}]{battisti2012}
{Battisti}, A.~J., {Meiring}, J.~D., {Tripp}, T.~M., {et~al.} 2012, \apj, 744,
  93

\bibitem[{{Behroozi} {et~al.}(2019){Behroozi}, {Wechsler}, {Hearin}, \&
  {Conroy}}]{behroozi2019}
{Behroozi}, P., {Wechsler}, R.~H., {Hearin}, A.~P., \& {Conroy}, C. 2019,
  \mnras, 488, 3143

\bibitem[{{Bellstedt} {et~al.}(2020){Bellstedt}, {Robotham}, {Driver},
  {Thorne}, {Davies}, {Lagos}, {Stevens}, {Taylor}, {Baldry}, {Moffett},
  {Hopkins}, \& {Phillipps}}]{bellstedt2020}
{Bellstedt}, S., {Robotham}, A. S.~G., {Driver}, S.~P., {et~al.} 2020, \mnras,
  498, 5581

\bibitem[{{Berg} {et~al.}(2023){Berg}, {Lehner}, {Howk}, {O'Meara}, {Schaye},
  {Straka}, {Cooksey}, {Tripp}, {Prochaska}, {Oppenheimer}, {Johnson},
  {Muzahid}, {Bordoloi}, {Werk}, {Fox}, {Katz}, {Wendt}, {Peeples}, {Ribaudo},
  \& {Tumlinson}}]{berg2023}
{Berg}, M.~A., {Lehner}, N., {Howk}, J.~C., {et~al.} 2023, \apj, 944, 101

\bibitem[{{Berg} {et~al.}(2016){Berg}, {Ellison}, {S{\'a}nchez-Ram{\'\i}rez},
  {Prochaska}, {Lopez}, {D'Odorico}, {Becker}, {Christensen}, {Cupani},
  {Denney}, \& {Worseck}}]{berg2016}
{Berg}, T.~A.~M., {Ellison}, S.~L., {S{\'a}nchez-Ram{\'\i}rez}, R., {et~al.}
  2016, \mnras, 463, 3021

\bibitem[{{Bertone} {et~al.}(2008){Bertone}, {Schaye}, \&
  {Dolag}}]{bertone2008}
{Bertone}, S., {Schaye}, J., \& {Dolag}, K. 2008, \ssr, 134, 295

\bibitem[{{Black}(1981)}]{black1981}
{Black}, J.~H. 1981, \mnras, 197, 553

\bibitem[{{Blackwell} {et~al.}(2022){Blackwell}, {Bregman}, \&
  {Snowden}}]{blackwell2022}
{Blackwell}, A.~E., {Bregman}, J.~N., \& {Snowden}, S.~L. 2022, \apj, 927, 104

\bibitem[{{Bocquet} {et~al.}(2016){Bocquet}, {Saro}, {Dolag}, \&
  {Mohr}}]{bocquet2016}
{Bocquet}, S., {Saro}, A., {Dolag}, K., \& {Mohr}, J.~J. 2016, \mnras, 456,
  2361

\bibitem[{{Boksenberg} \& {Sargent}(2015)}]{boksenberg2015}
{Boksenberg}, A., \& {Sargent}, W. L.~W. 2015, \apjs, 218, 7

\bibitem[{{Bollo} {et~al.}(2025){Bollo}, {P{\'e}roux}, {Zwaan}, {Hamanowicz},
  {Chen}, {Weng}, {Lagos}, {Bravo}, {Ivison}, \& {Biggs}}]{bollo2025}
{Bollo}, V., {P{\'e}roux}, C., {Zwaan}, M., {et~al.} 2025, arXiv e-prints,
  arXiv:2502.06778

\bibitem[{{Bond} {et~al.}(1996){Bond}, {Kofman}, \& {Pogosyan}}]{bond1996}
{Bond}, J.~R., {Kofman}, L., \& {Pogosyan}, D. 1996, \nat, 380, 603

\bibitem[{{Boogaard} {et~al.}(2020){Boogaard}, {van der Werf}, {Weiss},
  {Popping}, {Decarli}, {Walter}, {Aravena}, {Bouwens}, {Riechers},
  {Gonz{\'a}lez-L{\'o}pez}, {Smail}, {Carilli}, {Kaasinen}, {Daddi}, {Cox},
  {D{\'\i}az-Santos}, {Inami}, {Cortes}, \& {Wagg}}]{boogaard2020}
{Boogaard}, L.~A., {van der Werf}, P., {Weiss}, A., {et~al.} 2020, \apj, 902,
  109

\bibitem[{{Boogaard} {et~al.}(2023){Boogaard}, {Decarli}, {Walter}, {Wei{\ss}},
  {Popping}, {Neri}, {Aravena}, {Riechers}, {Ellis}, {Carilli}, {Cox}, \&
  {Pety}}]{Boogaard2023}
{Boogaard}, L.~A., {Decarli}, R., {Walter}, F., {et~al.} 2023, \apj, 945, 111

\bibitem[{{Bouch{\'e}} {et~al.}(2012){Bouch{\'e}}, {Hohensee}, {Vargas},
  {Kacprzak}, {Martin}, {Cooke}, \& {Churchill}}]{bouche2012}
{Bouch{\'e}}, N., {Hohensee}, W., {Vargas}, R., {et~al.} 2012, \mnras, 426, 801

\bibitem[{{Bouch{\'e}} {et~al.}(2007){Bouch{\'e}}, {Lehnert}, {Aguirre},
  {P{\'e}roux}, \& {Bergeron}}]{bouche2007}
{Bouch{\'e}}, N., {Lehnert}, M.~D., {Aguirre}, A., {P{\'e}roux}, C., \&
  {Bergeron}, J. 2007, \mnras, 378, 525

\bibitem[{{Bouch{\'e}} {et~al.}(2005){Bouch{\'e}}, {Lehnert}, \&
  {P{\'e}roux}}]{bouche2005}
{Bouch{\'e}}, N., {Lehnert}, M.~D., \& {P{\'e}roux}, C. 2005, \mnras, 364, 319

\bibitem[{{Bouch{\'e}} {et~al.}(2006){Bouch{\'e}}, {Murphy}, {P{\'e}roux},
  {Csabai}, \& {Wild}}]{bouche2006}
{Bouch{\'e}}, N., {Murphy}, M.~T., {P{\'e}roux}, C., {Csabai}, I., \& {Wild},
  V. 2006, \mnras, 371, 495

\bibitem[{{Bouwens} {et~al.}(2020){Bouwens}, {Gonz{\'a}lez-L{\'o}pez},
  {Aravena}, {Decarli}, {Novak}, {Stefanon}, {Walter}, {Boogaard}, {Carilli},
  {Dudzevi{\v{c}}i{\={u}}t{\.{e}}}, {Smail}, {Daddi}, {da Cunha}, {Ivison},
  {Nanayakkara}, {Cortes}, {Cox}, {Inami}, {Oesch}, {Popping}, {Riechers}, {van
  der Werf}, {Weiss}, {Fudamoto}, \& {Wagg}}]{bouwens2020}
{Bouwens}, R., {Gonz{\'a}lez-L{\'o}pez}, J., {Aravena}, M., {et~al.} 2020,
  \apj, 902, 112

\bibitem[{{Braun}(2012)}]{braun2012}
{Braun}, R. 2012, \apj, 749, 87

\bibitem[{{Bromm} \& {Yoshida}(2011)}]{bromm2011}
{Bromm}, V., \& {Yoshida}, N. 2011, \araa, 49, 373

\bibitem[{{Bromm} {et~al.}(2009){Bromm}, {Yoshida}, {Hernquist}, \&
  {McKee}}]{bromm2009}
{Bromm}, V., {Yoshida}, N., {Hernquist}, L., \& {McKee}, C.~F. 2009, \nat, 459,
  49

\bibitem[{{Bruzual} \& {Charlot}(2003)}]{bruzual2003}
{Bruzual}, G., \& {Charlot}, S. 2003, \mnras, 344, 1000

\bibitem[{{Buck} {et~al.}(2021){Buck}, {Rybizki}, {Buder}, {Obreja},
  {Macci{\`o}}, {Pfrommer}, {Steinmetz}, \& {Ness}}]{buck2021}
{Buck}, T., {Rybizki}, J., {Buder}, S., {et~al.} 2021, \mnras, 508, 3365

\bibitem[{Buitinck {et~al.}(2013)Buitinck, Louppe, Blondel, Pedregosa, Mueller,
  Grisel, Niculae, Prettenhofer, Gramfort, Grobler, Layton, VanderPlas, Joly,
  Holt, \& Varoquaux}]{sklearn_api}
Buitinck, L., Louppe, G., Blondel, M., {et~al.} 2013, in ECML PKDD Workshop:
  Languages for Data Mining and Machine Learning, 108--122

\bibitem[{{Burchett} {et~al.}(2020){Burchett}, {Elek}, {Tejos}, {Prochaska},
  {Tripp}, {Bordoloi}, \& {Forbes}}]{burchett2020}
{Burchett}, J.~N., {Elek}, O., {Tejos}, N., {et~al.} 2020, \apjl, 891, L35

\bibitem[{{Burchett} {et~al.}(2019){Burchett}, {Tripp}, {Prochaska}, {Werk},
  {Tumlinson}, {Howk}, {Willmer}, {Lehner}, {Meiring}, {Bowen}, {Bordoloi},
  {Peeples}, {Jenkins}, {O'Meara}, {Tejos}, \& {Katz}}]{burchett2019}
{Burchett}, J.~N., {Tripp}, T.~M., {Prochaska}, J.~X., {et~al.} 2019, \apjl,
  877, L20

\bibitem[{{Cai} {et~al.}(2017){Cai}, {Fan}, {Dave}, {Finlator}, \&
  {Oppenheimer}}]{cai2017}
{Cai}, Z., {Fan}, X., {Dave}, R., {Finlator}, K., \& {Oppenheimer}, B. 2017,
  \apjl, 849, L18

\bibitem[{{Carnall} {et~al.}(2022){Carnall}, {McLure}, {Dunlop}, {Hamadouche},
  {Cullen}, {McLeod}, {Begley}, {Amorin}, {Bolzonella}, {Castellano},
  {Cimatti}, {Fontanot}, {Gargiulo}, {Garilli}, {Mannucci}, {Pentericci},
  {Talia}, {Zamorani}, {Calabro}, {Cresci}, \& {Hathi}}]{carnall2022}
{Carnall}, A.~C., {McLure}, R.~J., {Dunlop}, J.~S., {et~al.} 2022, \apj, 929,
  131

\bibitem[{{Casavecchia} {et~al.}(2024){Casavecchia}, {Maio}, {P{\'e}roux}, \&
  {Ciardi}}]{casavecchia2024}
{Casavecchia}, B., {Maio}, U., {P{\'e}roux}, C., \& {Ciardi}, B. 2024, arXiv
  e-prints, arXiv:2406.01277

\bibitem[{{Cen} \& {Fang}(2006)}]{cen2006}
{Cen}, R., \& {Fang}, T. 2006, \apj, 650, 573

\bibitem[{{Chabrier}(2003)}]{chabrier2003}
{Chabrier}, G. 2003, \pasp, 115, 763

\bibitem[{{Chartab} {et~al.}(2024){Chartab}, {Newman}, {Rudie}, {Blanc}, \&
  {Kelson}}]{chartab2023}
{Chartab}, N., {Newman}, A.~B., {Rudie}, G.~C., {Blanc}, G.~A., \& {Kelson},
  D.~D. 2024, \apj, 960, 73

\bibitem[{{Chen} {et~al.}(2017{\natexlab{a}}){Chen}, {Johnson}, {Zahedy},
  {Rauch}, \& {Mulchaey}}]{chen2017a}
{Chen}, H.-W., {Johnson}, S.~D., {Zahedy}, F.~S., {Rauch}, M., \& {Mulchaey},
  J.~S. 2017{\natexlab{a}}, \apjl, 842, L19

\bibitem[{{Chen} \& {Zahedy}(2024)}]{chen2024}
{Chen}, H.-W., \& {Zahedy}, F.~S. 2024, arXiv e-prints, arXiv:2412.10579

\bibitem[{{Chen} {et~al.}(2017{\natexlab{b}}){Chen}, {Ho}, {Mandelbaum},
  {Bahcall}, {Brownstein}, {Freeman}, {Genovese}, {Schneider}, \&
  {Wasserman}}]{chen2017}
{Chen}, Y.-C., {Ho}, S., {Mandelbaum}, R., {et~al.} 2017{\natexlab{b}}, \mnras,
  466, 1880

\bibitem[{{Chiappini} {et~al.}(2008){Chiappini}, {Ekstr{\"o}m}, {Meynet},
  {Hirschi}, {Maeder}, \& {Charbonnel}}]{chiappini2008}
{Chiappini}, C., {Ekstr{\"o}m}, S., {Meynet}, G., {et~al.} 2008, \aap, 479, L9

\bibitem[{{Chieffi} \& {Limongi}(2004)}]{chieffi2004}
{Chieffi}, A., \& {Limongi}, M. 2004, \apj, 608, 405

\bibitem[{{Chiu} {et~al.}(2018){Chiu}, {Mohr}, {McDonald}, {Bocquet}, {Desai},
  {Klein}, {Israel}, {Ashby}, {Stanford}, {Benson}, {Brodwin}, {Abbott},
  {Abdalla}, {Allam}, {Annis}, {Bayliss}, {Benoit-L{\'e}vy}, {Bertin}, {Bleem},
  {Brooks}, {Buckley-Geer}, {Bulbul}, {Capasso}, {Carlstrom}, {Rosell},
  {Carretero}, {Castander}, {Cunha}, {D'Andrea}, {da Costa}, {Davis}, {Diehl},
  {Dietrich}, {Doel}, {Drlica-Wagner}, {Eifler}, {Evrard}, {Flaugher},
  {Garc{\'\i}a-Bellido}, {Garmire}, {Gaztanaga}, {Gerdes}, {Gonzalez}, {Gruen},
  {Gruendl}, {Gschwend}, {Gupta}, {Gutierrez}, {Hlavacek-L}, {Honscheid},
  {James}, {Jeltema}, {Kraft}, {Krause}, {Kuehn}, {Kuhlmann}, {Kuropatkin},
  {Lahav}, {Lima}, {Maia}, {Marshall}, {Melchior}, {Menanteau}, {Miquel},
  {Murray}, {Nord}, {Ogando}, {Plazas}, {Rapetti}, {Reichardt}, {Romer},
  {Roodman}, {Sanchez}, {Saro}, {Scarpine}, {Schindler}, {Schubnell}, {Sharon},
  {Smith}, {Smith}, {Soares-Santos}, {Sobreira}, {Stalder}, {Stern},
  {Strazzullo}, {Suchyta}, {Swanson}, {Tarle}, {Vikram}, {Walker}, {Weller}, \&
  {Zhang}}]{chiu2018}
{Chiu}, I., {Mohr}, J.~J., {McDonald}, M., {et~al.} 2018, \mnras, 478, 3072

\bibitem[{{Choi} {et~al.}(2020){Choi}, {Brennan}, {Somerville}, {Ostriker},
  {Hirschmann}, \& {Naab}}]{choi2020}
{Choi}, E., {Brennan}, R., {Somerville}, R.~S., {et~al.} 2020, \apj, 904, 8

\bibitem[{{Choi} {et~al.}(2017){Choi}, {Conroy}, \& {Byler}}]{choi2017}
{Choi}, J., {Conroy}, C., \& {Byler}, N. 2017, \apj, 838, 159

\bibitem[{{Chowdhury} {et~al.}(2024){Chowdhury}, {Kanekar}, \&
  {Chengalur}}]{chowdhury2024}
{Chowdhury}, A., {Kanekar}, N., \& {Chengalur}, J.~N. 2024, \apjl, 966, L39

\bibitem[{Collaboration {et~al.}(2018)Collaboration, Price-Whelan, Sip{\H o}cz,
  G{\"u}nther, Lim, Crawford, Conseil, Shupe, Craig, Dencheva, Ginsburg,
  VanderPlas, Bradley, P{\'e}rez-Su{\'a}rez, de~Val-Borro, Contributors),
  Aldcroft, Cruz, Robitaille, Tollerud, Committee), Ardelean, Babej, Bach,
  Bachetti, Bakanov, Bamford, Barentsen, Barmby, Baumbach, Berry, Biscani,
  Boquien, Bostroem, Bouma, Brammer, Bray, Breytenbach, Buddelmeijer, Burke,
  Calderone, Rodr{\'\i}guez, Cara, Cardoso, Cheedella, Copin, Corrales,
  Crichton, D'Avella, Deil, Depagne, Dietrich, Donath, Droettboom, Earl, Erben,
  Fabbro, Ferreira, Finethy, Fox, Garrison, Gibbons, Goldstein, Gommers, Greco,
  Greenfield, Groener, Grollier, Hagen, Hirst, Homeier, Horton, Hosseinzadeh,
  Hu, Hunkeler, Ivezi{\'c}, Jain, Jenness, Kanarek, Kendrew, Kern, Kerzendorf,
  Khvalko, King, Kirkby, Kulkarni, Kumar, Lee, Lenz, Littlefair, Ma, Macleod,
  Mastropietro, McCully, Montagnac, Morris, Mueller, Mumford, Muna, Murphy,
  Nelson, Nguyen, Ninan, N{\"o}the, Ogaz, Oh, Parejko, Parley, Pascual, Patil,
  Patil, Plunkett, Prochaska, Rastogi, Janga, Sabater, Sakurikar, Seifert,
  Sherbert, Sherwood-Taylor, Shih, Sick, Silbiger, Singanamalla, Singer,
  Sladen, Sooley, Sornarajah, Streicher, Teuben, Thomas, Tremblay, Turner,
  Terr{\'o}n, van Kerkwijk, de~la Vega, Watkins, Weaver, Whitmore, Woillez,
  Zabalza, \& Contributors)}]{astropy2018}
Collaboration, T.~A., Price-Whelan, A.~M., Sip{\H o}cz, B.~M., {et~al.} 2018,
  The Astronomical Journal, 156, 123.
\newblock \url{http://stacks.iop.org/1538-3881/156/i=3/a=123}

\bibitem[{{Cooksey} {et~al.}(2013){Cooksey}, {Kao}, {Simcoe}, {O'Meara}, \&
  {Prochaska}}]{cooksey2013}
{Cooksey}, K.~L., {Kao}, M.~M., {Simcoe}, R.~A., {O'Meara}, J.~M., \&
  {Prochaska}, J.~X. 2013, \apj, 763, 37

\bibitem[{{Cooksey} {et~al.}(2010){Cooksey}, {Thom}, {Prochaska}, \&
  {Chen}}]{cooksey2010}
{Cooksey}, K.~L., {Thom}, C., {Prochaska}, J.~X., \& {Chen}, H.-W. 2010, \apj,
  708, 868

\bibitem[{{Corbelli} \& {Bandiera}(2002)}]{corbelli2002}
{Corbelli}, E., \& {Bandiera}, R. 2002, \apj, 567, 712

\bibitem[{{Cousin} {et~al.}(2016){Cousin}, {Buat}, {Boissier}, {Bethermin},
  {Roehlly}, \& {G{\'e}nois}}]{cousin2016}
{Cousin}, M., {Buat}, V., {Boissier}, S., {et~al.} 2016, \aap, 589, A109

\bibitem[{{Crighton} {et~al.}(2015){Crighton}, {Hennawi}, {Simcoe}, {Cooksey},
  {Murphy}, {Fumagalli}, {Prochaska}, \& {Shanks}}]{crighton2015}
{Crighton}, N. H.~M., {Hennawi}, J.~F., {Simcoe}, R.~A., {et~al.} 2015, \mnras,
  446, 18

\bibitem[{{Croton} {et~al.}(2016){Croton}, {Stevens}, {Tonini}, {Garel},
  {Bernyk}, {Bibiano}, {Hodkinson}, {Mutch}, {Poole}, \&
  {Shattow}}]{croton2016}
{Croton}, D.~J., {Stevens}, A. R.~H., {Tonini}, C., {et~al.} 2016, \apjs, 222,
  22

\bibitem[{{Cullen} {et~al.}(2019){Cullen}, {McLure}, {Dunlop}, {Khochfar},
  {Dav{\'e}}, {Amor{\'\i}n}, {Bolzonella}, {Carnall}, {Castellano}, {Cimatti},
  {Cirasuolo}, {Cresci}, {Fynbo}, {Fontanot}, {Gargiulo}, {Garilli}, {Guaita},
  {Hathi}, {Hibon}, {Mannucci}, {Marchi}, {McLeod}, {Pentericci}, {Pozzetti},
  {Shapley}, {Talia}, \& {Zamorani}}]{cullen2019}
{Cullen}, F., {McLure}, R.~J., {Dunlop}, J.~S., {et~al.} 2019, \mnras, 487,
  2038

\bibitem[{{Cyburt} {et~al.}(2016){Cyburt}, {Fields}, {Olive}, \&
  {Yeh}}]{cyburt2016}
{Cyburt}, R.~H., {Fields}, B.~D., {Olive}, K.~A., \& {Yeh}, T.-H. 2016, Reviews
  of Modern Physics, 88, 015004

\bibitem[{{Danforth} \& {Shull}(2008)}]{danforth2008}
{Danforth}, C.~W., \& {Shull}, J.~M. 2008, \apj, 679, 194

\bibitem[{{Danforth} {et~al.}(2016){Danforth}, {Keeney}, {Tilton}, {Shull},
  {Stocke}, {Stevans}, {Pieri}, {Savage}, {France}, {Syphers}, {Smith},
  {Green}, {Froning}, {Penton}, \& {Osterman}}]{danforth2016}
{Danforth}, C.~W., {Keeney}, B.~A., {Tilton}, E.~M., {et~al.} 2016, \apj, 817,
  111

\bibitem[{{Dav{\'e}} {et~al.}(2010{\natexlab{a}}){Dav{\'e}}, {Oppenheimer},
  {Katz}, {Kollmeier}, \& {Weinberg}}]{romeel2010}
{Dav{\'e}}, R., {Oppenheimer}, B.~D., {Katz}, N., {Kollmeier}, J.~A., \&
  {Weinberg}, D.~H. 2010{\natexlab{a}}, \mnras, 408, 2051

\bibitem[{{Dav{\'e}} {et~al.}(2010{\natexlab{b}}){Dav{\'e}}, {Oppenheimer},
  {Katz}, {Kollmeier}, \& {Weinberg}}]{dave2010}
---. 2010{\natexlab{b}}, \mnras, 408, 2051

\bibitem[{{Dav{\'e}} {et~al.}(2017){Dav{\'e}}, {Rafieferantsoa}, {Thompson}, \&
  {Hopkins}}]{dave2017}
{Dav{\'e}}, R., {Rafieferantsoa}, M.~H., {Thompson}, R.~J., \& {Hopkins}, P.~F.
  2017, \mnras, 467, 115

\bibitem[{{Davies} \& {Cummings}(1975)}]{davies1975}
{Davies}, R.~D., \& {Cummings}, E.~R. 1975, \mnras, 170, 95

\bibitem[{{Davies} {et~al.}(2023){Davies}, {Ryan-Weber}, {D'Odorico}, {Bosman},
  {Meyer}, {Becker}, {Cupani}, {Keating}, {Bischetti}, {Davies}, {Eilers},
  {Farina}, {Haehnelt}, {Pallottini}, \& {Zhu}}]{davies2023}
{Davies}, R.~L., {Ryan-Weber}, E., {D'Odorico}, V., {et~al.} 2023, \mnras, 521,
  314

\bibitem[{{De Cia} {et~al.}(2016){De Cia}, {Ledoux}, {Mattsson}, {Petitjean},
  {Srianand}, {Gavignaud}, \& {Jenkins}}]{decia2016}
{De Cia}, A., {Ledoux}, C., {Mattsson}, L., {et~al.} 2016, \aap, 596, A97

\bibitem[{{De Cia} {et~al.}(2018){De Cia}, {Ledoux}, {Petitjean}, \&
  {Savaglio}}]{decia2018}
{De Cia}, A., {Ledoux}, C., {Petitjean}, P., \& {Savaglio}, S. 2018, \aap, 611,
  A76

\bibitem[{{Decarli} {et~al.}(2019){Decarli}, {Walter},
  {G{\'o}nzalez-L{\'o}pez}, {Aravena}, {Boogaard}, {Carilli}, {Cox}, {Daddi},
  {Popping}, {Riechers}, {Uzgil}, {Weiss}, {Assef}, {Bacon}, {Bauer},
  {Bertoldi}, {Bouwens}, {Contini}, {Cortes}, {da Cunha}, {D{\'\i}az-Santos},
  {Elbaz}, {Inami}, {Hodge}, {Ivison}, {Le F{\`e}vre}, {Magnelli}, {Novak},
  {Oesch}, {Rix}, {Sargent}, {Smail}, {Swinbank}, {Somerville}, {van der Werf},
  {Wagg}, \& {Wisotzki}}]{decarli2019}
{Decarli}, R., {Walter}, F., {G{\'o}nzalez-L{\'o}pez}, J., {et~al.} 2019, \apj,
  882, 138

\bibitem[{{Decarli} {et~al.}(2020){Decarli}, {Aravena}, {Boogaard}, {Carilli},
  {Gonz{\'a}lez-L{\'o}pez}, {Walter}, {Cortes}, {Cox}, {da Cunha}, {Daddi},
  {D{\'\i}az-Santos}, {Hodge}, {Inami}, {Neeleman}, {Novak}, {Oesch},
  {Popping}, {Riechers}, {Smail}, {Uzgil}, {van der Werf}, {Wagg}, \&
  {Weiss}}]{decarli2020}
{Decarli}, R., {Aravena}, M., {Boogaard}, L., {et~al.} 2020, \apj, 902, 110

\bibitem[{{Dickey} \& {Lockman}(1990)}]{dickey1990}
{Dickey}, J.~M., \& {Lockman}, F.~J. 1990, \araa, 28, 215

\bibitem[{{D'Odorico} {et~al.}(2013){D'Odorico}, {Cupani}, {Cristiani},
  {Maiolino}, {Molaro}, {Nonino}, {Centuri{\'o}n}, {Cimatti}, {di Serego
  Alighieri}, {Fiore}, {Fontana}, {Gallerani}, {Giallongo}, {Mannucci},
  {Marconi}, {Pentericci}, {Viel}, \& {Vladilo}}]{dodorico2013}
{D'Odorico}, V., {Cupani}, G., {Cristiani}, S., {et~al.} 2013, \mnras, 435,
  1198

\bibitem[{{D'Odorico} {et~al.}(2022){D'Odorico}, {Finlator}, {Cristiani},
  {Cupani}, {Perrotta}, {Calura}, {C{\`e}nturion}, {Becker}, {Berg}, {Lopez},
  {Ellison}, \& {Pomante}}]{dodorico2022}
{D'Odorico}, V., {Finlator}, K., {Cristiani}, S., {et~al.} 2022, \mnras, 512,
  2389

\bibitem[{{Draine}(2011)}]{draine2011}
{Draine}, B.~T. 2011, {Physics of the Interstellar and Intergalactic Medium}

\bibitem[{{Draine} \& {Li}(2007)}]{draine2007}
{Draine}, B.~T., \& {Li}, A. 2007, \apj, 657, 810

\bibitem[{{Draine} {et~al.}(2014){Draine}, {Aniano}, {Krause}, {Groves},
  {Sandstrom}, {Braun}, {Leroy}, {Klaas}, {Linz}, {Rix}, {Schinnerer},
  {Schmiedeke}, \& {Walter}}]{draine2014}
{Draine}, B.~T., {Aniano}, G., {Krause}, O., {et~al.} 2014, \apj, 780, 172

\bibitem[{{Driver} {et~al.}(2011){Driver}, {Hill}, {Kelvin}, {Robotham},
  {Liske}, {Norberg}, {Baldry}, {Bamford}, {Hopkins}, {Loveday}, {Peacock},
  {Andrae}, {Bland-Hawthorn}, {Brough}, {Brown}, {Cameron}, {Ching}, {Colless},
  {Conselice}, {Croom}, {Cross}, {de Propris}, {Dye}, {Drinkwater}, {Ellis},
  {Graham}, {Grootes}, {Gunawardhana}, {Jones}, {van Kampen}, {Maraston},
  {Nichol}, {Parkinson}, {Phillipps}, {Pimbblet}, {Popescu}, {Prescott},
  {Roseboom}, {Sadler}, {Sansom}, {Sharp}, {Smith}, {Taylor}, {Thomas},
  {Tuffs}, {Wijesinghe}, {Dunne}, {Frenk}, {Jarvis}, {Madore}, {Meyer},
  {Seibert}, {Staveley-Smith}, {Sutherland}, \& {Warren}}]{driver2011}
{Driver}, S.~P., {Hill}, D.~T., {Kelvin}, L.~S., {et~al.} 2011, \mnras, 413,
  971

\bibitem[{{Driver} {et~al.}(2018){Driver}, {Andrews}, {da Cunha}, {Davies},
  {Lagos}, {Robotham}, {Vinsen}, {Wright}, {Alpaslan}, {Bland-Hawthorn},
  {Bourne}, {Brough}, {Bremer}, {Cluver}, {Colless}, {Conselice}, {Dunne},
  {Eales}, {Gomez}, {Holwerda}, {Hopkins}, {Kafle}, {Kelvin}, {Loveday},
  {Liske}, {Maddox}, {Phillipps}, {Pimbblet}, {Rowlands}, {Sansom}, {Taylor},
  {Wang}, \& {Wilkins}}]{driver2018}
{Driver}, S.~P., {Andrews}, S.~K., {da Cunha}, E., {et~al.} 2018, \mnras, 475,
  2891

\bibitem[{{D'Silva} {et~al.}(2023){D'Silva}, {Driver}, {Lagos}, {Robotham},
  {Bellstedt}, {Davies}, {Thorne}, {Bland-Hawthorn}, {Bravo}, {Holwerda},
  {Phillipps}, {Seymour}, {Siudek}, \& {Windhorst}}]{dsilva2023}
{D'Silva}, J. C.~J., {Driver}, S.~P., {Lagos}, C. D.~P., {et~al.} 2023, \mnras,
  524, 1448

\bibitem[{{Eldridge} {et~al.}(2017){Eldridge}, {Stanway}, {Xiao}, {McClelland},
  {Taylor}, {Ng}, {Greis}, \& {Bray}}]{eldridge2017}
{Eldridge}, J.~J., {Stanway}, E.~R., {Xiao}, L., {et~al.} 2017, \pasa, 34, e058

\bibitem[{{Estrada-Carpenter} {et~al.}(2019){Estrada-Carpenter}, {Papovich},
  {Momcheva}, {Brammer}, {Long}, {Quadri}, {Bridge}, {Dickinson}, {Ferguson},
  {Finkelstein}, {Giavalisco}, {Gosmeyer}, {Lotz}, {Salmon}, {Skelton},
  {Trump}, \& {Weiner}}]{estrada2019}
{Estrada-Carpenter}, V., {Papovich}, C., {Momcheva}, I., {et~al.} 2019, \apj,
  870, 133

\bibitem[{{Feigelson} \& {Nelson}(1985)}]{feigelson1985}
{Feigelson}, E.~D., \& {Nelson}, P.~I. 1985, \apj, 293, 192

\bibitem[{{Ferland} {et~al.}(2013){Ferland}, {Porter}, {van Hoof}, {Williams},
  {Abel}, {Lykins}, {Shaw}, {Henney}, \& {Stancil}}]{ferland2013}
{Ferland}, G.~J., {Porter}, R.~L., {van Hoof}, P.~A.~M., {et~al.} 2013, \rmxaa,
  49, 137

\bibitem[{{Ferrara} {et~al.}(2005){Ferrara}, {Scannapieco}, \&
  {Bergeron}}]{ferrara2005}
{Ferrara}, A., {Scannapieco}, E., \& {Bergeron}, J. 2005, \apjl, 634, L37

\bibitem[{{Fields} {et~al.}(2000){Fields}, {Freese}, \& {Graff}}]{fields2000}
{Fields}, B.~D., {Freese}, K., \& {Graff}, D.~S. 2000, \apj, 534, 265

\bibitem[{{Flores} {et~al.}(2021){Flores}, {Mantz}, {Allen}, {Morris},
  {Canning}, {Bleem}, {Calzadilla}, {Floyd}, {McDonald}, \&
  {Ruppin}}]{flores2021}
{Flores}, A.~M., {Mantz}, A.~B., {Allen}, S.~W., {et~al.} 2021, \mnras, 507,
  5195

\bibitem[{{Fontanot} {et~al.}(2017){Fontanot}, {De Lucia}, {Hirschmann},
  {Bruzual}, {Charlot}, \& {Zibetti}}]{fontanot2017}
{Fontanot}, F., {De Lucia}, G., {Hirschmann}, M., {et~al.} 2017, \mnras, 464,
  3812

\bibitem[{{Fox}(2011)}]{fox2011}
{Fox}, A.~J. 2011, \apj, 730, 58

\bibitem[{{Fox} {et~al.}(2007){Fox}, {Petitjean}, {Ledoux}, \&
  {Srianand}}]{fox2007}
{Fox}, A.~J., {Petitjean}, P., {Ledoux}, C., \& {Srianand}, R. 2007, \aap, 465,
  171

\bibitem[{{Fox} {et~al.}(2009){Fox}, {Prochaska}, {Ledoux}, {Petitjean},
  {Wolfe}, \& {Srianand}}]{fox2009}
{Fox}, A.~J., {Prochaska}, J.~X., {Ledoux}, C., {et~al.} 2009, \aap, 503, 731

\bibitem[{{Frank} {et~al.}(2018){Frank}, {Pieri}, {Mathur}, {Danforth}, \&
  {Shull}}]{frank2018}
{Frank}, S., {Pieri}, M.~M., {Mathur}, S., {Danforth}, C.~W., \& {Shull}, J.~M.
  2018, \mnras, 476, 1356

\bibitem[{{Fukugita} \& {Peebles}(2004)}]{fukugita2004}
{Fukugita}, M., \& {Peebles}, P.~J.~E. 2004, \apj, 616, 643

\bibitem[{{Fumagalli} {et~al.}(2016){Fumagalli}, {O'Meara}, \&
  {Prochaska}}]{fumagalli2016}
{Fumagalli}, M., {O'Meara}, J.~M., \& {Prochaska}, J.~X. 2016, \mnras, 455,
  4100

\bibitem[{{Gallazzi} {et~al.}(2014){Gallazzi}, {Bell}, {Zibetti}, {Brinchmann},
  \& {Kelson}}]{gallazzi2014}
{Gallazzi}, A., {Bell}, E.~F., {Zibetti}, S., {Brinchmann}, J., \& {Kelson},
  D.~D. 2014, \apj, 788, 72

\bibitem[{{Gallazzi} {et~al.}(2008){Gallazzi}, {Brinchmann}, {Charlot}, \&
  {White}}]{gallazzi2008}
{Gallazzi}, A., {Brinchmann}, J., {Charlot}, S., \& {White}, S. D.~M. 2008,
  \mnras, 383, 1439

\bibitem[{{Gallazzi} {et~al.}(2005){Gallazzi}, {Charlot}, {Brinchmann},
  {White}, \& {Tremonti}}]{gallazzi2005}
{Gallazzi}, A., {Charlot}, S., {Brinchmann}, J., {White}, S. D.~M., \&
  {Tremonti}, C.~A. 2005, \mnras, 362, 41

\bibitem[{{Gao} {et~al.}(2025){Gao}, {Peng}, {Wang}, {Ho}, {Renzini},
  {Gallazzi}, {Mannucci}, {Mo}, {Jing}, {Yang}, {Wang}, {Zhao}, {Dou}, {Gu},
  {Lyu}, {Maiolino}, {Wang}, {Wang}, {Xu}, {Yuan}, \& {Zhu}}]{gao2025}
{Gao}, Z., {Peng}, Y., {Wang}, K., {et~al.} 2025, \apj, 979, 66

\bibitem[{{Gastaldello} {et~al.}(2021){Gastaldello}, {Simionescu}, {Mernier},
  {Biffi}, {Gaspari}, {Sato}, \& {Matsushita}}]{gastaldello2021}
{Gastaldello}, F., {Simionescu}, A., {Mernier}, F., {et~al.} 2021, Universe, 7,
  208

\bibitem[{{Ghizzardi} {et~al.}(2021){Ghizzardi}, {Molendi}, {van der Burg}, {De
  Grandi}, {Bartalucci}, {Gastaldello}, {Rossetti}, {Biffi}, {Borgani},
  {Eckert}, {Ettori}, {Gaspari}, {Ghirardini}, \& {Rasia}}]{ghizzardi2021}
{Ghizzardi}, S., {Molendi}, S., {van der Burg}, R., {et~al.} 2021, \aap, 646,
  A92

\bibitem[{{Gibson} {et~al.}(2022){Gibson}, {Lehner}, {Oppenheimer}, {Howk},
  {Cooksey}, \& {Fox}}]{gibson2022}
{Gibson}, J.~L., {Lehner}, N., {Oppenheimer}, B.~D., {et~al.} 2022, \aj, 164, 9

\bibitem[{{Girichidis} {et~al.}(2020){Girichidis}, {Offner}, {Kritsuk},
  {Klessen}, {Hennebelle}, {Kruijssen}, {Krause}, {Glover}, \&
  {Padovani}}]{girichidis2020}
{Girichidis}, P., {Offner}, S. S.~R., {Kritsuk}, A.~G., {et~al.} 2020, \ssr,
  216, 68

\bibitem[{{Glover} \& {Clark}(2012)}]{glover2012}
{Glover}, S. C.~O., \& {Clark}, P.~C. 2012, \mnras, 421, 9

\bibitem[{{Gnat}(2017)}]{gnat2017}
{Gnat}, O. 2017, The Astrophysical Journal Supplement Series, 228, 11

\bibitem[{{Gottumukkala} {et~al.}(2024){Gottumukkala}, {Barrufet}, {Oesch},
  {Weibel}, {Allen}, {Alcalde Pampliega}, {Nelson}, {Williams}, {Brammer},
  {Fudamoto}, {Gonz{\'a}lez}, {Heintz}, {Illingworth}, {Magee}, {Naidu},
  {Shuntov}, {Stefanon}, {Toft}, {Valentino}, \& {Xiao}}]{Gottumukkala2024}
{Gottumukkala}, R., {Barrufet}, L., {Oesch}, P.~A., {et~al.} 2024, \mnras, 530,
  966

\bibitem[{{Guo} {et~al.}(2023){Guo}, {Wang}, {Jones}, \& {Behroozi}}]{guo2023}
{Guo}, H., {Wang}, J., {Jones}, M.~G., \& {Behroozi}, P. 2023, \apj, 955, 57

\bibitem[{{Haardt} \& {Madau}(1996)}]{haardt1996}
{Haardt}, F., \& {Madau}, P. 1996, \apj, 461, 20

\bibitem[{{Haardt} \& {Madau}(2012)}]{haardt2012}
---. 2012, \apj, 746, 125

\bibitem[{{Hafen} {et~al.}(2017){Hafen}, {Faucher-Gigu{\`e}re},
  {Angl{\'e}s-Alc{\'a}zar}, {Kere{\v s}}, {Feldmann}, {Chan}, {Quataert},
  {Murray}, \& {Hopkins}}]{hafen2017}
{Hafen}, Z., {Faucher-Gigu{\`e}re}, C.-A., {Angl{\'e}s-Alc{\'a}zar}, D.,
  {et~al.} 2017, \mnras, 469, 2292

\bibitem[{{Hamanowicz} {et~al.}(2020){Hamanowicz}, {P{\'e}roux}, {Zwaan},
  {Rahmani}, {Pettini}, {York}, {Klitsch}, {Augustin}, {Krogager}, {Kulkarni},
  {Fresco}, {Biggs}, {Milliard}, \& {Vernet}}]{hamanowicz2020}
{Hamanowicz}, A., {P{\'e}roux}, C., {Zwaan}, M.~A., {et~al.} 2020, \mnras, 492,
  2347

\bibitem[{{Hamanowicz} {et~al.}(2023){Hamanowicz}, {Zwaan}, {P{\'e}roux},
  {Lagos}, {Klitsch}, {Ivison}, {Biggs}, {Szakacs}, \&
  {Fresco}}]{Hamanowicz2023}
{Hamanowicz}, A., {Zwaan}, M.~A., {P{\'e}roux}, C., {et~al.} 2023, \mnras, 519,
  34

\bibitem[{{He} {et~al.}(2024){He}, {Wang}, {Jones}, {Treu}, {Glazebrook},
  {Malkan}, {Vulcani}, {Metha}, {Brada{\v{c}}}, {Brammer}, {Roberts-Borsani},
  {Strait}, {Bonchi}, {Castellano}, {Fontana}, {Mason}, {Merlin}, {Morishita},
  {Paris}, {Santini}, {Trenti}, {Boyett}, \& {Grasha}}]{He2024}
{He}, X., {Wang}, X., {Jones}, T., {et~al.} 2024, \apjl, 960, L13

\bibitem[{{Heiles}(2001)}]{heiles2001}
{Heiles}, C. 2001, \apjl, 551, L105

\bibitem[{{Heiles} \& {Troland}(2003)}]{heiles2003}
{Heiles}, C., \& {Troland}, T.~H. 2003, \apj, 586, 1067

\bibitem[{{Heintz} {et~al.}(2023){Heintz}, {Shapley}, {Sanders}, {Killi},
  {Watson}, {Magdis}, {Valentino}, {Ginolfi}, {Narayanan}, {Greve}, {Fynbo},
  {Vizgan}, \& {Wilson}}]{heintz2023}
{Heintz}, K.~E., {Shapley}, A.~E., {Sanders}, R.~L., {et~al.} 2023, \aap, 678,
  A30

\bibitem[{{Henriques} {et~al.}(2020){Henriques}, {Yates}, {Fu}, {Guo},
  {Kauffmann}, {Srisawat}, {Thomas}, \& {White}}]{henriques2020}
{Henriques}, B. M.~B., {Yates}, R.~M., {Fu}, J., {et~al.} 2020, \mnras, 491,
  5795

\bibitem[{{Hirschi} {et~al.}(2005){Hirschi}, {Meynet}, \&
  {Maeder}}]{hirchi2005}
{Hirschi}, R., {Meynet}, G., \& {Maeder}, A. 2005, \aap, 433, 1013

\bibitem[{{Hirschmann} {et~al.}(2016){Hirschmann}, {De Lucia}, \&
  {Fontanot}}]{hirschman2016}
{Hirschmann}, M., {De Lucia}, G., \& {Fontanot}, F. 2016, \mnras, 461, 1760

\bibitem[{{Hough} {et~al.}(2023){Hough}, {Rennehan}, {Kobayashi}, {Loubser},
  {Dav{\'e}}, {Babul}, \& {Cui}}]{hough2023}
{Hough}, R.~T., {Rennehan}, D., {Kobayashi}, C., {et~al.} 2023, \mnras, 525,
  1061

\bibitem[{{Howk} {et~al.}(2009){Howk}, {Ribaudo}, {Lehner}, {Prochaska}, \&
  {Chen}}]{howk2009}
{Howk}, J.~C., {Ribaudo}, J.~S., {Lehner}, N., {Prochaska}, J.~X., \& {Chen},
  H.-W. 2009, \mnras, 396, 1875

\bibitem[{{Hunter}(2007)}]{hunter2007}
{Hunter}, J.~D. 2007, Computing in Science and Engineering, 9, 90

\bibitem[{{Hussain} {et~al.}(2015){Hussain}, {Muzahid}, {Narayanan},
  {Srianand}, {Wakker}, {Charlton}, \& {Pathak}}]{hussain2015}
{Hussain}, T., {Muzahid}, S., {Narayanan}, A., {et~al.} 2015, \mnras, 446, 2444

\bibitem[{{Janknecht} {et~al.}(2006){Janknecht}, {Reimers}, {Lopez}, \&
  {Tytler}}]{janknecht2006}
{Janknecht}, E., {Reimers}, D., {Lopez}, S., \& {Tytler}, D. 2006, \aap, 458,
  427

\bibitem[{{Jenkins}(2009)}]{jenkins2009}
{Jenkins}, E.~B. 2009, \apj, 700, 1299

\bibitem[{{Jenkins} \& {Wallerstein}(2017)}]{jenkins2017}
{Jenkins}, E.~B., \& {Wallerstein}, G. 2017, \apj, 838, 85

\bibitem[{{Johnson} {et~al.}(2013){Johnson}, {Chen}, \&
  {Mulchaey}}]{johnson2013}
{Johnson}, S.~D., {Chen}, H.-W., \& {Mulchaey}, J.~S. 2013, \mnras, 434, 1765

\bibitem[{{Jones} {et~al.}(2018){Jones}, {Haynes}, {Giovanelli}, \&
  {Moorman}}]{jones2018}
{Jones}, M.~G., {Haynes}, M.~P., {Giovanelli}, R., \& {Moorman}, C. 2018,
  \mnras, 477, 2

\bibitem[{{Kalberla} \& {Haud}(2018)}]{kalberla2018}
{Kalberla}, P.~M.~W., \& {Haud}, U. 2018, \aap, 619, A58

\bibitem[{{Kashino} {et~al.}(2021){Kashino}, {Lilly}, {Silverman}, {Renzini},
  {Daddi}, {Bardelli}, {Cucciati}, {Kartaltepe}, {Mainieri}, {Pell{\'o}},
  {Peng}, {Sanders}, \& {Zucca}}]{kashino2021}
{Kashino}, D., {Lilly}, S.~J., {Silverman}, J.~D., {et~al.} 2021, \apj, 909,
  213

\bibitem[{{Kashino} {et~al.}(2022){Kashino}, {Lilly}, {Renzini}, {Daddi},
  {Zamorani}, {Silverman}, {Ilbert}, {Peng}, {Mainieri}, {Bardelli}, {Zucca},
  {Kartaltepe}, \& {Sanders}}]{kashino2022}
{Kashino}, D., {Lilly}, S.~J., {Renzini}, A., {et~al.} 2022, \apj, 925, 82

\bibitem[{{Kim} {et~al.}(2021){Kim}, {Wakker}, {Nasir}, {Carswell}, {Savage},
  {Bolton}, {Fox}, {Viel}, {Haehnelt}, {Charlton}, \& {Rosenwasser}}]{kim2021}
{Kim}, T.~S., {Wakker}, B.~P., {Nasir}, F., {et~al.} 2021, \mnras, 501, 5811

\bibitem[{{Kirby} {et~al.}(2013){Kirby}, {Cohen}, {Guhathakurta}, {Cheng},
  {Bullock}, \& {Gallazzi}}]{kirby2013}
{Kirby}, E.~N., {Cohen}, J.~G., {Guhathakurta}, P., {et~al.} 2013, \apj, 779,
  102

\bibitem[{{Kirkpatrick} {et~al.}(2011){Kirkpatrick}, {McNamara}, \&
  {Cavagnolo}}]{kirkpatrick2011}
{Kirkpatrick}, C.~C., {McNamara}, B.~R., \& {Cavagnolo}, K.~W. 2011, \apjl,
  731, L23

\bibitem[{{Klessen} \& {Glover}(2016)}]{klessen2016}
{Klessen}, R.~S., \& {Glover}, S. C.~O. 2016, Saas-Fee Advanced Course, 43, 85

\bibitem[{{Klitsch} {et~al.}(2019){Klitsch}, {P{\'e}roux}, {Zwaan}, {Smail},
  {Nelson}, {Popping}, {Chen}, {Diemer}, {Ivison}, {Allison}, {Muller},
  {Swinbank}, {Hamanowicz}, {Biggs}, \& {Dutta}}]{klitsch2019}
{Klitsch}, A., {P{\'e}roux}, C., {Zwaan}, M.~A., {et~al.} 2019, \mnras, 490,
  1220

\bibitem[{{Konstantopoulou} {et~al.}(2024){Konstantopoulou}, {De Cia},
  {Ledoux}, {Krogager}, {Mattsson}, {Watson}, {Heintz}, {P{\'e}roux},
  {Noterdaeme}, {Andersen}, {Fynbo}, {Jermann}, \&
  {Ramburuth-Hurt}}]{konstantopoulou2024}
{Konstantopoulou}, C., {De Cia}, A., {Ledoux}, C., {et~al.} 2024, \aap, 681,
  A64

\bibitem[{{Kriek} {et~al.}(2019){Kriek}, {Price}, {Conroy}, {Suess}, {Mowla},
  {Pasha}, {Bezanson}, {van Dokkum}, \& {Barro}}]{kriek2019}
{Kriek}, M., {Price}, S.~H., {Conroy}, C., {et~al.} 2019, \apjl, 880, L31

\bibitem[{{Kroupa}(2001)}]{kroupa2001}
{Kroupa}, P. 2001, \mnras, 322, 231

\bibitem[{{Krumholz} {et~al.}(2009){Krumholz}, {McKee}, \&
  {Tumlinson}}]{krumholz2009}
{Krumholz}, M.~R., {McKee}, C.~F., \& {Tumlinson}, J. 2009, \apj, 693, 216

\bibitem[{{Lagos} {et~al.}(2018){Lagos}, {Tobar}, {Robotham}, {Obreschkow},
  {Mitchell}, {Power}, \& {Elahi}}]{lagos2018}
{Lagos}, C. d.~P., {Tobar}, R.~J., {Robotham}, A. S.~G., {et~al.} 2018, \mnras,
  481, 3573

\bibitem[{{Lanzetta} {et~al.}(1995){Lanzetta}, {Bowen}, {Tytler}, \&
  {Webb}}]{lanzetta1995}
{Lanzetta}, K.~M., {Bowen}, D.~V., {Tytler}, D., \& {Webb}, J.~K. 1995, \apj,
  442, 538

\bibitem[{{Lehner} {et~al.}(2014){Lehner}, {O'Meara}, {Fox}, {Howk},
  {Prochaska}, {Burns}, \& {Armstrong}}]{lehner2014}
{Lehner}, N., {O'Meara}, J.~M., {Fox}, A.~J., {et~al.} 2014, \apj, 788, 119

\bibitem[{{Lehner} {et~al.}(2016){Lehner}, {O'Meara}, {Howk}, {Prochaska}, \&
  {Fumagalli}}]{lehner2016}
{Lehner}, N., {O'Meara}, J.~M., {Howk}, J.~C., {Prochaska}, J.~X., \&
  {Fumagalli}, M. 2016, \apj, 833, 283

\bibitem[{{Lehner} {et~al.}(2018){Lehner}, {Wotta}, {Howk}, {O'Meara},
  {Oppenheimer}, \& {Cooksey}}]{lehner2018a}
{Lehner}, N., {Wotta}, C.~B., {Howk}, J.~C., {et~al.} 2018, \apj, 866, 33

\bibitem[{{Lehner} {et~al.}(2019){Lehner}, {Wotta}, {Howk}, {O'Meara},
  {Oppenheimer}, \& {Cooksey}}]{lehner2019}
---. 2019, \apj, 887, 5

\bibitem[{{Lehner} {et~al.}(2013){Lehner}, {Howk}, {Tripp}, {Tumlinson},
  {Prochaska}, {O'Meara}, {Thom}, {Werk}, {Fox}, \& {Ribaudo}}]{lehner2013}
{Lehner}, N., {Howk}, J.~C., {Tripp}, T.~M., {et~al.} 2013, \apj, 770, 138

\bibitem[{{Lehner} {et~al.}(2022){Lehner}, {Kopenhafer}, {O'Meara}, {Howk},
  {Fumagalli}, {Prochaska}, {Acharyya}, {O'Shea}, {Peeples}, {Tumlinson}, \&
  {Hummels}}]{lehner2022}
{Lehner}, N., {Kopenhafer}, C., {O'Meara}, J.~M., {et~al.} 2022, \apj, 936, 156

\bibitem[{{Leitherer} {et~al.}(1999){Leitherer}, {Schaerer}, {Goldader},
  {Delgado}, {Robert}, {Kune}, {de Mello}, {Devost}, \&
  {Heckman}}]{Leitherer1999}
{Leitherer}, C., {Schaerer}, D., {Goldader}, J.~D., {et~al.} 1999, \apjs, 123,
  3

\bibitem[{{Leja} {et~al.}(2020){Leja}, {Speagle}, {Johnson}, {Conroy}, {van
  Dokkum}, \& {Franx}}]{leja2020}
{Leja}, J., {Speagle}, J.~S., {Johnson}, B.~D., {et~al.} 2020, \apj, 893, 111

\bibitem[{{Leung} {et~al.}(2024){Leung}, {Wild}, {Papathomas}, {Carnall},
  {Zheng}, {Boardman}, {Wang}, \& {Johansson}}]{leung2024}
{Leung}, H.-H., {Wild}, V., {Papathomas}, M., {et~al.} 2024, \mnras, 528, 4029

\bibitem[{{Li} {et~al.}(2013){Li}, {de Grijs}, \& {Deng}}]{li2013}
{Li}, C., {de Grijs}, R., \& {Deng}, L. 2013, \mnras, 436, 1497

\bibitem[{{Li} {et~al.}(2023){Li}, {Cai}, {Bian}, {Lin}, {Li}, {Wu}, {Sun},
  {Zhang}, {Golden-Marx}, {Sun}, {Zou}, {Fan}, {Egami}, {Charlot}, {Bruzual},
  \& {Chevallard}}]{Li2023}
{Li}, M., {Cai}, Z., {Bian}, F., {et~al.} 2023, \apjl, 955, L18

\bibitem[{{Lilly} {et~al.}(2007){Lilly}, {Le F{\`e}vre}, {Renzini}, {Zamorani},
  {Scodeggio}, {Contini}, {Carollo}, {Hasinger}, {Kneib}, {Iovino}, {Le Brun},
  {Maier}, {Mainieri}, {Mignoli}, {Silverman}, {Tasca}, {Bolzonella},
  {Bongiorno}, {Bottini}, {Capak}, {Caputi}, {Cimatti}, {Cucciati}, {Daddi},
  {Feldmann}, {Franzetti}, {Garilli}, {Guzzo}, {Ilbert}, {Kampczyk}, {Kovac},
  {Lamareille}, {Leauthaud}, {Le Borgne}, {McCracken}, {Marinoni}, {Pello},
  {Ricciardelli}, {Scarlata}, {Vergani}, {Sanders}, {Schinnerer}, {Scoville},
  {Taniguchi}, {Arnouts}, {Aussel}, {Bardelli}, {Brusa}, {Cappi}, {Ciliegi},
  {Finoguenov}, {Foucaud}, {Franceschini}, {Halliday}, {Impey}, {Knobel},
  {Koekemoer}, {Kurk}, {Maccagni}, {Maddox}, {Marano}, {Marconi}, {Meneux},
  {Mobasher}, {Moreau}, {Peacock}, {Porciani}, {Pozzetti}, {Scaramella},
  {Schiminovich}, {Shopbell}, {Smail}, {Thompson}, {Tresse}, {Vettolani},
  {Zanichelli}, \& {Zucca}}]{lilly2007}
{Lilly}, S.~J., {Le F{\`e}vre}, O., {Renzini}, A., {et~al.} 2007, \apjs, 172,
  70

\bibitem[{{Liu} {et~al.}(2020){Liu}, {Tozzi}, {Ettori}, {De Grandi},
  {Gastaldello}, {Rosati}, \& {Norman}}]{liu2020}
{Liu}, A., {Tozzi}, P., {Ettori}, S., {et~al.} 2020, \aap, 637, A58

\bibitem[{{Liu} {et~al.}(2025){Liu}, {Dunlop}, {McLure}, {McLeod}, {Barrufet},
  {Carnall}, {Begley}, {P{\'e}rez-Gonz{\'a}lez}, {Donnan}, {Ellis}, {Grogin},
  {Magee}, {Illingworth}, {Cullen}, {Stevenson}, {Koekemoer}, {Fontana}, \&
  {Bowler}}]{liu2025}
{Liu}, F.-Y., {Dunlop}, J.~S., {McLure}, R.~J., {et~al.} 2025, arXiv e-prints,
  arXiv:2503.07774

\bibitem[{{Lundgren} {et~al.}(2021){Lundgren}, {Creech}, {Brammer}, {Kirse},
  {Peek}, {Wake}, {York}, {Chisholm}, {Erb}, {Kulkarni}, {Straka}, {Tremonti},
  \& {van Dokkum}}]{lundgren2021}
{Lundgren}, B.~F., {Creech}, S., {Brammer}, G., {et~al.} 2021, \apj, 913, 50

\bibitem[{{Madau} \& {Dickinson}(2014)}]{madau2014}
{Madau}, P., \& {Dickinson}, M. 2014, \araa, 52, 415

\bibitem[{{Maeder}(1992)}]{maeder1992}
{Maeder}, A. 1992, \aap, 264, 105

\bibitem[{{Maiolino} \& {Mannucci}(2019)}]{maiolino2019}
{Maiolino}, R., \& {Mannucci}, F. 2019, \aapr, 27, 3

\bibitem[{{Mantz} {et~al.}(2017){Mantz}, {Allen}, {Morris}, {Simionescu},
  {Urban}, {Werner}, \& {Zhuravleva}}]{mantz2017}
{Mantz}, A.~B., {Allen}, S.~W., {Morris}, R.~G., {et~al.} 2017, \mnras, 472,
  2877

\bibitem[{{Manuwal} {et~al.}(2021){Manuwal}, {Narayanan}, {Udhwani},
  {Srianand}, {Savage}, {Charlton}, \& {Misawa}}]{Manuwal2021}
{Manuwal}, A., {Narayanan}, A., {Udhwani}, P., {et~al.} 2021, \mnras, 505, 3635

\bibitem[{{Mas-Ribas} {et~al.}(2017){Mas-Ribas}, {Miralda-Escud{\'e}},
  {P{\'e}rez-R{\`a}fols}, {Arinyo-i-Prats}, {Noterdaeme}, {Petitjean},
  {Schneider}, {York}, \& {Ge}}]{masribas2017}
{Mas-Ribas}, L., {Miralda-Escud{\'e}}, J., {P{\'e}rez-R{\`a}fols}, I., {et~al.}
  2017, \apj, 846, 4

\bibitem[{{Matthews} {et~al.}(2024){Matthews}, {Kelson}, {Newman}, {Camilo},
  {Condon}, {Cotton}, {Dickinson}, {Jarrett}, \& {Lacy}}]{matthews2024}
{Matthews}, A.~M., {Kelson}, D.~D., {Newman}, A.~B., {et~al.} 2024, \apj, 966,
  194

\bibitem[{{McLeod} {et~al.}(2021){McLeod}, {McLure}, {Dunlop}, {Cullen},
  {Carnall}, \& {Duncan}}]{mcleoud2021}
{McLeod}, D.~J., {McLure}, R.~J., {Dunlop}, J.~S., {et~al.} 2021, \mnras, 503,
  4413

\bibitem[{{McLure} {et~al.}(2018){McLure}, {Pentericci}, {Cimatti}, {Dunlop},
  {Elbaz}, {Fontana}, {Nandra}, {Amorin}, {Bolzonella}, {Bongiorno}, {Carnall},
  {Castellano}, {Cirasuolo}, {Cucciati}, {Cullen}, {De Barros}, {Finkelstein},
  {Fontanot}, {Franzetti}, {Fumana}, {Gargiulo}, {Garilli}, {Guaita},
  {Hartley}, {Iovino}, {Jarvis}, {Juneau}, {Karman}, {Maccagni}, {Marchi},
  {M{\'a}rmol-Queralt{\'o}}, {Pompei}, {Pozzetti}, {Scodeggio}, {Sommariva},
  {Talia}, {Almaini}, {Balestra}, {Bardelli}, {Bell}, {Bourne}, {Bowler},
  {Brusa}, {Buitrago}, {Caputi}, {Cassata}, {Charlot}, {Citro}, {Cresci},
  {Cristiani}, {Curtis-Lake}, {Dickinson}, {Fazio}, {Ferguson}, {Fiore},
  {Franco}, {Fynbo}, {Galametz}, {Georgakakis}, {Giavalisco}, {Grazian},
  {Hathi}, {Jung}, {Kim}, {Koekemoer}, {Khusanova}, {Le F{\`e}vre}, {Lotz},
  {Mannucci}, {Maltby}, {Matsuoka}, {McLeod}, {Mendez-Hernandez},
  {Mendez-Abreu}, {Mignoli}, {Moresco}, {Mortlock}, {Nonino}, {Pannella},
  {Papovich}, {Popesso}, {Rosario}, {Salvato}, {Santini}, {Schaerer},
  {Schreiber}, {Stark}, {Tasca}, {Thomas}, {Treu}, {Vanzella}, {Wild},
  {Williams}, {Zamorani}, \& {Zucca}}]{mclure2018}
{McLure}, R.~J., {Pentericci}, L., {Cimatti}, A., {et~al.} 2018, \mnras, 479,
  25

\bibitem[{{McQuinn}(2016)}]{mcquinn2016}
{McQuinn}, M. 2016, \araa, 54, 313

\bibitem[{{Meiksin}(2009)}]{meiksin2009}
{Meiksin}, A.~A. 2009, Reviews of Modern Physics, 81, 1405

\bibitem[{{Meiring} {et~al.}(2009){Meiring}, {Lauroesch}, {Kulkarni},
  {P{\'e}roux}, {Khare}, \& {York}}]{meiring2009}
{Meiring}, J.~D., {Lauroesch}, J.~T., {Kulkarni}, V.~P., {et~al.} 2009, \mnras,
  397, 2037

\bibitem[{{Mo} {et~al.}(2010){Mo}, {van den Bosch}, \& {White}}]{mo2010}
{Mo}, H., {van den Bosch}, F.~C., \& {White}, S. 2010, {Galaxy Formation and
  Evolution}

\bibitem[{{Molendi} {et~al.}(2016){Molendi}, {Eckert}, {De Grandi}, {Ettori},
  {Gastaldello}, {Ghizzardi}, {Pratt}, \& {Rossetti}}]{molendi2016}
{Molendi}, S., {Eckert}, D., {De Grandi}, S., {et~al.} 2016, \aap, 586, A32

\bibitem[{{Molendi} {et~al.}(2024){Molendi}, {Ghizzardi}, {De Grandi},
  {Balboni}, {Bartalucci}, {Eckert}, {Gastaldello}, {Lovisari}, {Riva}, \&
  {Rossetti}}]{molendi2024}
{Molendi}, S., {Ghizzardi}, S., {De Grandi}, S., {et~al.} 2024, \aap, 685, A88

\bibitem[{{Moser} {et~al.}(2022){Moser}, {Battaglia}, {Nagai}, {Lau}, {Machado
  Poletti Valle}, {Villaescusa-Navarro}, {Amodeo}, {Angl{\'e}s-Alc{\'a}zar},
  {Bryan}, {Dave}, {Hernquist}, \& {Vogelsberger}}]{moser2022}
{Moser}, E., {Battaglia}, N., {Nagai}, D., {et~al.} 2022, \apj, 933, 133

\bibitem[{{Nandra} {et~al.}(2013){Nandra}, {Barret}, {Barcons}, {Fabian}, {den
  Herder}, {Piro}, {Watson}, {Adami}, {Aird}, {Afonso}, {Alexander},
  {Argiroffi}, {Amati}, {Arnaud}, {Atteia}, {Audard}, {Badenes}, {Ballet},
  {Ballo}, {Bamba}, {Bhardwaj}, {Stefano Battistelli}, {Becker}, {De Becker},
  {Behar}, {Bianchi}, {Biffi}, {B{\^\i}rzan}, {Bocchino}, {Bogdanov}, {Boirin},
  {Boller}, {Borgani}, {Borm}, {Bouch{\'e}}, {Bourdin}, {Bower}, {Braito},
  {Branchini}, {Branduardi-Raymont}, {Bregman}, {Brenneman}, {Brightman},
  {Br{\"u}ggen}, {Buchner}, {Bulbul}, {Brusa}, {Bursa}, {Caccianiga},
  {Cackett}, {Campana}, {Cappelluti}, {Cappi}, {Carrera}, {Ceballos},
  {Christensen}, {Chu}, {Churazov}, {Clerc}, {Corbel}, {Corral}, {Comastri},
  {Costantini}, {Croston}, {Dadina}, {D'Ai}, {Decourchelle}, {Della Ceca},
  {Dennerl}, {Dolag}, {Done}, {Dovciak}, {Drake}, {Eckert}, {Edge}, {Ettori},
  {Ezoe}, {Feigelson}, {Fender}, {Feruglio}, {Finoguenov}, {Fiore}, {Galeazzi},
  {Gallagher}, {Gandhi}, {Gaspari}, {Gastaldello}, {Georgakakis},
  {Georgantopoulos}, {Gilfanov}, {Gitti}, {Gladstone}, {Goosmann}, {Gosset},
  {Grosso}, {Guedel}, {Guerrero}, {Haberl}, {Hardcastle}, {Heinz}, {Alonso
  Herrero}, {Herv{\'e}}, {Holmstrom}, {Iwasawa}, {Jonker}, {Kaastra}, {Kara},
  {Karas}, {Kastner}, {King}, {Kosenko}, {Koutroumpa}, {Kraft}, {Kreykenbohm},
  {Lallement}, {Lanzuisi}, {Lee}, {Lemoine-Goumard}, {Lobban}, {Lodato},
  {Lovisari}, {Lotti}, {McCharthy}, {McNamara}, {Maggio}, {Maiolino}, {De
  Marco}, {de Martino}, {Mateos}, {Matt}, {Maughan}, {Mazzotta}, {Mendez},
  {Merloni}, {Micela}, {Miceli}, {Mignani}, {Miller}, {Miniutti}, {Molendi},
  {Montez}, {Moretti}, {Motch}, {Naz{\'e}}, {Nevalainen}, {Nicastro}, {Nulsen},
  {Ohashi}, {O'Brien}, {Osborne}, {Oskinova}, {Pacaud}, {Paerels}, {Page},
  {Papadakis}, {Pareschi}, {Petre}, {Petrucci}, {Piconcelli}, {Pillitteri},
  {Pinto}, {de Plaa}, {Pointecouteau}, {Ponman}, {Ponti}, {Porquet}, {Pounds},
  {Pratt}, {Predehl}, {Proga}, {Psaltis}, {Rafferty}, {Ramos-Ceja}, {Ranalli},
  {Rasia}, {Rau}, {Rauw}, {Rea}, {Read}, {Reeves}, {Reiprich}, {Renaud},
  {Reynolds}, {Risaliti}, {Rodriguez}, {Rodriguez Hidalgo}, {Roncarelli},
  {Rosario}, {Rossetti}, {Rozanska}, {Rovilos}, {Salvaterra}, {Salvato}, {Di
  Salvo}, {Sanders}, {Sanz-Forcada}, {Schawinski}, {Schaye}, {Schwope},
  {Sciortino}, {Severgnini}, {Shankar}, {Sijacki}, {Sim}, {Schmid}, {Smith},
  {Steiner}, {Stelzer}, {Stewart}, {Strohmayer}, {Str{\"u}der}, {Sun}, {Takei},
  {Tatischeff}, {Tiengo}, {Tombesi}, {Trinchieri}, {Tsuru}, {Ud-Doula},
  {Ursino}, {Valencic}, {Vanzella}, {Vaughan}, {Vignali}, {Vink}, {Vito},
  {Volonteri}, {Wang}, {Webb}, {Willingale}, {Wilms}, {Wise}, {Worrall},
  {Young}, {Zampieri}, {In't Zand}, {Zane}, {Zezas}, {Zhang}, \&
  {Zhuravleva}}]{nandra2013}
{Nandra}, K., {Barret}, D., {Barcons}, X., {et~al.} 2013, arXiv e-prints,
  arXiv:1306.2307

\bibitem[{{Narayanan} {et~al.}(2012){Narayanan}, {Savage}, \&
  {Wakker}}]{narayanan2012}
{Narayanan}, A., {Savage}, B.~D., \& {Wakker}, B.~P. 2012, \apj, 752, 65

\bibitem[{{Nasir} {et~al.}(2021){Nasir}, {Cain}, {D'Aloisio}, {Gangolli}, \&
  {McQuinn}}]{nasir2021}
{Nasir}, F., {Cain}, C., {D'Aloisio}, A., {Gangolli}, N., \& {McQuinn}, M.
  2021, \apj, 923, 161

\bibitem[{{Neeleman} {et~al.}(2015){Neeleman}, {Prochaska}, \&
  {Wolfe}}]{neeleman2015}
{Neeleman}, M., {Prochaska}, J.~X., \& {Wolfe}, A.~M. 2015, \apj, 800, 7

\bibitem[{{Nicastro} {et~al.}(2021){Nicastro}, {Kaastra}, {Argiroffi}, {Behar},
  {Bianchi}, {Bocchino}, {Borgani}, {Branduardi-Raymont}, {Bregman},
  {Churazov}, {Diaz-Trigo}, {Done}, {Drake}, {Fang}, {Grosso}, {Luminari},
  {Mehdipour}, {Paerels}, {Piconcelli}, {Pinto}, {Porquet}, {Reeves}, {Schaye},
  {Sciortino}, {Smith}, {Spiga}, {Tomaru}, {Tombesi}, {Wijers}, \&
  {Zappacosta}}]{nicastro2021}
{Nicastro}, F., {Kaastra}, J., {Argiroffi}, C., {et~al.} 2021, Experimental
  Astronomy, 51, 1013

\bibitem[{{Nomoto} {et~al.}(2013){Nomoto}, {Kobayashi}, \&
  {Tominaga}}]{nomoto2013}
{Nomoto}, K., {Kobayashi}, C., \& {Tominaga}, N. 2013, \araa, 51, 457

\bibitem[{{Noterdaeme} {et~al.}(2012){Noterdaeme}, {Petitjean}, {Carithers},
  {P{\^a}ris}, {Font-Ribera}, {Bailey}, {Aubourg}, {Bizyaev}, {Ebelke},
  {Finley}, {Ge}, {Malanushenko}, {Malanushenko}, {Miralda-Escud{\'e}},
  {Myers}, {Oravetz}, {Pan}, {Pieri}, {Ross}, {Schneider}, {Simmons}, \&
  {York}}]{noterdaeme2012}
{Noterdaeme}, P., {Petitjean}, P., {Carithers}, W.~C., {et~al.} 2012, \aap,
  547, L1

\bibitem[{{Oliveira} {et~al.}(2014){Oliveira}, {Sembach}, {Tumlinson},
  {O'Meara}, \& {Thom}}]{oliveira2014}
{Oliveira}, C.~M., {Sembach}, K.~R., {Tumlinson}, J., {O'Meara}, J., \& {Thom},
  C. 2014, \apj, 783, 22

\bibitem[{{O'Meara} {et~al.}(2017){O'Meara}, {Lehner}, {Howk}, {Prochaska},
  {Fox}, {Peeples}, {Tumlinson}, \& {O'Shea}}]{omeara2017}
{O'Meara}, J.~M., {Lehner}, N., {Howk}, J.~C., {et~al.} 2017, \aj, 154, 114

\bibitem[{{O'Meara} {et~al.}(2013){O'Meara}, {Prochaska}, {Worseck}, {Chen}, \&
  {Madau}}]{omeara2013}
{O'Meara}, J.~M., {Prochaska}, J.~X., {Worseck}, G., {Chen}, H.-W., \& {Madau},
  P. 2013, \apj, 765, 137

\bibitem[{{O'Meara} {et~al.}(2015){O'Meara}, {Lehner}, {Howk}, {Prochaska},
  {Fox}, {Swain}, {Gelino}, {Berriman}, \& {Tran}}]{omeara2015}
{O'Meara}, J.~M., {Lehner}, N., {Howk}, J.~C., {et~al.} 2015, \aj, 150, 111

\bibitem[{{Oppenheimer} \& {Dav{\'e}}(2006)}]{oppenheimer2006}
{Oppenheimer}, B.~D., \& {Dav{\'e}}, R. 2006, \mnras, 373, 1265

\bibitem[{{Oppenheimer} {et~al.}(2016){Oppenheimer}, {Crain}, {Schaye},
  {Rahmati}, {Richings}, {Trayford}, {Tumlinson}, {Bower}, {Schaller}, \&
  {Theuns}}]{oppenheimer2016}
{Oppenheimer}, B.~D., {Crain}, R.~A., {Schaye}, J., {et~al.} 2016, \mnras, 460,
  2157

\bibitem[{{Ostriker} \& {Vishniac}(1986)}]{ostriker1986}
{Ostriker}, J.~P., \& {Vishniac}, E.~T. 1986, \apjl, 306, L51

\bibitem[{{Pachat} {et~al.}(2017){Pachat}, {Narayanan}, {Khaire}, {Savage},
  {Muzahid}, \& {Wakker}}]{pachat2017}
{Pachat}, S., {Narayanan}, A., {Khaire}, V., {et~al.} 2017, \mnras, 471, 792

\bibitem[{{Pagel}(1999)}]{pagel1999}
{Pagel}, B.~E.~J. 1999, arXiv e-prints, astro

\bibitem[{{Patnaude} {et~al.}(2023){Patnaude}, {Kraft}, {Kilbourne}, {Bandler},
  {Bogdan}, {Cumbee}, {Eckart}, {Garraffo}, {Hodges-Kluck}, {Kelley},
  {Markevitch}, {Ogorzalek}, {Plucinsky}, {Porter}, {ZuHone}, {Zhuravleva},
  {Drake}, {Leutenegger}, {Kenyon}, {Smith}, {Zhang}, {DePalo}, {Li},
  {Williams}, {Amatucci}, {Houston}, {Apostolou}, {Kanner}, {Coderre},
  {Hayden}, {Martin}, {Osborne}, {Olson}, {Ramm}, \&
  {Richardson}}]{patnaude2023}
{Patnaude}, D.~J., {Kraft}, R.~P., {Kilbourne}, C., {et~al.} 2023, Journal of
  Astronomical Telescopes, Instruments, and Systems, 9, 041008

\bibitem[{Pedregosa {et~al.}(2011)Pedregosa, Varoquaux, Gramfort, Michel,
  Thirion, Grisel, Blondel, Prettenhofer, Weiss, Dubourg, Vanderplas, Passos,
  Cournapeau, Brucher, Perrot, \& Duchesnay}]{scikit-learn}
Pedregosa, F., Varoquaux, G., Gramfort, A., {et~al.} 2011, Journal of Machine
  Learning Research, 12, 2825

\bibitem[{{Peeples} \& {Somerville}(2013)}]{peeples2013}
{Peeples}, M.~S., \& {Somerville}, R.~S. 2013, \mnras, 428, 1766

\bibitem[{{Peeples} {et~al.}(2014){Peeples}, {Werk}, {Tumlinson},
  {Oppenheimer}, {Prochaska}, {Katz}, \& {Weinberg}}]{peeples2014}
{Peeples}, M.~S., {Werk}, J.~K., {Tumlinson}, J., {et~al.} 2014, \apj, 786, 54

\bibitem[{{Peeples} {et~al.}(2019){Peeples}, {Corlies}, {Tumlinson}, {O'Shea},
  {Lehner}, {O'Meara}, {Howk}, {Earl}, {Smith}, {Wise}, \&
  {Hummels}}]{peeples2019}
{Peeples}, M.~S., {Corlies}, L., {Tumlinson}, J., {et~al.} 2019, \apj, 873, 129

\bibitem[{{Pentericci} {et~al.}(2018){Pentericci}, {McLure}, {Garilli},
  {Cucciati}, {Franzetti}, {Iovino}, {Amorin}, {Bolzonella}, {Bongiorno},
  {Carnall}, {Castellano}, {Cimatti}, {Cirasuolo}, {Cullen}, {De Barros},
  {Dunlop}, {Elbaz}, {Finkelstein}, {Fontana}, {Fontanot}, {Fumana},
  {Gargiulo}, {Guaita}, {Hartley}, {Jarvis}, {Juneau}, {Karman}, {Maccagni},
  {Marchi}, {Marmol-Queralto}, {Nandra}, {Pompei}, {Pozzetti}, {Scodeggio},
  {Sommariva}, {Talia}, {Almaini}, {Balestra}, {Bardelli}, {Bell}, {Bourne},
  {Bowler}, {Brusa}, {Buitrago}, {Caputi}, {Cassata}, {Charlot}, {Citro},
  {Cresci}, {Cristiani}, {Curtis-Lake}, {Dickinson}, {Fazio}, {Ferguson},
  {Fiore}, {Franco}, {Fynbo}, {Galametz}, {Georgakakis}, {Giavalisco},
  {Grazian}, {Hathi}, {Jung}, {Kim}, {Koekemoer}, {Khusanova}, {Le F{\`e}vre},
  {Lotz}, {Mannucci}, {Maltby}, {Matsuoka}, {McLeod}, {Mendez-Hernandez},
  {Mendez-Abreu}, {Mignoli}, {Moresco}, {Mortlock}, {Nonino}, {Pannella},
  {Papovich}, {Popesso}, {Rosario}, {Salvato}, {Santini}, {Schaerer},
  {Schreiber}, {Stark}, {Tasca}, {Thomas}, {Treu}, {Vanzella}, {Wild},
  {Williams}, {Zamorani}, \& {Zucca}}]{pentericci2018}
{Pentericci}, L., {McLure}, R.~J., {Garilli}, B., {et~al.} 2018, \aap, 616,
  A174

\bibitem[{{P{\'e}roux} {et~al.}(2023){P{\'e}roux}, {De Cia}, \&
  {Howk}}]{peroux2023}
{P{\'e}roux}, C., {De Cia}, A., \& {Howk}, J.~C. 2023, \mnras, 522, 4852

\bibitem[{{P{\'e}roux} \& {Howk}(2020)}]{peroux2020}
{P{\'e}roux}, C., \& {Howk}, J.~C. 2020, \araa, 58, 363

\bibitem[{{P{\'e}roux} {et~al.}(2003){P{\'e}roux}, {McMahon},
  {Storrie-Lombardi}, \& {Irwin}}]{peroux2003}
{P{\'e}roux}, C., {McMahon}, R.~G., {Storrie-Lombardi}, L.~J., \& {Irwin},
  M.~J. 2003, \mnras, 346, 1103

\bibitem[{{Peroux} {et~al.}(2023){Peroux}, {Merloni}, {Liske}, {Salvato},
  {Augustin}, {Balzer}, {Cioni}, {Comparat}, {Driver}, {Fresco}, {Garzilli},
  {Hamanowicz}, {Klitsch}, {Kneib}, {Krogager}, {Nelson}, {Richard}, {Schady},
  {Shen}, {Szakacs}, {Weng}, {Yang}, \& {ByCycle Team}}]{peroux2023b}
{Peroux}, C., {Merloni}, A., {Liske}, J., {et~al.} 2023, The Messenger, 190, 42

\bibitem[{{Persic} \& {Salucci}(1992)}]{persic1992}
{Persic}, M., \& {Salucci}, P. 1992, \mnras, 258, 14P

\bibitem[{{Pettini}(1999)}]{pettini1999}
{Pettini}, M. 1999, in Chemical Evolution from Zero to High Redshift, ed. J.~R.
  {Walsh} \& M.~R. {Rosa}, 233

\bibitem[{{Pillepich} {et~al.}(2018){Pillepich}, {Springel}, {Nelson}, {Genel},
  {Naiman}, {Pakmor}, {Hernquist}, {Torrey}, {Vogelsberger}, {Weinberger}, \&
  {Marinacci}}]{pillepich2018}
{Pillepich}, A., {Springel}, V., {Nelson}, D., {et~al.} 2018, \mnras, 473, 4077

\bibitem[{{Planck Collaboration} {et~al.}(2020){Planck Collaboration},
  {Aghanim}, {Akrami}, {Ashdown}, {Aumont}, {Baccigalupi}, {Ballardini},
  {Banday}, {Barreiro}, {Bartolo}, {Basak}, {Battye}, {Benabed}, {Bernard},
  {Bersanelli}, {Bielewicz}, {Bock}, {Bond}, {Borrill}, {Bouchet}, {Boulanger},
  {Bucher}, {Burigana}, {Butler}, {Calabrese}, {Cardoso}, {Carron},
  {Challinor}, {Chiang}, {Chluba}, {Colombo}, {Combet}, {Contreras}, {Crill},
  {Cuttaia}, {de Bernardis}, {de Zotti}, {Delabrouille}, {Delouis}, {Di
  Valentino}, {Diego}, {Dor{\'e}}, {Douspis}, {Ducout}, {Dupac}, {Dusini},
  {Efstathiou}, {Elsner}, {En{\ss}lin}, {Eriksen}, {Fantaye}, {Farhang},
  {Fergusson}, {Fernandez-Cobos}, {Finelli}, {Forastieri}, {Frailis},
  {Fraisse}, {Franceschi}, {Frolov}, {Galeotta}, {Galli}, {Ganga},
  {G{\'e}nova-Santos}, {Gerbino}, {Ghosh}, {Gonz{\'a}lez-Nuevo}, {G{\'o}rski},
  {Gratton}, {Gruppuso}, {Gudmundsson}, {Hamann}, {Handley}, {Hansen},
  {Herranz}, {Hildebrandt}, {Hivon}, {Huang}, {Jaffe}, {Jones}, {Karakci},
  {Keih{\"a}nen}, {Keskitalo}, {Kiiveri}, {Kim}, {Kisner}, {Knox},
  {Krachmalnicoff}, {Kunz}, {Kurki-Suonio}, {Lagache}, {Lamarre}, {Lasenby},
  {Lattanzi}, {Lawrence}, {Le Jeune}, {Lemos}, {Lesgourgues}, {Levrier},
  {Lewis}, {Liguori}, {Lilje}, {Lilley}, {Lindholm}, {L{\'o}pez-Caniego},
  {Lubin}, {Ma}, {Mac{\'\i}as-P{\'e}rez}, {Maggio}, {Maino}, {Mandolesi},
  {Mangilli}, {Marcos-Caballero}, {Maris}, {Martin}, {Martinelli},
  {Mart{\'\i}nez-Gonz{\'a}lez}, {Matarrese}, {Mauri}, {McEwen}, {Meinhold},
  {Melchiorri}, {Mennella}, {Migliaccio}, {Millea}, {Mitra},
  {Miville-Desch{\^e}nes}, {Molinari}, {Montier}, {Morgante}, {Moss}, {Natoli},
  {N{\o}rgaard-Nielsen}, {Pagano}, {Paoletti}, {Partridge}, {Patanchon},
  {Peiris}, {Perrotta}, {Pettorino}, {Piacentini}, {Polastri}, {Polenta},
  {Puget}, {Rachen}, {Reinecke}, {Remazeilles}, {Renzi}, {Rocha}, {Rosset},
  {Roudier}, {Rubi{\~n}o-Mart{\'\i}n}, {Ruiz-Granados}, {Salvati}, {Sandri},
  {Savelainen}, {Scott}, {Shellard}, {Sirignano}, {Sirri}, {Spencer},
  {Sunyaev}, {Suur-Uski}, {Tauber}, {Tavagnacco}, {Tenti}, {Toffolatti},
  {Tomasi}, {Trombetti}, {Valenziano}, {Valiviita}, {Van Tent}, {Vibert},
  {Vielva}, {Villa}, {Vittorio}, {Wandelt}, {Wehus}, {White}, {White},
  {Zacchei}, \& {Zonca}}]{planck2020}
{Planck Collaboration}, {Aghanim}, N., {Akrami}, Y., {et~al.} 2020, \aap, 641,
  A6

\bibitem[{P{\"o}lsterl(2020)}]{sksurv}
P{\"o}lsterl, S. 2020, Journal of Machine Learning Research, 21, 1.
\newblock \url{http://jmlr.org/papers/v21/20-729.html}

\bibitem[{{Portinari} {et~al.}(1998){Portinari}, {Chiosi}, \&
  {Bressan}}]{portinari1998}
{Portinari}, L., {Chiosi}, C., \& {Bressan}, A. 1998, \aap, 334, 505

\bibitem[{{Poudel} {et~al.}(2018){Poudel}, {Kulkarni}, {Morrison},
  {P{\'e}roux}, {Som}, {Rahmani}, \& {Quiret}}]{poudel2018}
{Poudel}, S., {Kulkarni}, V.~P., {Morrison}, S., {et~al.} 2018, \mnras, 473,
  3559

\bibitem[{{Prochaska} {et~al.}(2003){Prochaska}, {Gawiser}, {Wolfe}, {Castro},
  \& {Djorgovski}}]{prochaska2003}
{Prochaska}, J.~X., {Gawiser}, E., {Wolfe}, A.~M., {Castro}, S., \&
  {Djorgovski}, S.~G. 2003, \apjl, 595, L9

\bibitem[{{Prochaska} {et~al.}(2014){Prochaska}, {Madau}, {O'Meara}, \&
  {Fumagalli}}]{prochaska2014}
{Prochaska}, J.~X., {Madau}, P., {O'Meara}, J.~M., \& {Fumagalli}, M. 2014,
  \mnras, 438, 476

\bibitem[{{Prochaska} {et~al.}(2015){Prochaska}, {O'Meara}, {Fumagalli},
  {Bernstein}, \& {Burles}}]{prochaska2015}
{Prochaska}, J.~X., {O'Meara}, J.~M., {Fumagalli}, M., {Bernstein}, R.~A., \&
  {Burles}, S.~M. 2015, \apjs, 221, 2

\bibitem[{{Qu} {et~al.}(2022){Qu}, {Chen}, {Rudie}, {Zahedy}, {Johnson},
  {Boettcher}, {Cantalupo}, {Chen}, {Cooksey}, {DePalma},
  {Faucher-Gigu{\`e}re}, {Rauch}, {Schaye}, \& {Simcoe}}]{qu2022}
{Qu}, Z., {Chen}, H.-W., {Rudie}, G.~C., {et~al.} 2022, \mnras, 516, 4882

\bibitem[{{Qu} {et~al.}(2023){Qu}, {Chen}, {Rudie}, {Johnson}, {Zahedy},
  {DePalma}, {Boettcher}, {Cantalupo}, {Chen}, {Cooksey},
  {Faucher-Gigu{\`e}re}, {Li}, {Lopez}, {Schaye}, \& {Simcoe}}]{qu2023}
---. 2023, \mnras, 524, 512

\bibitem[{{Qu} {et~al.}(2024){Qu}, {Chen}, {Johnson}, {Rudie}, {Zahedy},
  {DePalma}, {Schaye}, {Boettcher}, {Cantalupo}, {Chen}, {Faucher-Gigu{\`e}re},
  {Li}, {Mulchaey}, {Petitjean}, \& {Rafelski}}]{qu2024}
{Qu}, Z., {Chen}, H.-W., {Johnson}, S.~D., {et~al.} 2024, \apj, 968, 8

\bibitem[{{Rahmani} {et~al.}(2010){Rahmani}, {Srianand}, {Noterdaeme}, \&
  {Petitjean}}]{rahmani2010}
{Rahmani}, H., {Srianand}, R., {Noterdaeme}, P., \& {Petitjean}, P. 2010,
  \mnras, 409, L59

\bibitem[{{Rahmani} {et~al.}(2016){Rahmani}, {P{\'e}roux}, {Turnshek}, {Rao},
  {Quiret}, {Hamilton}, {Kulkarni}, {Monier}, \& {Zafar}}]{quiret2016}
{Rahmani}, H., {P{\'e}roux}, C., {Turnshek}, D.~A., {et~al.} 2016, \mnras, 463,
  980

\bibitem[{{Ramesh} \& {Nelson}(2024)}]{ramesh2024}
{Ramesh}, R., \& {Nelson}, D. 2024, \mnras, 528, 3320

\bibitem[{{Rauch}(1998)}]{rauch1998}
{Rauch}, M. 1998, \araa, 36, 267

\bibitem[{{Rhee} {et~al.}(2023){Rhee}, {Meyer}, {Popping}, {Bellstedt},
  {Driver}, {Robotham}, {Whiting}, {Baldry}, {Brough}, {Brown}, {Bunton},
  {Dodson}, {Holwerda}, {Hopkins}, {Koribalski}, {Lee-Waddell},
  {L{\'o}pez-S{\'a}nchez}, {Loveday}, {Mahony}, {Roychowdhury}, {Rozgonyi}, \&
  {Staveley-Smith}}]{rhee2023}
{Rhee}, J., {Meyer}, M., {Popping}, A., {et~al.} 2023, \mnras, 518, 4646

\bibitem[{{Robotham} {et~al.}(2020){Robotham}, {Bellstedt}, {Lagos}, {Thorne},
  {Davies}, {Driver}, \& {Bravo}}]{robotham2020}
{Robotham}, A.~S.~G., {Bellstedt}, S., {Lagos}, C. d.~P., {et~al.} 2020,
  \mnras, 495, 905

\bibitem[{{Roman-Duval} {et~al.}(2022){Roman-Duval}, {Jenkins}, {Tchernyshyov},
  {Clark}, {De Cia}, {Gordon}, {Hamanowicz}, {Lebouteiller}, {Rafelski},
  {Sandstrom}, {Werk}, \& {Yanchulova Merica-Jones}}]{roman-duval2022}
{Roman-Duval}, J., {Jenkins}, E.~B., {Tchernyshyov}, K., {et~al.} 2022, \apj,
  928, 90

\bibitem[{{Romano} {et~al.}(2010){Romano}, {Karakas}, {Tosi}, \&
  {Matteucci}}]{romano2010}
{Romano}, D., {Karakas}, A.~I., {Tosi}, M., \& {Matteucci}, F. 2010, \aap, 522,
  A32

\bibitem[{{Roy} {et~al.}(2013){Roy}, {Kanekar}, {Braun}, \&
  {Chengalur}}]{roy2013}
{Roy}, N., {Kanekar}, N., {Braun}, R., \& {Chengalur}, J.~N. 2013, \mnras, 436,
  2352

\bibitem[{{Saeedzadeh} {et~al.}(2023){Saeedzadeh}, {Jung}, {Rennehan}, {Babul},
  {Tremmel}, {Quinn}, {Shao}, {Sharma}, {Mayer}, {O'Sullivan}, \&
  {Loubser}}]{saeedzadeh2023}
{Saeedzadeh}, V., {Jung}, S.~L., {Rennehan}, D., {et~al.} 2023, \mnras, 525,
  5677

\bibitem[{{Salpeter}(1955)}]{salpeter1955}
{Salpeter}, E.~E. 1955, \apj, 121, 161

\bibitem[{{Sameer} {et~al.}(2024){Sameer}, {Lehner}, {Howk}, {Fox}, {O'Meara},
  \& {Oppenheimer}}]{sameer2024}
{Sameer}, {Lehner}, N., {Howk}, J.~C., {et~al.} 2024, \apj, 975, 264

\bibitem[{{Sana} {et~al.}(2012){Sana}, {de Mink}, {de Koter}, {Langer},
  {Evans}, {Gieles}, {Gosset}, {Izzard}, {Le Bouquin}, \&
  {Schneider}}]{sana2012}
{Sana}, H., {de Mink}, S.~E., {de Koter}, A., {et~al.} 2012, Science, 337, 444

\bibitem[{{Sana} {et~al.}(2014){Sana}, {Le Bouquin}, {Lacour}, {Berger},
  {Duvert}, {Gauchet}, {Norris}, {Olofsson}, {Pickel}, {Zins}, {Absil}, {de
  Koter}, {Kratter}, {Schnurr}, \& {Zinnecker}}]{sana2014}
{Sana}, H., {Le Bouquin}, J.~B., {Lacour}, S., {et~al.} 2014, \apjs, 215, 15

\bibitem[{{Sanders} {et~al.}(2023){Sanders}, {Shapley}, {Jones}, {Shivaei},
  {Popping}, {Reddy}, {Dav{\'e}}, {Price}, {Mobasher}, {Kriek}, {Coil}, \&
  {Siana}}]{sanders2023}
{Sanders}, R.~L., {Shapley}, A.~E., {Jones}, T., {et~al.} 2023, \apj, 942, 24

\bibitem[{{Saracco} {et~al.}(2023){Saracco}, {Barbera}, {De Propris},
  {Bevacqua}, {Marchesini}, {De Lucia}, {Fontanot}, {Hirschmann}, {Nonino},
  {Pasquali}, {Spiniello}, \& {Tortora}}]{saracco2023}
{Saracco}, P., {Barbera}, F.~L., {De Propris}, R., {et~al.} 2023, \mnras, 520,
  3027

\bibitem[{{Sarkar} {et~al.}(2022){Sarkar}, {Su}, {Truong}, {Randall},
  {Mernier}, {Gastaldello}, {Biffi}, \& {Kraft}}]{sarkar2022}
{Sarkar}, A., {Su}, Y., {Truong}, N., {et~al.} 2022, \mnras, 516, 3068

\bibitem[{{Savage} {et~al.}(2014){Savage}, {Kim}, {Wakker}, {Keeney}, {Shull},
  {Stocke}, \& {Green}}]{savage2014}
{Savage}, B.~D., {Kim}, T.~S., {Wakker}, B.~P., {et~al.} 2014, \apjs, 212, 8

\bibitem[{{Savage} {et~al.}(2005){Savage}, {Lehner}, {Wakker}, {Sembach}, \&
  {Tripp}}]{savage2005}
{Savage}, B.~D., {Lehner}, N., {Wakker}, B.~P., {Sembach}, K.~R., \& {Tripp},
  T.~M. 2005, \apj, 626, 776

\bibitem[{{Scannapieco} {et~al.}(2006){Scannapieco}, {Pichon}, {Aracil},
  {Petitjean}, {Thacker}, {Pogosyan}, {Bergeron}, \&
  {Couchman}}]{scannapieco2006}
{Scannapieco}, E., {Pichon}, C., {Aracil}, B., {et~al.} 2006, \mnras, 365, 615

\bibitem[{{Schaye}(2001)}]{schaye2001}
{Schaye}, J. 2001, \apj, 559, 507

\bibitem[{{Schaye} {et~al.}(2003){Schaye}, {Aguirre}, {Kim}, {Theuns}, {Rauch},
  \& {Sargent}}]{schaye2003}
{Schaye}, J., {Aguirre}, A., {Kim}, T.-S., {et~al.} 2003, \apj, 596, 768

\bibitem[{{Schaye} {et~al.}(2015){Schaye}, {Crain}, {Bower}, {Furlong},
  {Schaller}, {Theuns}, {Dalla Vecchia}, {Frenk}, {McCarthy}, {Helly},
  {Jenkins}, {Rosas-Guevara}, {White}, {Baes}, {Booth}, {Camps}, {Navarro},
  {Qu}, {Rahmati}, {Sawala}, {Thomas}, \& {Trayford}}]{schaye2015}
{Schaye}, J., {Crain}, R.~A., {Bower}, R.~G., {et~al.} 2015, \mnras, 446, 521

\bibitem[{{Schaye} {et~al.}(2023){Schaye}, {Kugel}, {Schaller}, {Helly},
  {Braspenning}, {Elbers}, {McCarthy}, {van Daalen}, {Vandenbroucke}, {Frenk},
  {Kwan}, {Salcido}, {Bah{\'e}}, {Borrow}, {Chaikin}, {Hahn}, {Hu{\v{s}}ko},
  {Jenkins}, {Lacey}, \& {Nobels}}]{schaye2023}
{Schaye}, J., {Kugel}, R., {Schaller}, M., {et~al.} 2023, \mnras, 526, 4978

\bibitem[{{Schechter}(1976)}]{schechter1976}
{Schechter}, P. 1976, \apj, 203, 297

\bibitem[{{Schimd} {et~al.}(2024){Schimd}, {Kraljic}, {Dav{\'e}}, \&
  {Pichon}}]{schimd2024}
{Schimd}, C., {Kraljic}, K., {Dav{\'e}}, R., \& {Pichon}, C. 2024, arXiv
  e-prints, arXiv:2406.04430

\bibitem[{{Sextl} {et~al.}(2023){Sextl}, {Kudritzki}, {Zahid}, \&
  {Ho}}]{sextl2023}
{Sextl}, E., {Kudritzki}, R.-P., {Zahid}, H.~J., \& {Ho}, I.~T. 2023, \apj,
  949, 60

\bibitem[{{Sharda} {et~al.}(2024){Sharda}, {Ginzburg}, {Krumholz}, {Forbes},
  {Wisnioski}, {Mingozzi}, {Zovaro}, \& {Dekel}}]{sharda2024}
{Sharda}, P., {Ginzburg}, O., {Krumholz}, M.~R., {et~al.} 2024, \mnras, 528,
  2232

\bibitem[{{Sharma} {et~al.}(2024){Sharma}, {Page}, {Symeonidis}, \&
  {Ferreras}}]{sharma2024}
{Sharma}, M., {Page}, M.~J., {Symeonidis}, M., \& {Ferreras}, I. 2024, \mnras,
  528, 1997

\bibitem[{{Shull} {et~al.}(2014){Shull}, {Danforth}, \& {Tilton}}]{shull2014}
{Shull}, J.~M., {Danforth}, C.~W., \& {Tilton}, E.~M. 2014, \apj, 796, 49

\bibitem[{{Shull} {et~al.}(2017){Shull}, {Danforth}, {Tilton}, {Moloney}, \&
  {Stevans}}]{shull2017}
{Shull}, J.~M., {Danforth}, C.~W., {Tilton}, E.~M., {Moloney}, J., \&
  {Stevans}, M.~L. 2017, \apj, 849, 106

\bibitem[{{Shull} {et~al.}(2012){Shull}, {Smith}, \& {Danforth}}]{shull2012}
{Shull}, J.~M., {Smith}, B.~D., \& {Danforth}, C.~W. 2012, \apj, 759, 23

\bibitem[{{Simcoe}(2011)}]{simcoe2011b}
{Simcoe}, R.~A. 2011, \apj, 738, 159

\bibitem[{{Simcoe} {et~al.}(2002){Simcoe}, {Sargent}, \& {Rauch}}]{simcoe2002}
{Simcoe}, R.~A., {Sargent}, W. L.~W., \& {Rauch}, M. 2002, \apj, 578, 737

\bibitem[{{Simcoe} {et~al.}(2004){Simcoe}, {Sargent}, \& {Rauch}}]{simcoe2004}
---. 2004, \apj, 606, 92

\bibitem[{{Slob} {et~al.}(2024){Slob}, {Kriek}, {Beverage}, {Suess}, {Barro},
  {Bezanson}, {Cheng}, {Conroy}, {de Graaff}, {F{\"o}rster Schreiber}, {Franx},
  {Lorenz}, {Mancera Pi{\~n}a}, {Marchesini}, {Muzzin}, {Newman}, {Price},
  {Shapley}, {Stefanon}, {van Dokkum}, \& {Weisz}}]{slob2024}
{Slob}, M., {Kriek}, M., {Beverage}, A.~G., {et~al.} 2024, arXiv e-prints,
  arXiv:2404.12432

\bibitem[{{Som} {et~al.}(2015){Som}, {Kulkarni}, {Meiring}, {York},
  {P{\'e}roux}, {Lauroesch}, {Aller}, \& {Khare}}]{som2015}
{Som}, D., {Kulkarni}, V.~P., {Meiring}, J., {et~al.} 2015, \apj, 806, 25

\bibitem[{{Somerville} \& {Dav{\'e}}(2015)}]{somerville2015b}
{Somerville}, R.~S., \& {Dav{\'e}}, R. 2015, \araa, 53, 51

\bibitem[{{Somerville} {et~al.}(2015){Somerville}, {Popping}, \&
  {Trager}}]{somerville2015}
{Somerville}, R.~S., {Popping}, G., \& {Trager}, S.~C. 2015, \mnras, 453, 4337

\bibitem[{{Songaila}(2005)}]{songaila2005}
{Songaila}, A. 2005, \aj, 130, 1996

\bibitem[{{Songaila} \& {Cowie}(2001)}]{songaila2001}
{Songaila}, A., \& {Cowie}, L. 2001, in The Extragalactic Infrared Background
  and its Cosmological Implications, ed. M.~{Harwit} \& M.~G. {Hauser}, Vol.
  204, 323

\bibitem[{{Stern} {et~al.}(2018){Stern}, {Faucher-Gigu{\`e}re}, {Hennawi},
  {Hafen}, {Johnson}, \& {Fielding}}]{stern2018}
{Stern}, J., {Faucher-Gigu{\`e}re}, C.-A., {Hennawi}, J.~F., {et~al.} 2018,
  \apj, 865, 91

\bibitem[{{Stocke} {et~al.}(2014){Stocke}, {Keeney}, {Danforth}, {Syphers},
  {Yamamoto}, {Shull}, {Green}, {Froning}, {Savage}, {Wakker}, {Kim},
  {Ryan-Weber}, \& {Kacprzak}}]{stocke2014}
{Stocke}, J.~T., {Keeney}, B.~A., {Danforth}, C.~W., {et~al.} 2014, \apj, 791,
  128

\bibitem[{{Strawn} {et~al.}(2024){Strawn}, {Roca-F{\`a}brega}, {Primack},
  {Kim}, {Genina}, {Hausammann}, {Kim}, {Lupi}, {Nagamine}, {Powell}, {Revaz},
  {Shimizu}, {Vel{\'a}zquez}, {Abel}, {Ceverino}, {Dong}, {Jung}, {Quinn},
  {Shin}, {Barrow}, {Dekel}, {Oh}, {Mandelker}, {Teyssier}, {Hummels}, {Maji},
  {Man}, {Mayerhofer}, \& {The Agora Collaboration}}]{strawn2024}
{Strawn}, C., {Roca-F{\`a}brega}, S., {Primack}, J.~R., {et~al.} 2024, \apj,
  962, 29

\bibitem[{{Tacconi} {et~al.}(2020){Tacconi}, {Genzel}, \&
  {Sternberg}}]{tacconi2020}
{Tacconi}, L.~J., {Genzel}, R., \& {Sternberg}, A. 2020, \araa, 58, 157

\bibitem[{{Tchernyshyov} {et~al.}(2022){Tchernyshyov}, {Werk}, {Wilde},
  {Prochaska}, {Tripp}, {Burchett}, {Bordoloi}, {Howk}, {Lehner}, {O'Meara},
  {Tejos}, \& {Tumlinson}}]{tchernyshyov2022}
{Tchernyshyov}, K., {Werk}, J.~K., {Wilde}, M.~C., {et~al.} 2022, \apj, 927,
  147

\bibitem[{{Tejos} {et~al.}(2012){Tejos}, {Morris}, {Crighton}, {Theuns},
  {Altay}, \& {Finn}}]{tejos2012}
{Tejos}, N., {Morris}, S.~L., {Crighton}, N. H.~M., {et~al.} 2012, \mnras, 425,
  245

\bibitem[{{Tejos} {et~al.}(2016){Tejos}, {Prochaska}, {Crighton}, {Morris},
  {Werk}, {Theuns}, {Padilla}, {Bielby}, \& {Finn}}]{tejos2016}
{Tejos}, N., {Prochaska}, J.~X., {Crighton}, N. H.~M., {et~al.} 2016, \mnras,
  455, 2662

\bibitem[{{Tilton} {et~al.}(2012){Tilton}, {Danforth}, {Shull}, \&
  {Ross}}]{tilton2012}
{Tilton}, E.~M., {Danforth}, C.~W., {Shull}, J.~M., \& {Ross}, T.~L. 2012,
  \apj, 759, 112

\bibitem[{{Tinsley} \& {Danly}(1980)}]{tinsley1980}
{Tinsley}, B.~M., \& {Danly}, L. 1980, \apj, 242, 435

\bibitem[{{Tripp} {et~al.}(2008){Tripp}, {Sembach}, {Bowen}, {Savage},
  {Jenkins}, {Lehner}, \& {Richter}}]{tripp2008}
{Tripp}, T.~M., {Sembach}, K.~R., {Bowen}, D.~V., {et~al.} 2008, \apjs, 177, 39

\bibitem[{{Tumlinson} {et~al.}(2017){Tumlinson}, {Peeples}, \&
  {Werk}}]{tumlinson2017}
{Tumlinson}, J., {Peeples}, M.~S., \& {Werk}, J.~K. 2017, \araa, 55, 389

\bibitem[{{Tumlinson} {et~al.}(2011{\natexlab{a}}){Tumlinson}, {Thom}, {Werk},
  {Prochaska}, {Tripp}, {Weinberg}, {Peeples}, {O'Meara}, {Oppenheimer},
  {Meiring}, {Katz}, {Dav{\'e}}, {Ford}, \& {Sembach}}]{tumlinson2011Sci}
{Tumlinson}, J., {Thom}, C., {Werk}, J.~K., {et~al.} 2011{\natexlab{a}},
  Science, 334, 948

\bibitem[{{Tumlinson} {et~al.}(2011{\natexlab{b}}){Tumlinson}, {Thom}, {Werk},
  {Prochaska}, {Tripp}, {Weinberg}, {Peeples}, {O'Meara}, {Oppenheimer},
  {Meiring}, {Katz}, {Dav{\'e}}, {Ford}, \& {Sembach}}]{tumlinson2011}
---. 2011{\natexlab{b}}, Science, 334, 948

\bibitem[{{Tumlinson} {et~al.}(2013){Tumlinson}, {Thom}, {Werk}, {Prochaska},
  {Tripp}, {Katz}, {Dav{\'e}}, {Oppenheimer}, {Meiring}, {Ford}, {O'Meara},
  {Peeples}, {Sembach}, \& {Weinberg}}]{tumlinson2013}
---. 2013, \apj, 777, 59

\bibitem[{{Tuominen} {et~al.}(2023){Tuominen}, {Nevalainen},
  {Hein{\"a}m{\"a}ki}, {Tempel}, {Wijers}, {Bonamente}, {Aragon-Calvo}, \&
  {Finoguenov}}]{tuominen2023}
{Tuominen}, T., {Nevalainen}, J., {Hein{\"a}m{\"a}ki}, P., {et~al.} 2023, \aap,
  671, A103

\bibitem[{{Venkatesan} {et~al.}(1999){Venkatesan}, {Olinto}, \&
  {Truran}}]{venkatesan1999}
{Venkatesan}, A., {Olinto}, A.~V., \& {Truran}, J.~W. 1999, \apj, 516, 863

\bibitem[{{Vincenzo} {et~al.}(2016){Vincenzo}, {Matteucci}, {Belfiore}, \&
  {Maiolino}}]{vincenzo2016}
{Vincenzo}, F., {Matteucci}, F., {Belfiore}, F., \& {Maiolino}, R. 2016,
  \mnras, 455, 4183

\bibitem[{{Vladilo} {et~al.}(2001){Vladilo}, {Centuri{\'o}n}, {Bonifacio}, \&
  {Howk}}]{vladilo2001}
{Vladilo}, G., {Centuri{\'o}n}, M., {Bonifacio}, P., \& {Howk}, J.~C. 2001,
  \apj, 557, 1007

\bibitem[{{Walter} {et~al.}(2020){Walter}, {Carilli}, {Neeleman}, {Decarli},
  {Popping}, {Somerville}, {Aravena}, {Bertoldi}, {Boogaard}, {Cox}, {da
  Cunha}, {Magnelli}, {Obreschkow}, {Riechers}, {Rix}, {Smail}, {Weiss},
  {Assef}, {Bauer}, {Bouwens}, {Contini}, {Cortes}, {Daddi}, {Diaz-Santos},
  {Gonz{\'a}lez-L{\'o}pez}, {Hennawi}, {Hodge}, {Inami}, {Ivison}, {Oesch},
  {Sargent}, {van der Werf}, {Wagg}, \& {Yung}}]{walter2020}
{Walter}, F., {Carilli}, C., {Neeleman}, M., {et~al.} 2020, \apj, 902, 111

\bibitem[{{Wang} \& {Wei}(2023)}]{wang2023}
{Wang}, B., \& {Wei}, J.-J. 2023, \apj, 944, 50

\bibitem[{{Weaver} {et~al.}(2023){Weaver}, {Davidzon}, {Toft}, {Ilbert},
  {McCracken}, {Gould}, {Jespersen}, {Steinhardt}, {Lagos}, {Capak}, {Casey},
  {Chartab}, {Faisst}, {Hayward}, {Kartaltepe}, {Kauffmann}, {Koekemoer},
  {Kokorev}, {Laigle}, {Liu}, {Long}, {Magdis}, {McPartland}, {Milvang-Jensen},
  {Mobasher}, {Moneti}, {Peng}, {Sanders}, {Shuntov}, {Sneppen}, {Valentino},
  {Zalesky}, \& {Zamorani}}]{weaver2022}
{Weaver}, J.~R., {Davidzon}, I., {Toft}, S., {et~al.} 2023, \aap, 677, A184

\bibitem[{{Weng} {et~al.}(2024){Weng}, {P{\'e}roux}, {Ramesh}, {Nelson},
  {Sadler}, {Zwaan}, {Bollo}, \& {Casavecchia}}]{weng2024}
{Weng}, S., {P{\'e}roux}, C., {Ramesh}, R., {et~al.} 2024, \mnras, 527, 3494

\bibitem[{{Weng} {et~al.}(2023{\natexlab{a}}){Weng}, {P{\'e}roux}, {Karki},
  {Augustin}, {Kulkarni}, {Szakacs}, {Zwaan}, {Klitsch}, {Hamanowicz},
  {Sadler}, {Biggs}, {Fresco}, {Hayes}, {Howk}, {Kacprzak}, {Kuntschner},
  {Nelson}, \& {Pettini}}]{weng2023}
{Weng}, S., {P{\'e}roux}, C., {Karki}, A., {et~al.} 2023{\natexlab{a}}, \mnras,
  519, 931

\bibitem[{{Weng} {et~al.}(2023{\natexlab{b}}){Weng}, {P{\'e}roux}, {Karki},
  {Augustin}, {Kulkarni}, {Hamanowicz}, {Zwaan}, {Sadler}, {Nelson}, {Hayes},
  {Kacprzak}, {Fox}, {Bollo}, {Casavecchia}, \& {Szakacs}}]{weng2023b}
---. 2023{\natexlab{b}}, \mnras, 523, 676

\bibitem[{{Werk} {et~al.}(2014){Werk}, {Prochaska}, {Tumlinson}, {Peeples},
  {Tripp}, {Fox}, {Lehner}, {Thom}, {O'Meara}, {Ford}, {Bordoloi}, {Katz},
  {Tejos}, {Oppenheimer}, {Dav{\'e}}, \& {Weinberg}}]{werk2014}
{Werk}, J.~K., {Prochaska}, J.~X., {Tumlinson}, J., {et~al.} 2014, \apj, 792, 8

\bibitem[{{Werk} {et~al.}(2016){Werk}, {Prochaska}, {Cantalupo}, {Fox},
  {Oppenheimer}, {Tumlinson}, {Tripp}, {Lehner}, \& {McQuinn}}]{werk2016}
{Werk}, J.~K., {Prochaska}, J.~X., {Cantalupo}, S., {et~al.} 2016, \apj, 833,
  54

\bibitem[{{Werner} {et~al.}(2008){Werner}, {Durret}, {Ohashi}, {Schindler}, \&
  {Wiersma}}]{werner2008}
{Werner}, N., {Durret}, F., {Ohashi}, T., {Schindler}, S., \& {Wiersma},
  R.~P.~C. 2008, \ssr, 134, 337

\bibitem[{{Wetzel} {et~al.}(2023){Wetzel}, {Hayward}, {Sanderson}, {Ma},
  {Angl{\'e}s-Alc{\'a}zar}, {Feldmann}, {Chan}, {El-Badry}, {Wheeler},
  {Garrison-Kimmel}, {Nikakhtar}, {Panithanpaisal}, {Arora}, {Gurvich},
  {Samuel}, {Sameie}, {Pandya}, {Hafen}, {Hummels}, {Loebman},
  {Boylan-Kolchin}, {Bullock}, {Faucher-Gigu{\`e}re}, {Kere{\v{s}}},
  {Quataert}, \& {Hopkins}}]{wetzel2023}
{Wetzel}, A., {Hayward}, C.~C., {Sanderson}, R.~E., {et~al.} 2023, \apjs, 265,
  44

\bibitem[{{Whitaker} {et~al.}(2017){Whitaker}, {Pope}, {Cybulski}, {Casey},
  {Popping}, \& {Yun}}]{whitaker2017}
{Whitaker}, K.~E., {Pope}, A., {Cybulski}, R., {et~al.} 2017, \apj, 850, 208

\bibitem[{{Wilkins} {et~al.}(2019){Wilkins}, {Lovell}, \&
  {Stanway}}]{wilkins2019}
{Wilkins}, S.~M., {Lovell}, C.~C., \& {Stanway}, E.~R. 2019, \mnras, 490, 5359

\bibitem[{{Wolfe}(2005)}]{wolfe2005a}
{Wolfe}, A.~M. 2005, Highlights of Astronomy, 13, 572

\bibitem[{{Wolfe} {et~al.}(2005){Wolfe}, {Gawiser}, \&
  {Prochaska}}]{wolfe2005b}
{Wolfe}, A.~M., {Gawiser}, E., \& {Prochaska}, J.~X. 2005, \araa, 43, 861

\bibitem[{{Wolfe} {et~al.}(1986){Wolfe}, {Turnshek}, {Smith}, \&
  {Cohen}}]{wolfe1986}
{Wolfe}, A.~M., {Turnshek}, D.~A., {Smith}, H.~E., \& {Cohen}, R.~D. 1986,
  \apjs, 61, 249

\bibitem[{{Woosley} \& {Weaver}(1995)}]{woosley1995}
{Woosley}, S.~E., \& {Weaver}, T.~A. 1995, \apjs, 101, 181

\bibitem[{{Wotta} {et~al.}(2019){Wotta}, {Lehner}, {Howk}, {O'Meara},
  {Oppenheimer}, \& {Cooksey}}]{wotta2019}
{Wotta}, C.~B., {Lehner}, N., {Howk}, J.~C., {et~al.} 2019, \apj, 872, 81

\bibitem[{{Wotta} {et~al.}(2016){Wotta}, {Lehner}, {Howk}, {O'Meara}, \&
  {Prochaska}}]{wotta2016}
{Wotta}, C.~B., {Lehner}, N., {Howk}, J.~C., {O'Meara}, J.~M., \& {Prochaska},
  J.~X. 2016, \apj, 831, 95

\bibitem[{{Wright} {et~al.}(2018){Wright}, {Driver}, \&
  {Robotham}}]{wright2018}
{Wright}, A.~H., {Driver}, S.~P., \& {Robotham}, A.~S.~G. 2018, \mnras, 480,
  3491

\bibitem[{{Yates} {et~al.}(2021){Yates}, {P{\'e}roux}, \& {Nelson}}]{yates2021}
{Yates}, R.~M., {P{\'e}roux}, C., \& {Nelson}, D. 2021, \mnras, 508, 3535

\bibitem[{{Yates} {et~al.}(2017){Yates}, {Thomas}, \& {Henriques}}]{yates2017}
{Yates}, R.~M., {Thomas}, P.~A., \& {Henriques}, B. M.~B. 2017, \mnras, 464,
  3169

\bibitem[{{Zahid} {et~al.}(2017){Zahid}, {Kudritzki}, {Conroy}, {Andrews}, \&
  {Ho}}]{zahid2017}
{Zahid}, H.~J., {Kudritzki}, R.-P., {Conroy}, C., {Andrews}, B., \& {Ho}, I.~T.
  2017, \apj, 847, 18

\bibitem[{{Zhang} {et~al.}(2024{\natexlab{a}}){Zhang}, {Comparat}, {Ponti},
  {Meloni}, {Nandra}, {Haberl}, {Locatelli}, {Zhang}, {Sanders}, {Zheng},
  {Liu}, {Popesso}, {Liu}, {Truong}, {Pillepich}, {Predehl}, \&
  {Salvato}}]{zhang2024a}
{Zhang}, Y., {Comparat}, J., {Ponti}, G., {et~al.} 2024{\natexlab{a}}, arXiv
  e-prints, arXiv:2401.17308

\bibitem[{{Zhang} {et~al.}(2024{\natexlab{b}}){Zhang}, {Comparat}, {Ponti},
  {Meloni}, {Nandra}, {Haberl}, {Truong}, {Pillepich}, {Locatelli}, {Zhang},
  {Sanders}, {Zheng}, {Liu}, {Popesso}, {Liu}, {Predehl}, \&
  {Salvato}}]{zhang2024b}
---. 2024{\natexlab{b}}, arXiv e-prints, arXiv:2401.17309

\bibitem[{{Zwaan} {et~al.}(2005){Zwaan}, {van der Hulst}, {Briggs},
  {Verheijen}, \& {Ryan-Weber}}]{zwaan2005}
{Zwaan}, M.~A., {van der Hulst}, J.~M., {Briggs}, F.~H., {Verheijen}, M.~A.~W.,
  \& {Ryan-Weber}, E.~V. 2005, \mnras, 364, 1467

\end{thebibliography}

\end{document}